\documentclass[useAMS,usenatbib]{mn2e}

\usepackage{graphicx,amssymb,verbatim,times,color,ulem,listings,epsfig,url}
\newcommand\lya{Ly$\alpha$}

\newcommand\apj{ApJ}
\newcommand\aj{AJ}
\newcommand\apjl{ApJL}
\newcommand\apjs{ApJS}
\newcommand\aap{A\&A}
\newcommand\mnras{MNRAS}
\newcommand\fcp{FCPh}
\newcommand\pasp{PASP}

\newcommand\nat{Nature}
\newcommand\aaps{A\&AS}
\newcommand\memsai{Memorie della Societ\'a Astronomica Italiana}
\newcommand\procspie{SPIE}

\title[The Millennium Run Observatory: First Light]{The Millennium Run Observatory: First Light}

\author[Overzier et al.]{R. Overzier$^{1,\dagger}$\thanks{E-mail:
    overzier@astro.as.utexas.edu (RO)}, G.  Lemson$^2$,
  R.E. Angulo$^{3,2}$, E. Bertin$^4$, J. Blaizot$^5$,
  B.M.B. Henriques$^2$,
  \newauthor G.-D. Marleau$^{6,7}$, S.D.M. White$^2$\\
  $^1$Department of Astronomy, The University of Texas at Austin, 2515 Speedway, Stop C1400, Austin, TX 78712-1205, USA\\
  $^2$Max-Planck-Institut f\"ur Astrophysik, D-85748 Garching, Germany\\
  $^3$Kavli Institute for Particle Astrophysics and Cosmology, Stanford University, 452 Lomita Mall, Stanford, CA 94305-4085, USA\\
  $^4$Institut d'Astrophysique de Paris, 98 bis Boulevard Arago, 75014, Paris, France\\
  $^5$Universit\'e de Lyon, Lyon F-69003, France; Universit\'e Lyon 1,
  Observatoire de Lyon, 9 Avenue Charles Andr\'e, Saint-Genis Laval F--69230, France;\\ 
CNRS, UMR 5574, Centre de Recherche Astrophysique de Lyon, France; Ecole Normale Sup\'erieure de Lyon, Lyon F-69007, France\\
  $^6$Max-Planck Institut f\"ur Astronomie, K\"onigstuhl 17, D-69117 Heidelberg, Germany\\
  $^7$Department of Physics, McGill University, 3600 Rue University, Montr\'eal, QC H3A 2T8, Canada\\
  $^\dagger$ UT Prize Fellow}

\begin{document}

\date{}

\pagerange{\pageref{firstpage}--\pageref{lastpage}} \pubyear{2012}

\maketitle

\label{firstpage}

\begin{abstract} Simulations of galaxy evolution aim to capture our
  current understanding as well as to make predictions for testing by
  future experiments.  Simulations and observations are often compared
  in an indirect fashion: physical quantities are estimated from the
  observational data and compared to models.  However, many
  applications can benefit from a more direct approach, where the
  observing process is also simulated, so that the models are seen
  fully from the observer's perspective. To facilitate this, we have
  developed the Millennium Run Observatory (MRObs), a theoretical
  virtual observatory which uses virtual telescopes to `observe'
  semi-analytic galaxy formation simulations based on the suite of
  Millennium Run (MR) dark matter simulations. The MRObs produces data
  that can be processed and analyzed using the standard observational
  software packages developed for real observations. At present, we
  produce images in forty filters covering the rest-frame UV to
  infrared for two stellar population synthesis models, for three
  different models of absorption by the intergalactic medium, and in
  two cosmologies (WMAP1 and 7). Galaxy distributions for a large
  number of mock lightcones can be `observed' using models of major
  ground- and space-based telescopes.  The data include lightcone
  catalogues linked to structural properties of galaxies,
  pre-observation model images, mock telescope images, and Source
  Extractor products that can all be traced back to the higher-level
  dark matter, semi-analytic galaxy, and lightcone catalogues
  available in the MR database.  Here, we describe our methods and
  announce a first public release of simulated observations that
  emulate a large number of extra-galactic surveys (e.g. SDSS,
    CFHT-LS, GOODS, GOODS/ERS, CANDELS, and HUDF). The MRObs browser,
  an online tool, further facilitates exploration of the simulated
  data. We demonstrate the benefits of a direct approach through a
  number of example applications: (1) deep galaxy number counts in the CANDELS survey; (2) observed properties of galaxy
  clusters; (3) structural parameters of galaxies; and (4)
  identification of drop-out galaxies. The MRObs enhances the range of
  questions that can be asked of semi-analytic models, allowing
  observers and theorists to work toward each other with virtually
  complete freedom of where to meet.  \end{abstract}


\section{Introduction}

\noindent
Understanding formation and evolution of galaxies is one of the main
goals of extra-galactic astrophysics. This study is approached from
two sides, an observational one and a theoretical one. On the one
hand, observations become more and more detailed, producing ever
larger images and catalogues that need to be analyzed.  On the other
hand, theoretical research produces ever more refined models
describing the formation and evolutionary processes in ever greater
detail, often using sophisticated cosmological computer simulations
that create enormous, physically motivated data sets. The increasing
specialization and technical sophistication required means that it
becomes a problem to successfully match these two approaches as few
scientists are familiar with all the details on both the observational
and the theoretical side. For example, it is often difficult for
non-experts to understand detailed galaxy formation models or to
predict how model parameter changes affect the predictions. Likewise,
theorists are often unfamiliar with the extensive processing and the
inverse methods that need to be applied to observations in order to
derive physical properties that can be matched to the model
predictions.

From an observational perspective, the Sloan Digital Sky Survey (SDSS)
consortium played a pivotal role in opening up the results of one of
the most sophisticated observational programs ever performed to the
community. Through a public database of raw measurements, processed
results, and `value-added' products, a great many hurdles were
removed for using the results of the survey.  From a theoretical
perspective, the Millennium Run Database (MRDB) was the first to make
the results of large scale cosmological simulations widely accessible
to a broad-based audience. Analogous to the SDSS data, the richness of
the theoretical data sets available in the MRDB has allowed a wide
variety of scientific queries to be performed.

The comparison between cosmological model predictions and observations
has historically been performed mostly in one direction only: physical
quantities estimated from observations are compared with theoretical
predictions. The latter are not affected by the issues that
affect the observations, such as incompleteness, contamination, cosmic
variance, finite signal-to-noise, or limited resolution.  These are
typically assumed to be corrected for in full during the processing of
the observations, although some authors have included some of these
effects in a number of ways in order to be able to more realistically compare
data and predictions \citep[e.g.][]{baugh98,blaizot04,marinoni05,pollo06,guo09,overzier09a,delatorre11,cucciati12,pforr12}.  

We propose that the comparison between models and observations
should also be performed in the opposite direction.  The strength of
this method lies in the fact that one can never be sure to extract the
truth out of observations, but one will always know what the true
answer is in a set of synthetic observations based on the
simulations. Synthetic observations like we propose would also
  allow one to explore the uniqueness of possible solutions that are
  found, as different models or different parameter sets may produce
  indistinguishable synthetic observations. In this paper we present
an extension of the MR cosmological simulations project,
which we will henceforth refer to as the {\it Millennium Run
  Observatory} (MRObs). It aims to bridge the gap between the two
approaches by making the final step from realistic simulations to the
observational plane. MRObs consists of a fully connected set of
synthetic data products combined into a unique online framework that
ranges from the most fundamental simulations to realistic, synthetic
observations.

With the introduction of `lightcones', the comparison between
simulations and observations has been greatly enhanced. This technique
allows one to project the galaxy distribution predicted for a set of
discrete simulation snapshots along a virtual observer's line of
sight, mimicking the main geometric and photometric effects present in
deep galaxy surveys
\citep{davis82,davis85,cole98,diaferio99,blaizot05,kitzbichler07}.  However,
even the lightcone approach to model-data comparisons is still very
much idealized.  To illustrate this, let us consider a typical
observational scenario of determining the stellar mass function of
high redshift galaxies in a multi-wavelength imaging survey. Such an
analysis typically begins with the extraction of sources and their
photometric properties across a set of calibrated and registered
filter images. Due to missed light, it is often necessary to make
corrections to the measured magnitudes. Then, photometric redshifts
and physical parameters of
the galaxies (e.g., stellar mass, age, SFR) are estimated by fitting
the photometry with a set of template 
galaxy spectra. It is important to
note that the results often depend on, e.g., the source detection and
photometry method, the choice of template spectra, and the fitting
method. In order to calculate the number of galaxies detected in
different stellar mass bins over different redshift intervals, it is
often required to calculate the `effective volume' of the
survey. The latter is an estimate of the completeness of the sample,
and usually depends on redshift, limiting magnitude, galaxy colour or
size in complicated ways. This last step can be performed by
estimating the probability of recovering certain sources at a given
survey depth. Such estimates typically depend on the true source
population which is a priori unknown. At the end of the process, the
stellar mass function estimate is used for comparison with other
observational studies, or to constrain certain theoretical models or
simulations of galaxy formation.  It should be clear from the process
outlined above that a great number of non-trivial steps need to be
performed before any comparison with theory can be made. How better
could we test all these steps than by processing the output from the
simulations, for which all quantities are exactly known, through the
same kind of data analysis pipeline as the real observations?

\subsection{Goals of the MRObs}

\noindent
We will take the process of simulating the galaxy population for
comparison with observations into largely unexplored
territory by simulating the observational process applied to the MR simulations.  
The main aims of the MRObs are as follows:\\

\noindent$\bullet$ Extend the MR project approach by producing data
    products most directly corresponding to observations, namely
    synthetic images and extracted source catalogs\\

\noindent$\bullet$ Aid theorists in testing analytical models to observations\\

\noindent$\bullet$ Aid observers in making detailed predictions for observations
    and better analyses of observational data\\

\noindent$\bullet$ Allow the community to subject the models to new kinds of tests\\

\noindent$\bullet$ Allow observers and theorists to work toward each other from
    either direction with the freedom of where to meet\\

\noindent$\bullet$ Allow detailed comparisons with synthetic observations produced
    by other groups performing cosmological simulations\\

\noindent$\bullet$ Allow calibration of observational analysis
    methods by making available synthetic data for which the entire
    underlying `reality' is known\\

\noindent$\bullet$ Extend the realism with which semi-analytic
    models can address questions such as what is the probability that a $z\sim10$ galaxy
    will be detected within a particular observational data set?\\

\noindent$\bullet$ Provide a framework for future virtual theoretical observatories\\

\subsection{Connection to previous work}

\noindent
Only recently have simulations become sophisticated enough to allow
realistic visualizations of the galaxy population on a cosmological
scale. In order to illustrate the particular place that the MRObs
occupies within this simulations landscape, we give a short overview
of related work in the literature. Astronomical image simulation
software has been developed and used previously, mostly to aid in the
development of data processing pipelines for new telescopes and
instruments, for proposal planning, or for testing the accuracy of
specific measurement tools \citep[e.g.][]{bertin09,dobke10}. Within
the gravitational lensing community, it has been standard practice to
use simulated data to assess the accuracy of cosmic shear measurements
\citep{erben01,heymans06,forero07}. Simple galaxy evolution models
have been coupled to image simulators to compare with observations
\citep[e.g.][]{bouwens99,bouwens06}, and mock telescope data based on
semi-analytic models (SAMs) are also currently being used to investigate the
significant data and science challenges posed by future surveys \citep[e.g., with the Large Synoptic
Survey Telescope (LSST);][]{connolly10,gibson11}.  The detailed
morphological and kinematical structures of gas and stars have been
modeled using high-resolution, hydrodynamical simulations (of dark
matter, gas and stars), coupled with radiative transfer models that
allow one to study the effects of dust and orientation as a function
of wavelength
\citep[e.g.][]{jonsson06,jonsson10,robertson08,wuyts09,chilingarian10,lotz08,lotz10}.
However, hydrodynamical simulations of sufficient resolution are currently too
small to construct lightcones on cosmological scales. Also, unlike
SAMs, it is a much more time-consuming process to match $N$-body hydro
simulations to observations after each change in the sub-grid physics
modeling. As a result current hydro simulations of the galaxy
population are substantially further from the observations than
semi-analytical models.

\citet{blaizot05} pioneered in the production of realistic artificial
telescope data based on lightcones extracted from their semi-analytic
model. That paper already laid out most of the workflow that we use
here (see Fig. \ref{fig:overview}): dark matter particle simulations
are used to construct halo merger trees on which a semi-analytic model
is run. The output from the SAM is used to construct galaxy lightcones
that are used as input for artificial telescope image
simulations. Galaxies are extracted from the artificial images using
standard observational tools \citep[e.g. SExtractor;][]{bertin96}, and the resulting
galaxy catalogs are compared to the original simulations at different
levels, or to actual observations. Unfortunately, however, the methods
of \citet{blaizot05} were never employed on a large scale, and in
subsequent years the comparison between SAMs and real observations has
been mostly performed at the lightcone level or even at the snapshot
level, thereby sidestepping many of the details involved in analyzing
real telescope data that observers typically have to go through. As we
shall show, however, numerous problems in the field of galaxy
evolution could benefit from a simulation that accounts for the entire
observational process. This leads to new insights involving details
that are missed by higher-level comparisons between data and
simulations.  By expanding on the basic ideas of \citet{blaizot05},
the MRObs aims at making this possible.

\subsection{Why the Millennium Simulations?}

\noindent
Although in this paper we lay out the motivation and framework for
producing synthetic data from cosmological simulations in general, the
MRObs is based around the suite of MR simulations.  Through the
combination of simulations volume and particle resolution, an active
development of semi-analytic models, and an online database providing
access to numerous data products, the MR is ideally suited for most of
our purposes, as follows.

(1) Volume and resolution: The MR has an almost ideal combination of
volume and particle mass resolution suitable for a wide range of
applications. The resolution\footnote{Full convergence between the MR
  and the much higher resolution MR-II simulation is near $10^{11}$
  $M_\odot$.} is sufficient to identify the $\gtrsim5\times10^{10}$ $M_\odot$
halos believed to host faint galaxies at very high redshifts 
\citep[][]{ouchi05,overzier06}, while probing significantly down the
stellar mass function with good statistics at lower redshifts. The
volume is large enough to probe a very wide range of environments. The
MR contains about 3,000 cluster-sized objects at $z=0$, of which about
25 are of the Coma-type (i.e., more massive than $10^{15}$
$M_\odot$). The formation of all these systems can be traced back to
very high redshift for detailed studies of cluster formation
\citep{overzier09a}. The large volume is also crucial for constructing
synthetic galaxy surveys covering many square degrees without
significant replications
\citep{kitzbichler07,guo09,overzier09a,henriques12}.

More recent dark matter simulations have been performed. The MultiDark
simulations span an $8\times$ larger volume but with a $10\times$
lower mass resolution compared to the MR \citep{prada11}. The Bolshoi
simulations have a $10\times$ higher mass resolution, but are
$8\times$ smaller \citep{klypin11}. Neither simulation has yet
released semi-analytic galaxy catalogs that can be used to compare
with actual observations. The somewhat limited mass resolution of the
MR has recently been extended by two orders of magnitude through the
MR-II simulation \citep{boylan09}. This simulation is extremely useful
for further improving the semi-analytic model that can then be
re-applied to the original MR simulation \citep{guo11}. The somewhat
limited volume of the MR has also recently been extended by two orders
of magnitude through the Millennium XXL (MXXL) simulation
\citep{angulo12}, useful for studies of the rarest, most massive
objects. However, for the generation of mock lightcones, the MR is
currently still our preferred simulation ($125\times$ larger
volume compared to the MR-II and $7\times$ higher resolution
compared to the MXXL). 

Recently, it has become possible to re-cast the suite of MR simulation
results in more updated cosmologies relative to WMAP1 thanks to the
re-scaling technique of \citet[][see \S\ref{sec:cosmology}]{angulo10}.

(2) Semi-analytic models: As we will show, the \citet{guo11}
semi-analytic model applied to the MR is key to producing our synthetic
observations. This model gives detailed predictions for the evolving
sizes and spin axes of the stellar mass in disks and/or bulges that
are crucial for calculating angular sizes, bulge-to-disk ratios,
inclinations and position angles.

(3) MR database: The dark matter and galaxy catalogs of the MR
project and related simulations have been made widely accessible to
the community through the MRDB \citep{lemson06}. Interested users can query the data in this database
through various online services using standard Structured Query
Language (SQL). Regular updates to the MRDB holdings provide public
access to the latest model results, ensuring that anyone can analyze
the MR data and use its results in their publications. We have now
added to this system our synthetic imaging data and extracted source
catalogs that can be cross-correlated with the underlying simulations
data in the MRDB.

In summary, despite the age of the original MR \citep{springel05}, more recent dark
matter simulations do not yet provide equivalent data sets or the
infrastructure required for developing a facility such as the MRObs.

\subsection{This paper}

In this paper, the first in a series comparing theory and observations
in the observational plane, we lay a framework for producing synthetic
data from cosmological simulations, describe our main methods for
future reference, present a number of user examples, and announce the
public release of a large number of simulated surveys (synthetic
images and catalogs). We also present various new online services that
allow one to interact with the synthetic observations and the
underlying lightcones, semi-analytic galaxy and dark matter catalogs
in the MRDB.  The structure of this paper is as follows. In \S2 we
will present a concise overview of the MRObs and describe in detail
all the steps that are needed in order to go from a pure dark matter
simulation and semi-analytic galaxy catalog to producing realistic
synthetic observations. In \S3 we present a detailed simulations
example focusing on our synthetic images produced for the on-going
CANDELS HST program. In \S4 we illustrate the new types of questions
that can be asked of the MRObs through a number of examples related to
galaxy and galaxy cluster evolution. In \S5 we present the public data
release and the interactive online tools we have developed. In \S6 and
\S7, respectively, we present a brief outlook to future developments
related to the MRObs project and summarize our results. 

\begin{figure*}
\begin{center}
\includegraphics[width=\textwidth]{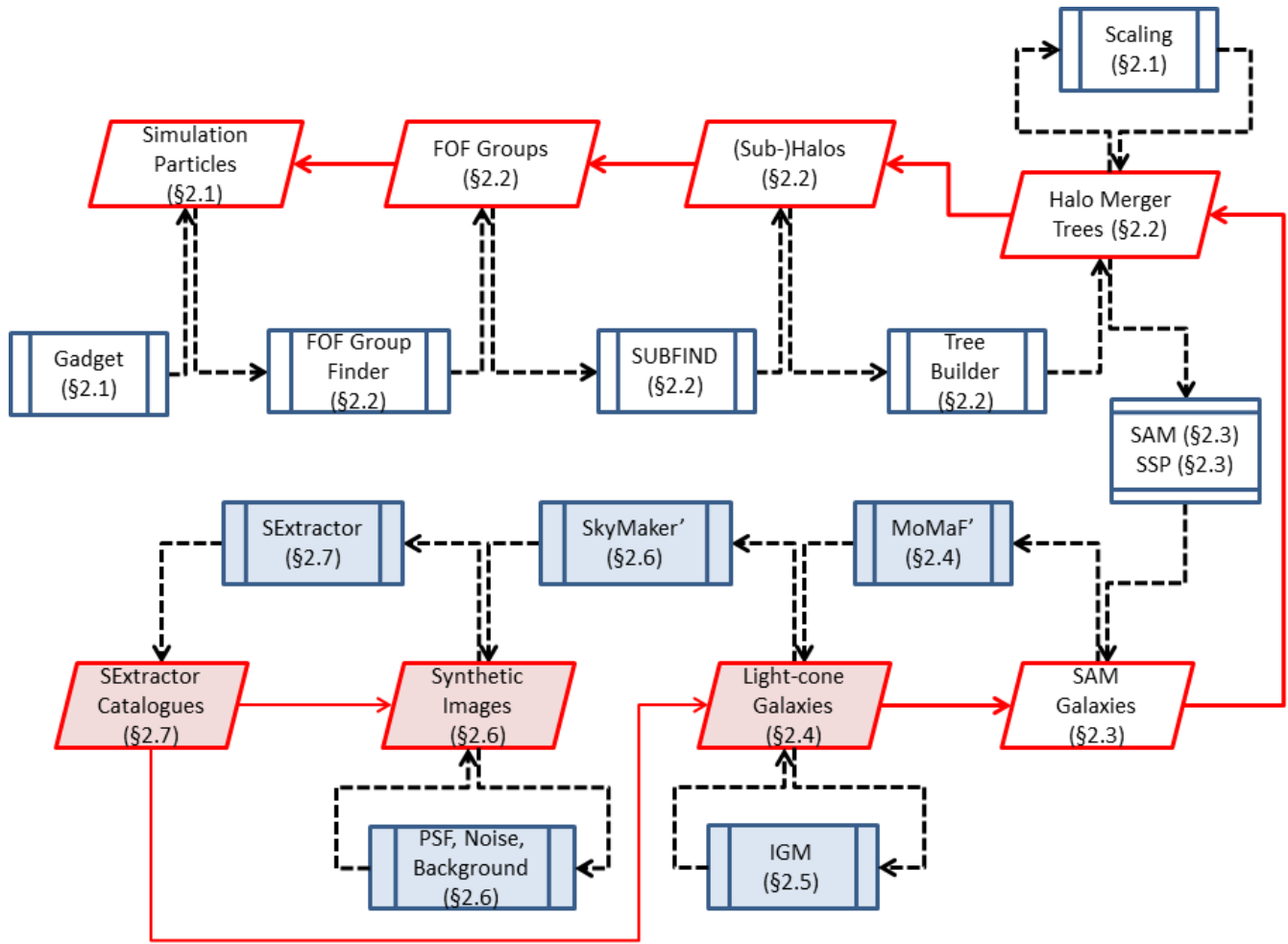}
\end{center}
\caption{\label{fig:overview}Schematic overview of the Millennium Run
  Observatory workflow. The blue rectangles indicate an action, while
  the red tilted rectangles represent data products that in each step
  can be linked to products elsewhere along the chain. Thick arrows
  indicate that there are direct links between data products, while
  thin arrows indicate that indirect links can be made using
  cross-correlation. Dashed lines link products to actions from which
  they result, or by which they are used. Shaded rectangles indicate
  products or actions that have been updated or are introduced in this
  paper for the first time. The workflow starts with an $N$-body
  dark-matter-only simulation (see \S\ref{sec:dm}). Dark matter
  particles are grouped together using a friends-of-friends group
  finder and decomposed into halos and sub-halos using a halo-finder
  algorithm (see \S\ref{sec:halos}). This results in positions,
  velocities, spin vectors and masses of dark matter halos in an
  evolving $\Lambda$CDM universe. A dark matter halo merger tree is
  constructed and stored in a database. Optionally, a scaling of the
  cosmological parameters can be applied to the halo merger tree (see
  \S\ref{sec:cosmology}). The merger tree forms the backbone for a
  semi-analytic model of galaxy formation that tracks the growth of
  galaxies inside halos based on simple recipes for, e.g., gas
  cooling, star formation, supernova and AGN heating, gas stripping,
  and merging between galaxies (see \S\ref{sec:sam}). In each time
  step or snapshot, the resulting physical properties of each galaxy
  in the semi-analytic galaxy population are used to select
  appropriate stellar population templates from a library of spectral
  energy distributions to model the rest-frame, dust-attenuated
  spectra or colours of each galaxy (see \S\ref{sec:ssps}). A pencil
  beam-shaped `lightcone' is carved out through the simulation
  volume using a modified version of the code MoMaF, selecting only
  galaxies from those snapshots that correspond to the cosmic time at
  the co-moving distances along the line-of-sight in the observer's
  frame of reference (see \S\ref{sec:lightcones}). Multi-band apparent
  magnitudes are calculated and corrected for absorption by neutral
  hydrogen in the intergalactic medium (see \S\ref{sec:igm}). This
  lightcone is then projected onto a plane giving virtual sky
  positions for each galaxy in terms of right ascension and
  declination. The positions, shapes, sizes and observed-frame
  apparent magnitudes are used to build a `perfect' pre-observation image of the sky
  using a modified version of SkyMaker (see \S\ref{sec:sims}). The
  perfect image is fed into the telescope simulator that applies a
  detector model (pixel scale, readout noise, dark current,
  sensitivity, gain), a sky background model, PSF convolution, and
  Poissonian object and sky noise for a particular survey description
  (see \S\ref{sec:od}). The MRObs produces a realistic, synthetic
  telescope image in \texttt{.fits} format for further scientific
  analysis. Source Extractor is run on the simulated image and the
  output catalogs can be analyzed analogous to the catalogs
  constructed from real observations (see \S\ref{sec:sextractor}).  }
\end{figure*}

\section{Structure of the Millennium Run Observatory}

The MRObs makes available a fully interconnected set of data products
covering the entire chain from dark matter simulations to synthetic
observations and extracted data.  In the MRObs, each subsequent step
uses data products produced by previous steps, and almost all the data
products are available for interrogation and public download for
further analysis. A schematic overview of this process is given in the
workflow diagram in Fig. \ref{fig:overview}, where rectangles indicate
an action and tilted rectangles represent data products that in each
step can be linked to products
elsewhere along the chain. The main steps are:\\

1. Dark matter particle simulation (DM density fields)\\

2. Identifying of friends-of-friends (FOF) groups\\

3. Identifying (sub-)halos\\

4. Constructing halo merger trees\\

5. Applying semi-analytic galaxy models\\

6. Observing galaxies on a synthetic light-cone\\

7. Producing synthetic telescope images\\

8. Extracting sources from synthetic images\\

In this section we will describe each of the steps in more detail, focusing on the
newly developed components that are most essential to bridge the gap to real
observations (steps 6--8), and refer to other work for 
the components described in detail elsewhere (steps 1--5). 

\subsection{The Millennium suite of dark matter simulations}
\label{sec:dm}

The evolution of the dark matter distribution with time is believed to
be mainly driven by the initial matter power spectrum, gravity, and
the expansion rate of the universe, and can be taken either from
direct $N$-body simulations
\citep[e.g.][]{davis85,jenkins98,springel05}, or from
(semi-)analytically constructed dark matter halo trees
\citep[e.g.][]{press74,kauffmannwhite93,laceycole1994,somervillekolatt99,sheth01,neistein08}.
In the suite of cosmological simulations centered around the
MR project, the dark matter simulation was performed with
versions of the cosmological simulation code Gadget
\citep{springel05}. The suite of simulations consist of (1) a 2160$^3$
particles simulation with particle mass $8.6\times10^8$ $h^{-1}$
$M_\odot$ and periodic box length of 500 $h^{-1}$ Mpc \citep[][]{springel05}, (2) a 2160$^3$ particles
simulation with mass $6.9\times10^6$ $h^{-1}$ $M_\odot$ and periodic
box length of 100 $h^{-1}$ Mpc \citep[the Millennium-II
(MS-II);][]{boylan09}, and (3) a 6720$^3$ particles simulation with
mass $6.2\times10^9$ $h^{-1}$ $M_\odot$ and periodic box length of 3
$h^{-1}$ Gpc \citep[the Millennium-XXL (MXXL);][]{angulo12}. All
simulations follow the gravitational growth as traced by these
particles from $z=127$ to 0 in a $\Lambda$CDM cosmology ($\Omega_m=
0.25$, $\Omega_\Lambda=0.75$, $h=0.73$, $n=1$, $\sigma_8=0.9$) most
consistent with the Wilkinson Microwave Anisotropy Probe (WMAP) year 1
data \citep{spergel03}.  The dark matter particle distributions were
stored at 64 discrete epochs (`snapshots').

\subsubsection{Scaling of cosmological parameters}
\label{sec:cosmology}

The suite of MR simulations were performed using the now disfavored
WMAP1 cosmology. While the lower value of $\sigma_8$ preferred by the
more recent WMAP7 data will cause the growth of dark matter structure
to be delayed with respect to a WMAP1 cosmology, its effect on galaxy
formation models is less straightforward to infer. Running simulations
with multiple cosmologies is a time-consuming process.  Instead, the
MRObs project uses a recent technique introduced by \citet{angulo10}
in which the output from a cosmological $N$-body simulation in one
cosmology (e.g., WMAP1) can be scaled to represent the growth of
structure in another cosmology (e.g., WMAP7). Tests comparing direct
$N$-body simulations done in two cosmologies with a simulation that
was scaled from one to another cosmology show that power spectra are
reproduced to better than 3\% at all scales. In the MRObs the
technique is applied to halo catalogues. Properties such as
mass, concentration, velocity dispersion and spin are scaled are reproduced at about the
10\% level or better \citep[][]{angulo10} \citep[see also][]{ruiz11}. 
\citet{guo12} give the properties of semi-analytic galaxies in the MR and MR-II
scaled to the WMAP7 cosmology.

\subsection{Dark matter halos}
\label{sec:halos} 

The MR simulations output the dark matter phase-space
distribution at 61 different epochs at $z<60$. The spacing between
these outputs is roughly equal in the log of the expansion factor,
specifically, $\approx$300 Myr for $z<2$ and $\approx$100 Myr for
$z>6$. In each of these snapshots, DM haloes are found using a FOF
algorithm \citep{davis85} with a linking length parameter equal to one
fifth of the mean inter-particle separation. Within each FOF halo,
self-bound substructures are identified using the SubFind algorithm
\citep{springel01}.

For each subhalo, at each output time, a unique descendant in
subsequent snapshots is assigned as the subhalo which contains the
majority of the most bound particles (slightly different definitions
have been used among the different Millennium simulations).  Finally,
these pointers are arranged in a tree-like data structure which allows
to access the full mass evolution of a given object across time. This
structure -- a halo merger tree -- represents the backbone and starting
point for our post-processing simulations of galaxy formation.

\subsection{Synthetic galaxy catalogues}
\label{sec:sam}

\subsubsection{Semi-analytical galaxy formation models}

The N-body simulations used in the MRObs follow dark matter particles
only.  To add predictions about the baryonic content of the model
universe, we rely on an approach that generally is referred to as
semi-analytical modelling (SAM)
\citep[e.g.][and references therein]{white91,kauffmann93,cole94,kauffmann99,somerville99,kauffmann00,somerville01,springel01,hatton03,
  kang05,cattaneo05,cattaneo06,delucia07,monaco07,guo11,somerville11}.  Using simplified descriptions
(`recipe') for the baryonic physics, these models follow the
evolution of the galaxies within the skeleton provided by dark matter
halo merging trees defined in the previous steps.  These recipes
include gas cooling, star formation, reionization heating, supernova
feedback, mergers, black hole growth, metal enrichment and feedback
from active galactic nuclei. The recipes are constrained by local
observations and by physical insight. 

This technique is much less computationally expensive than adding full
hydrodynamics to the basic simulations.  Once the backbone formed by
the dark matter structure has been established, the semi-analytic
modeling of the galaxies can be repeated many times in order to find
the recipes and parameters that are required to match the
observations.

The MRDB \citep[][]{lemson06} contains galaxy
catalogues from two SAMs, \texttt{L-GALAXIES}, created at the
Max-Planck-Institute for Astrophysics in Garching 
\citep{springel01,croton06,delucia07,bertone07,guo11}, and \texttt{GALFORM},
created by the University of Durham \citep{cole00,
  benson03,baugh05,bower06}.  Compared to earlier models, also stored
in the MRObs, the latest version of the Munich model by \citet{guo11}
that we focus on here, includes improved prescriptions for supernova
feedback, gas stripping, galaxy merging, and bulge formation
\citep[see][for successive versions of the Munich model applied to the
MR]{croton06,delucia07,bertone07,guo11,henriques12}.  The output of
the SAM is stored for each of the 64 snapshots, thus sampling the
evolution of the galaxy population every few 100 Myr. The SAM
calculations, however, are computed on a finer grid consisting of 20
steps of about 10 Myr each between each pair of snapshots. This
ensures that the properties of galaxies are modeled on time-scales
appropriate for a wide range of star formation histories, including
brief bursts of star formation that may happen in between snapshots.

\begin{figure}
\begin{center}
\includegraphics[width=\columnwidth]{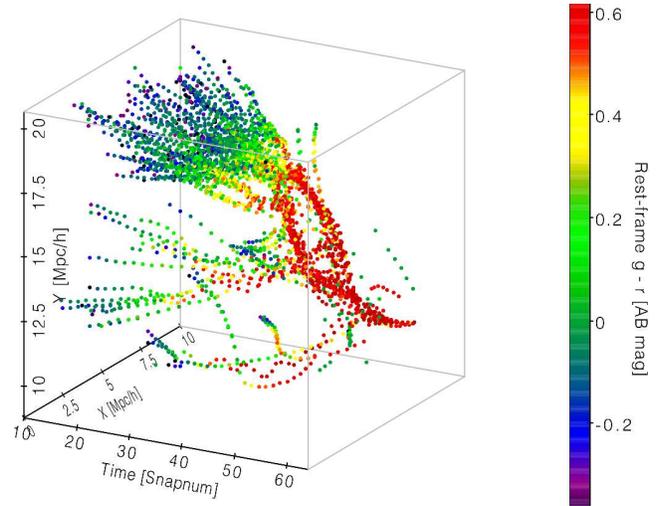}
\end{center}
\caption{\label{fig:evolution}The merger history of a single galaxy
  selected from the MR. The dark matter halo properties stored in a
  database are used as the backbone for a semi-analytic model of
  galaxy formation that tracks the growth of galaxies inside halos
  based on simple recipes for, e.g., gas cooling, star formation,
  supernova and AGN heating, gas stripping, and merging between
  galaxies. In each time step or snapshot of the simulation, the
  resulting physical properties of each galaxy in the semi-analytic
  galaxy are used to select appropriate stellar population templates
  from a library of spectral energy distributions to model the
  rest-frame spectra or colours of each galaxy. In the example shown
  here, the colour coding indicates the rest-frame
  $(g^\prime-r^\prime)_{AB}$ colour of all the galaxies that are part
  of the merger tree of a single galaxy selected in the simulation
  snapshot 63 ($z=0$), starting from 10 ($z\approx12$). The other two
  axes show the 2D positions of these galaxies in the simulations
  volume.}
\vspace{1cm}
\end{figure}

The galaxies resulting from the semi-analytic model naturally span a
wide variety in star formation histories (SFHs), corresponding to the
different gas accretion and merger histories of individual
galaxies. The relational database of the MRObs allows us to
reconstruct these SFHs in great detail.  It is important to keep in
mind the distinction between the SFH of the galaxy that forms the main
branch in a galaxy merger tree, and that of the stars in all the
progenitors of a descendant identified at some snapshot. As shown by
\citet{delucia07} this typically results in large differences between
the time it took for the stellar mass to be formed (`formation time')
and the time it took for that mass to assemble into a single galaxy
(`assembly time'). An example is shown in Fig. \ref{fig:evolution},
showing the stellar populations of all the different branches that
form the merger tree of a single galaxy at $z=0$. When observers study
the star formation history of a particular galaxy selected at some
redshift, they do thus not necessarily study the SFH of a single
galaxy, but rather the SFH of all its progenitors (weighted by stellar
mass).

\begin{figure}
\begin{center}
\includegraphics[width=0.8\columnwidth]{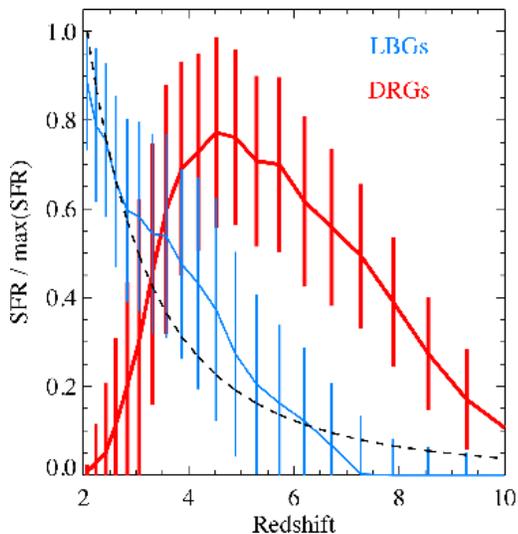}
\end{center}
\caption{\label{fig:sfhs}The average SFHs of star-forming and quiescent galaxies
  identified in the SAMs at $z\approx2$. The LBG-like systems (blue
  line) show a rising SFH, analogous to that derived from observations
\citep[][dashed line]{papovich11}. DRG-like systems have SFHs that
sharply decline after $z\approx5$.}
\end{figure}

Similar to real galaxies, galaxies in the MRObs span a very large
range in SFHs.  In Fig. \ref{fig:sfhs} we show the average SFHs for
star-forming and quiescent galaxies in the MRObs. These SFHs were
determined by summing the SFRs of all the progenitors of 100 galaxies
selected at $z\approx2$. For systems having SFRs of $>$10 $M_\odot$
yr$^{-1}$ and $M_*\sim10^{10}$ $M_\odot$ (e.g., similar to Lyman Break
Galaxies, LBGs), the SFHs are rising (blue line in
Fig. \ref{fig:sfhs}), roughly as derived from observations of LBGs
\citep{papovich11}. For systems having SFRs of $<$10 $M_\odot$
yr$^{-1}$ and $M_*\sim10^{11}$ $M_\odot$ (e.g., similar to Distant Red
Galaxies, DRGs), the average SFH rapidly declines after $z\sim5$ (red
line) analogous to the best-fit SFHs of DRGs observed
\citep[e.g.][]{kriek06}.

\subsubsection{Multi-wavelength model predictions}
\label{sec:ssps}

SAMs predict physical properties of galaxies, such as their stellar
masses, ages, metallicities, and gas content.  One common way of
testing the models is to compare them to the same physical properties
derived from the SEDs of observed galaxies. At the least, this
approach depends on having well-established measuring techniques and
accurate stellar population synthesis models
\citep[see][]{tinsley80}. In practice, this sort of analysis typically
includes numerous assumptions, and certain features of the galaxies
can never be recovered from the observations in full (e.g., their
exact star formation history or dust attenuation). 
In the MRObs the application of stellar population synthesis models
and dust recipes allow one to make detailed spectro-photometric
predictions for the model galaxies by adding up synthetic spectra
corresponding to the different generations of stars that these
galaxies consist of at any moment. The great predictive power of SAMs
in terms of the observable, photometric properties of galaxies is in
large part based on the spectral synthesis modeling of the stellar
populations being formed in the semi-analytic model galaxies according
to their SFRs at any given time (see Fig. \ref{fig:evolution} for an
example of a typical galaxy).

\begin{figure}
\begin{center}
\includegraphics[width=\columnwidth]{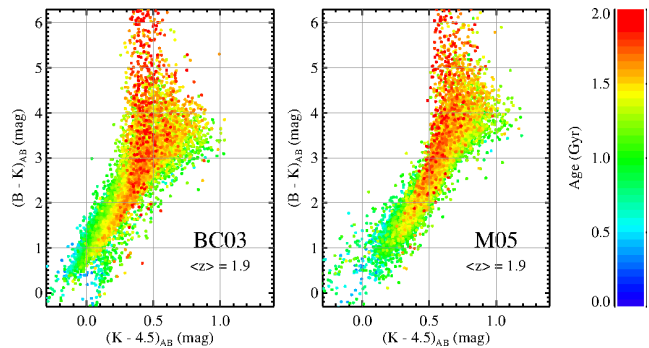}
\end{center}
\caption{\label{fig:colors_libraries}The choice of stellar population
  synthesis model affects the colour distributions of galaxies. We
  illustrate this by showing the optical-infrared colour-colour diagrams
  for galaxies at $z\sim1.9$ selected from the lightcones modeled
  using BC03 \citep[][left panel]{bc03} and using M05 \citep[][right
  panel]{maraston05}. Galaxies are colour-coded according to their
  mass-weighted age (see legend on the right). See also
  \citet[][Fig. 2 in that paper]{maraston06}. The MRObs offers the
  choice between different spectral synthesis models.}
\end{figure}

The predicted multi-wavelength properties of galaxies depend
on the spectral synthesis model used.  These models are currently
still affected by gaps in our understanding of stellar evolution
\citep[e.g., see][]{conroy09}, preventing us from making unambiguous
predictions for the main galaxy observables. 
Two well-known synthesis models implemented in the MRObs are those by
\citet[][`BC03']{bc03} and by \citet[][`M05']{maraston05}. These
models are illustrated in Fig. \ref{fig:colors_libraries}, where we show the
observed $K$--4.5$\mu$m vs. $B$--$K$ colour-colour diagram for galaxies
at $z\sim2$ in our simulations. The M05 model shown right predicts
significantly redder colours compared to the BC03 model
shown left, especially for galaxies between 1 and 2 Gyr in age
\citep[for discussion, see][and references
therein]{tonini10,henriques11}. 

We model the effect of dust on the predicted colours and magnitudes
using the dust treatment recipe detailed in \citet{guo09} \citep[see
also][]{delucia07,kitzbichler07,guo11,henriques12}, and we note that
in calculating the extinctions, a random inclination was assigned to
every galaxy to determine the extinction relative to the face-on value
(assuming a slab geometry). It is important to note that the dust
modeling in semi-analytic models is currently performed in a highly
simplistic manner. Even though models have shown to be fairly successful in
reproducing the overall colours of the evolving galaxy population in
some bands, a more realistic treatment of the effects of dust could
easily alter the spectral energy distributions of certain galaxies at any given
time significantly.

The significant uncertainties in the stellar population synthesis
models and the dust modeling techniques limit us in producing accurate
models of the real universe. For the same reasons, however, they also limit our ability to
derive exact physical quantities from real observations. At the very least, the
MRObs will thus allow one to test the accuracy of current techniques
designed to extract physical quantities from a given set of broadband
magnitudes. In the future, the highly modular approach of the MRObs (see Fig. \ref{fig:overview})
will make it straightforward to add alternative or improved models.

\begin{figure}
\begin{center}
\includegraphics[width=\columnwidth]{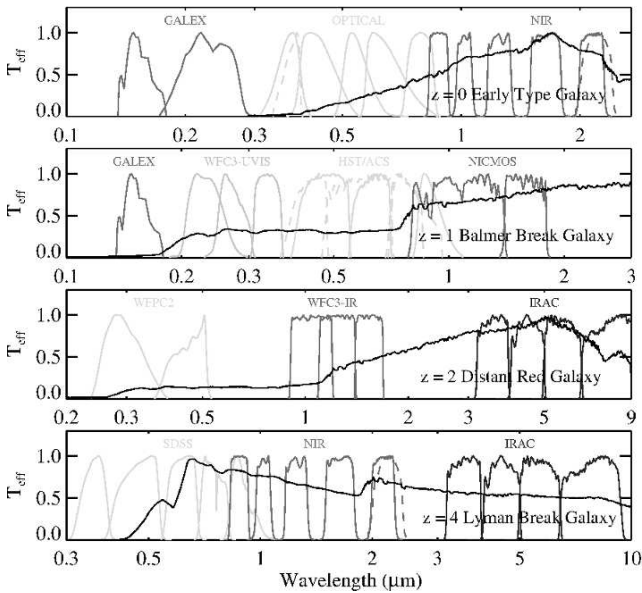}
\end{center}
\caption{\label{fig:filters}Examples of filter sets currently
  available in the MRObs: space-based UV (GALEX,
  HST WFC3-UVIS), ground-based optical (Johnson, SDSS, VIMOS),
  space-based optical (HST WFPC2, ACS), ground-based near-IR (Johnson,
  VISTA), space-based near-IR (HST NICMOS, WFC3-IR), and mid-IR
  (Spitzer/IRAC). Typical model galaxy spectra at $z=0$, $z=1$, $z=2$,
  and $z=4$ are shown for reference (black curves).}
\end{figure}

In the current version of the MRObs, the SFHs, stellar synthesis models, and dust extinction models
mentioned above are used to generate multi-wavelength filter catalogs
\citep[see][]{henriques11,henriques12}. These catalogs cover the FUV to the
mid-IR as observed by major telescopes and instruments (Table
\ref{tab:filters}). The filter bandpasses are illustrated in Fig. \ref{fig:filters}.

\subsection{Lightcone construction}
\label{sec:lightcones}

\begin{figure*}
\begin{center}
\includegraphics[height=4.5cm]{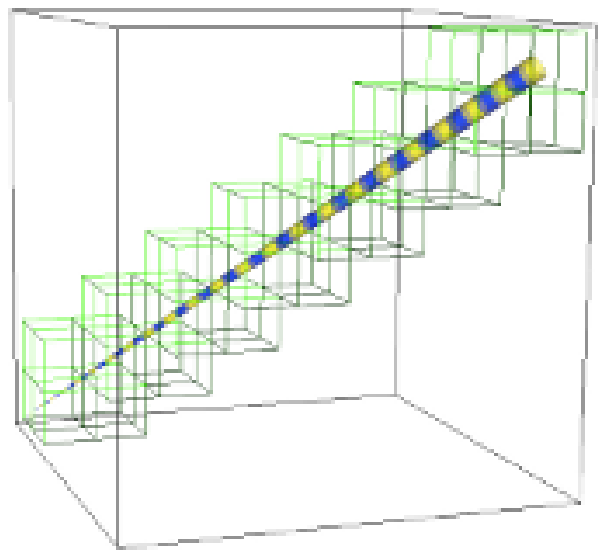}\hspace{1cm}
\includegraphics[height=5.cm]{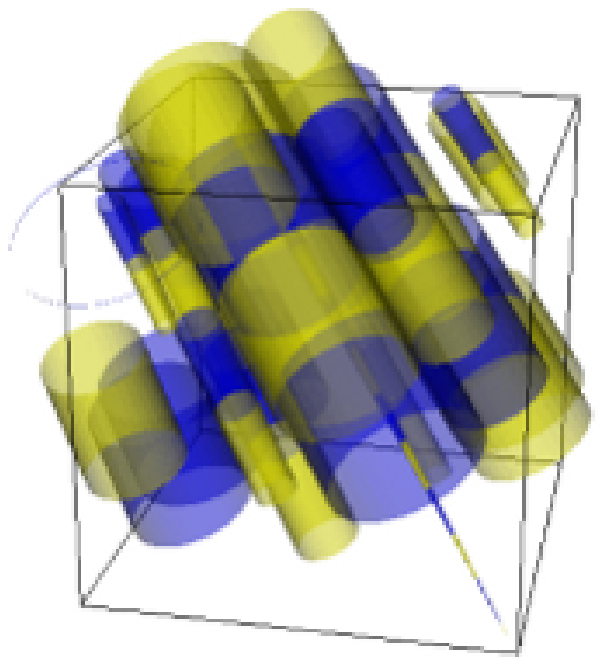}\hspace{1cm}
\includegraphics[height=5.cm]{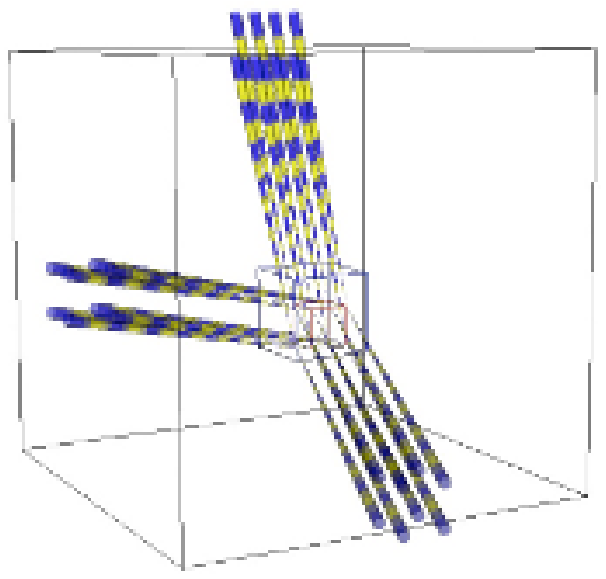}
\end{center}
\caption{\label{fig:cones}The construction of the lightcones. {\it
    Left panel:} The lightcone is constructed by replicating the
  simulation box until the co-moving distance corresponding to the
  desired limiting redshift is reached. In this example, the original
  co-moving size of the MR simulation is extended from 500 Mpc/h to
  $\sim$7000 Mpc/h, corresponding to $z\approx10$ for $h=0.73$. A
  conical volume is carved out from the volume that has now been
  expanded through the box replication process, and galaxies are
  selected from the overlap between the cone and the replicated
  volume. In order to model the relation between co-moving distance
  and redshift, at any point along the cone galaxies are selected only
  from that snapshot that is closest in redshift to the one
  corresponding to the co-moving distance along the line-of-sight. We
  do not interpolate over physical properties of the galaxies (they
  are assumed to be relatively constant between two consecutive
  snapshots), but apparent magnitudes and colours are interpolated, to
  make sure that galaxies have the right values for their
  redshifts. {\it Middle panel:} The starting position and orientation
  of the lightcone through the simulation box can be chosen such that
  the entire replicated volume can be constructed out of conical
  segments (drawn in blue and yellow) drawn from the original volume
  without passing through any region twice.  Different `views' of the
  simulated universe can be created by changing the starting points or
  orientations of the cones. Narrow pencil beams can be constructed
  out to very high redshift without replications, while very
  wide-field surveys can be made by keeping the limiting redshift
  low. Much larger volume surveys can be generated if the scientific
  application of interest allows some degree of replication. {\it
    Right panel:} Multiple lightcones can be
  extracted from the simulations box by choosing different starting
  positions (the position of the observer, $z=0$) and orientations
  (specified by the two angles $\theta,\phi$). The 24 `field'
  lightcones from \citet{henriques12} each with an opening angle of
  $1.4\degr\times1.4\degr$ are indicated.}
\end{figure*}

The snapshots of data (in time or in redshift) that are produced by
numerical simulations present an idealized view of the evolving
universe that is different from data resulting from observations of
the extra-galactic sky. In order to allow for more realistic and
direct comparisons between the model predictions and observations, we
construct so-called `lightcones' in which galaxies that were
simulated at discrete snapshots are re-arranged in order to mimic the
relation between the distance along an observer's line of sight and
cosmic time as accurately as possible \citep[e.g.,
see][]{hamana01,blaizot05,kitzbichler07,sousbie08,guo09,overzier09a,teyssier09,carlson10,henriques12}. 

In this paper we use lightcones introduced in \citet{henriques12}, to
which we add structural properties, and a set of new lightcones
pointed at specific objects.  These lightcones were built using a
version of the Mock Map Facility (MoMaF) code of
\citet{blaizot05}. Because of its importance to the MRObs, here we
give a short review of the technique, and describe a use of MoMaF that
allows us to create lightcones aimed at specific objects of interest
in the simulations and which is specially developed for the MRObs.

\begin{figure*}
\begin{center}
\includegraphics[height=5.cm]{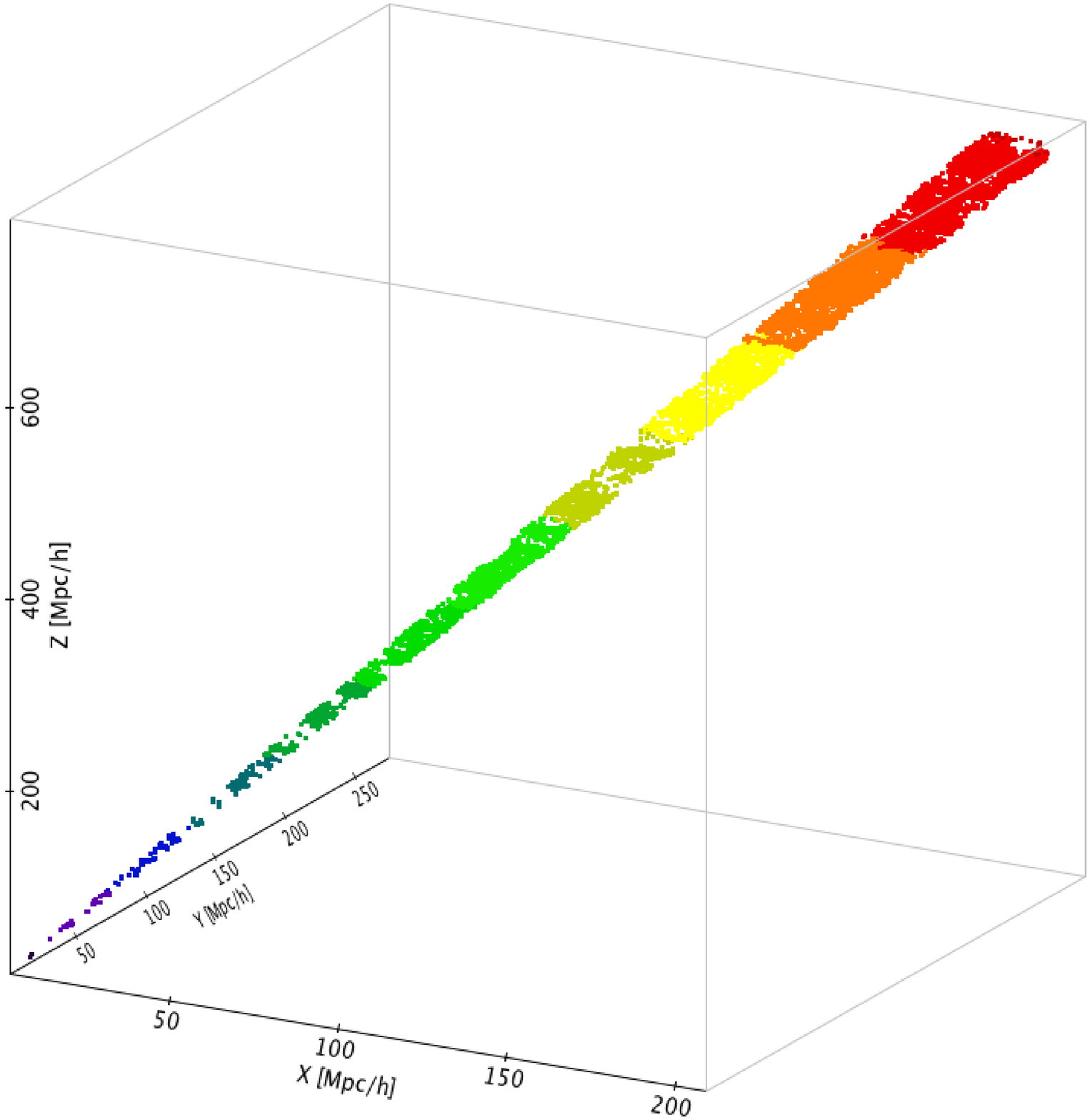}\hspace{5mm}
\includegraphics[height=5.cm]{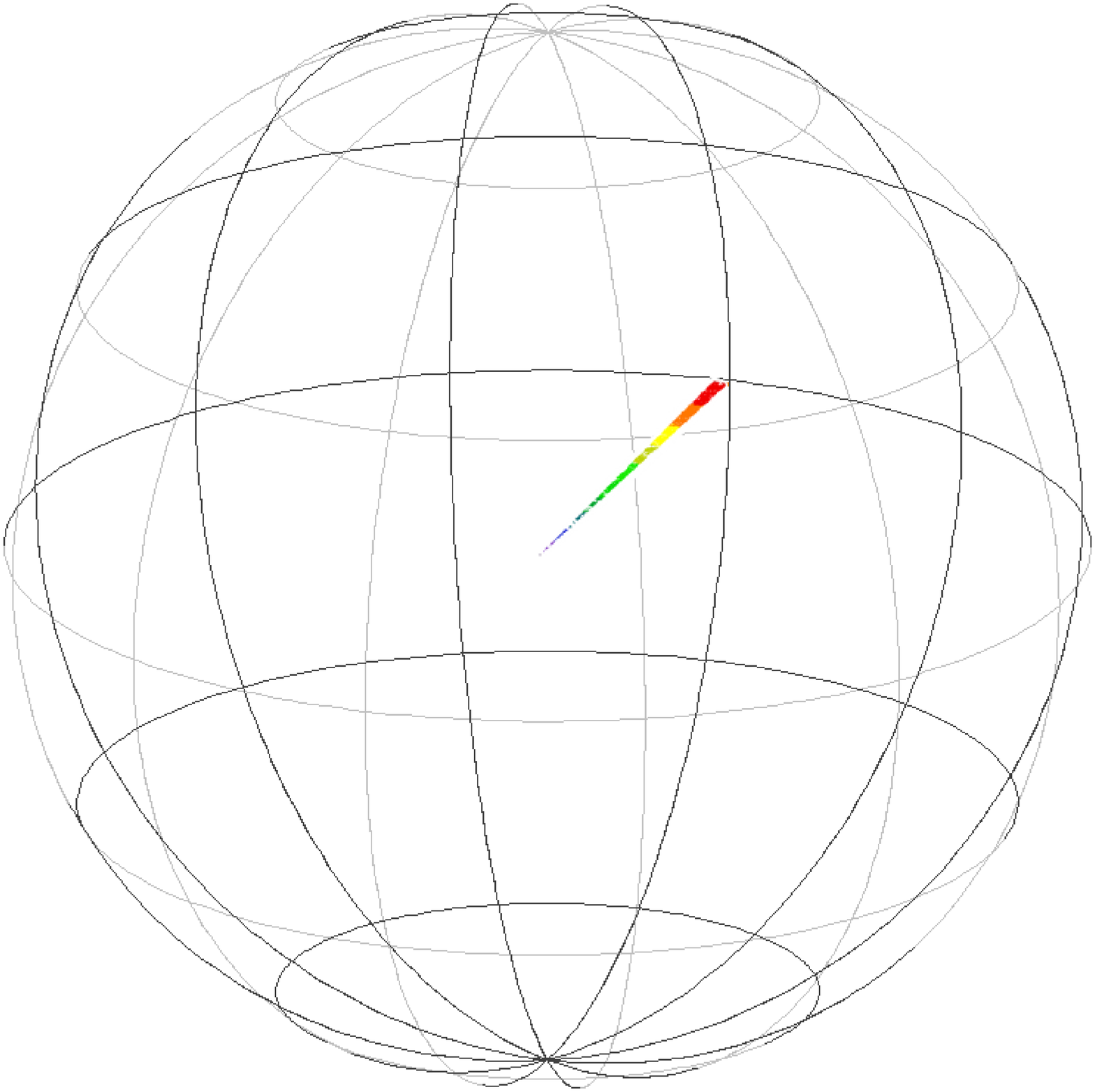}\hspace{5mm}
\includegraphics[height=5.cm]{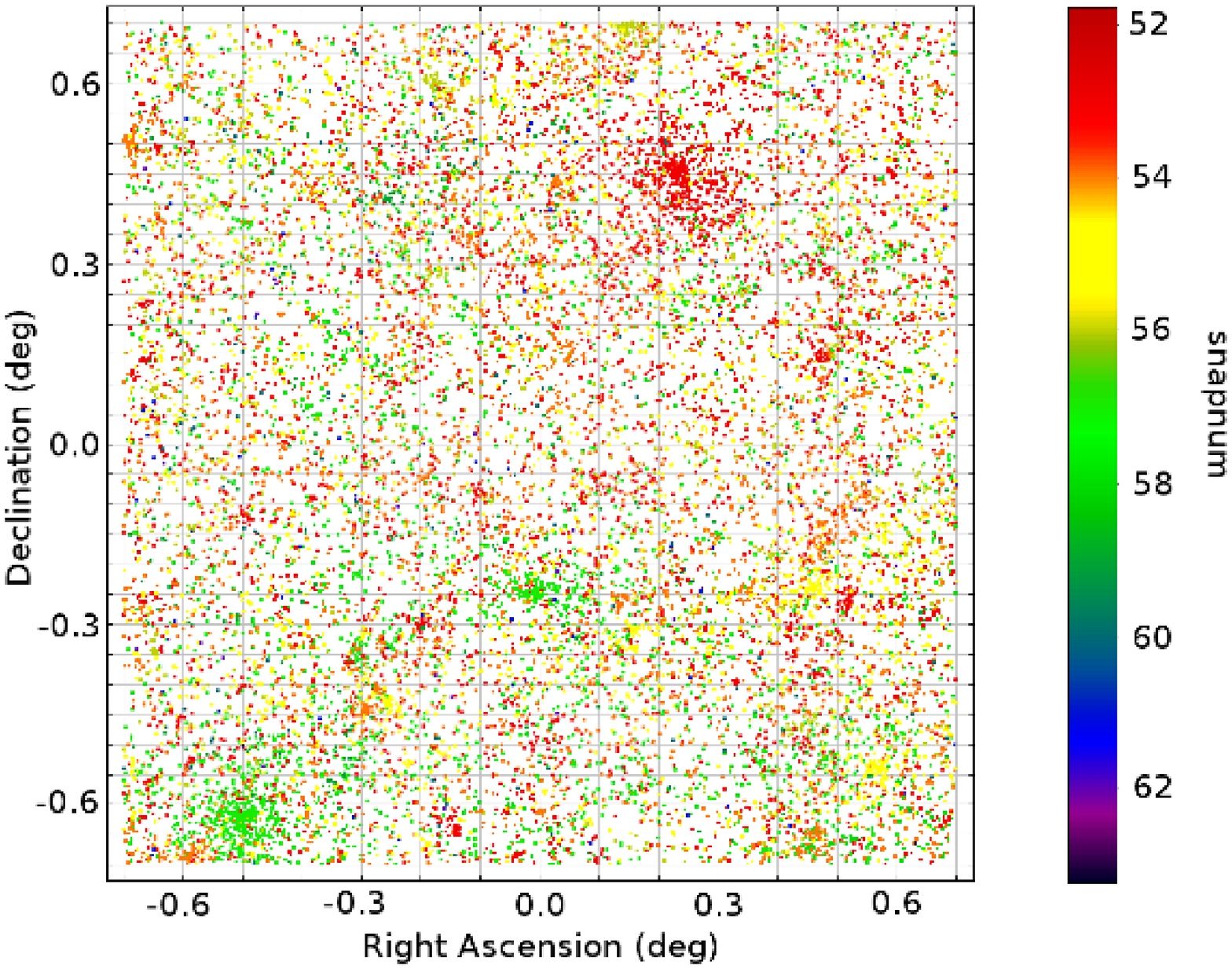}
\end{center}
\caption{\label{fig:conedetail}Lightcone construction further
  explained. {\it Left panel:} The lightcone in the expanded co-moving
  coordinate frame. {\it Middle panel:} Projection of the lightcone
  onto a virtual celestial sphere. {\it Right panel:} Galaxies in the
  lightcone as seen projected on the sky. The colour bar on the right
  illustrates which particular snapshot was used to populate each of
  the different sections along the lightcone. For clarity, we only
  plot the lightcones out to $z\approx0.3$ ($\textrm{snapnum}=52$). In reality
  our lightcones extend to beyond $z=10$ following the same
  procedure.}
\end{figure*}

\begin{figure}
\begin{center}
\includegraphics[width=\columnwidth]{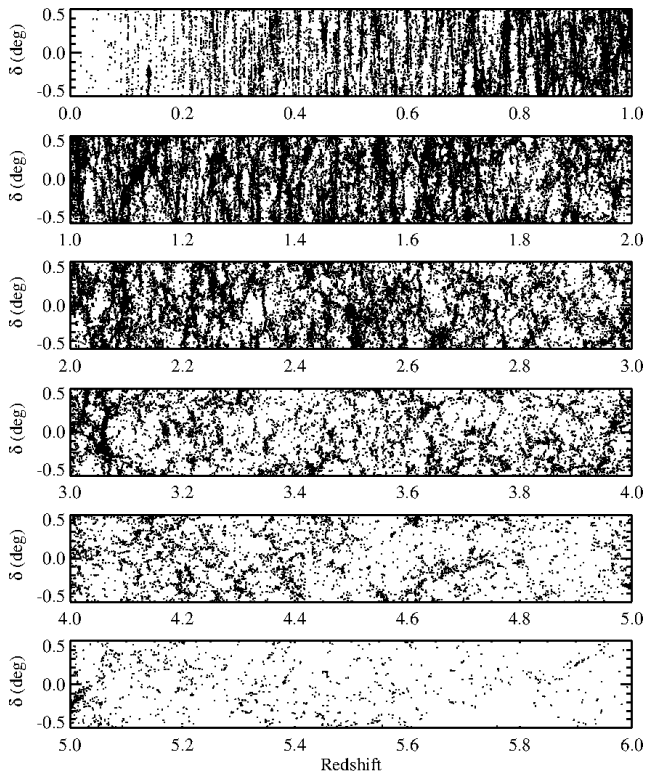}
\end{center}
\caption{\label{fig:cone1}Detail of a lightcone from
  \citet{henriques12} shown in the redshift versus
  declination plane. Galaxies plotted have SDSS z$^\prime$-band
  magnitudes of $<26.5$ (AB). The range in right ascension was limited
to $\pm0.1\deg$ for easier viewing.}
\end{figure}

\subsubsection{Summary of our lightcone method}
 
The MR predicts the detailed properties of the dark matter and the
galaxies it contains for a closely spaced set of snapshots that are
sufficient to compare with observations from $z=0$ to the highest
redshifts currently observed. In principle the simulations box probes
a sufficiently large volume to construct large pencil beam
surveys. For example, the total simulations volume of (500 $h^{-1}$
Mpc)$^3$ is equivalent to that probed in a pencil beam survey out to
$z=10$ and measuring 4 square degrees on the sky. On the other hand,
the co-moving distance to $z=10$ of $\sim$7,000 $h^{-1}$ Mpc is much
larger than the side of the simulations box of 500 $h^{-1}$ Mpc (900
$h^{-1}$ Mpc when taking the diagonal through the box).

\citet{blaizot05} suggested `replicating' the simulations
box along an artificial observer's line of sight until the maximum
co-moving distance desired is reached, and then extracting a conical
pencil beam out of the enlarged volume.  They explain that care must
be taken to avoid `perspective effects' caused by replication of the
same part of the universe in certain directions.  Whereas
\citet{blaizot05} solve this by adding random rotations and
translations of the boxes, thereby introducing discontinuities in the
galaxy distribution, \citet{kitzbichler07} showed that for certain
orientations of the lightcones through the MR box and for a small
enough opening angle of the cone, the lightcone can be constructed
without passing through any region of the simulations twice (or at
least ensuring that copies are widely separated in redshift if
replication occurs). It is the latter method that we use for all
pencil-beam light cones in the MRObs.

We illustrate the box-replication process in Fig. \ref{fig:cones}. In
the panel on the left, a virtual lightcone is drawn in a much enlarged
MR volume constructed using the box replication method. The opening
angle of the cone, its origin and angles of intersect with the
original MR box are chosen such that every cone segment (indicated by
the blue-yellow segments) can be extracted from the original MR volume
in such a way as to almost cover the complete simulation volume and
without passing through any region of the box twice (as illustrated in
the middle panel). Many pencil beam surveys can be
constructed from the MR by changing the angles or the origin of the
cone (right panel).

\begin{figure*}
\begin{center}
\includegraphics[width=0.7\textwidth]{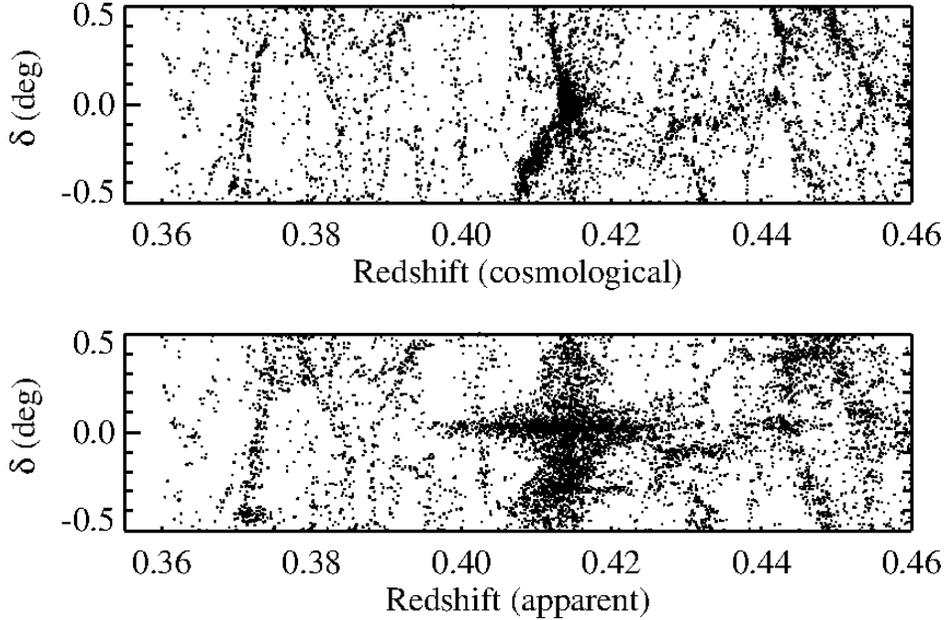}
\end{center}
\caption{\label{fig:pecvelocities}The effects of galaxy peculiar
  velocities on the apparent redshifts of galaxies in the
  lightcone. The top panel shows declination vs. the geometric or
  cosmological redshift for galaxies in and near a massive galaxy
  cluster at $z\approx0.4$. The bottom panel shows the redshift-space
  distortions an observer would see due to the peculiar velocities of
  galaxies moving through the gravitational potential well of the
  cluster.}
\end{figure*}

Besides these geometric considerations, one must take special care
that each galaxy is seen at the evolutionary phase and with the
photometric properties corresponding to its redshift along the
lightcone. In the MR, snapshots are separated by $\sim$100--400 Myr,
meaning that the evolving galaxy population is sampled at fairly
frequent intervals out to very high
redshifts. Fig. \ref{fig:conedetail} illustrates how we connect data
from many individual snapshots to obtain an evolving galaxy population
as a function of co-moving distance (or redshift) along the
lightcone. Each section consists of those galaxies having redshifts
$(z_i+z_{i+1})/2>z>(z_i+z_{i-1})/2)$, where $z_i$ is the redshift
corresponding to snapshot $i$. The physical properties that these
galaxies have are then also those they have in snapshot $i$. Because
the large-scale structure does not evolve rapidly between snapshots,
it is safe to neglect any changes occurring in the distributions of
galaxies. However, due to the relatively sparse number of
  snapshots, especially at very high redshifts, one has to be aware of
  the fact that there may be discontinuities or 'jumps' in the absolute
  numbers of sources as one moves from one lightcone section to the
  next. The physical properties of the galaxies can also fluctuate heavily
between snapshots, but as long as one is interested in the evolution
of the global population this can be safely ignored
\citep{kitzbichler07}. However, in order to ensure that `observed' 
galaxy properties are correctly related to redshift we perform small
interpolations of the observed-frame magnitudes, shifting each galaxy
in both redshift and luminosity distance from the snapshot
corresponding to redshift $z_i$ to the redshift at which it actually
appears on the lightcone \citep{blaizot05,kitzbichler07}. In addition
to this step, we make corrections to the observed magnitudes due to
absorption by the IGM (see \S\ref{sec:igm}).

The final step required is to project the cone onto a virtual sky seen
by a fictitious observer placed at the center of the celestial sphere
(middle panel of Fig. \ref{fig:conedetail}). It is now straightforward
to assign WCS coordinates (right ascension and declination) to every
object in the cone \citep[see][]{kitzbichler07}. The projected
large-scale structure can be seen in the sky distribution of galaxies
plotted in the right-hand panel of Fig. \ref{fig:conedetail}. Now that
we know both the sky coordinates and the redshifts to every object
along the lightcone, we can show the details of the large scale
structure that would be probed in a deep pencil beam survey as it
would appear in a large galaxy redshift survey out to $z\approx8$. In
Fig. \ref{fig:cone1} we plot the redshifts of objects versus their
declination on the sky for one of the \citet{henriques12}
lightcones. Points represent galaxies having $z^\prime$-band
magnitudes brighter than 26.5 (AB) mag.

\subsubsection{Aiming at a specific object}
\label{sec:aiming}

We have made a small modification to the MOMAF code that allows for
the construction of lightcones not only in arbitrary directions
as described above, but to also `aim' a lightcone such that it crosses
through a specific point of the MR box at a specific co-moving distance
(or redshift) from the origin. This new technique enables us to model
observations toward specific objects or regions by choosing a location
within the simulations volume at one redshift and `observing' it
within a lightcone with origin at another redshift. Obvious uses of
this technique are to study the appearance of a particular galaxy
cluster selected at $z=0$ and observed at $z=1$, or to study the $z=0$
descendant of a halo (or galaxy) selected at $z=6$. It is important to take into account
  the peculiar velocities of galaxies when constructing the lightcones as
  they can heavily distort the observed redshift distributions, 
  especially in the vicinity of massive objects such as galaxy clusters (see
  Fig. \ref{fig:pecvelocities}). 

  The MRDB allows one great flexibility in
  selecting such targets, and even allows one to the define the
  complete geometry of the light cone in a single SQL query.  
  Our new lightcone aiming technique thus greatly enhances the
  application of the MR to numerous new problems. Examples related to
  galaxy clusters will be shown in \S\ref{sec:examples_clusters}.

\subsubsection{Inclinations and position angles}
\label{sec:angles}

One of the unique features and key science drivers of the MRObs is
that it aims to produce detailed predictions for the observed galaxy
population without having to make assumptions that are not supported
or naturally accounted for by the model. The SAMs included in the
MRObs allow us not only to predict morphologies and sizes of galaxies,
but also their inclinations and position angles as seen by a virtual
observer.  The latter are derived from the orientation of the galaxy
as defined by the angular momentum vector of its stellar disk.  The
SAM that we use here tracks the change in the total angular momentum
vector of both gas and stellar disks. New gas condensing within a halo
is assumed to carry the specific angular momentum of that halo. The
total angular momentum change of gas disks in each time step is the
sum of the change in angular momentum due to gas condensation, gas
accretion and gas that is transformed into stars. The change in total
angular momentum of stellar disks is given by the change in angular
momentum due to gas that gets transformed into stars in each time
step.

As a consequence, the SAM predicts not only the spatial positions but
also the orientations of all galaxies with respect to the
three-dimensional, co-moving, Cartesian coordinate system of the
simulation box. From this we can then calculate the observed
inclinations and position angles of each galaxy based on the angles
between the line of sight of our lightcone and the simulation box. Our method ensures that
the orientations of galaxies in the MRObs are, at the very least,
physically motivated. This allows one to study in detail if the MR
predicts any observable correlations between the orientations of
galaxies, their parent halos or the large-scale structure. 
Such models are also suited for, e.g., conducting
completeness tests as a function of inclination, for testing galaxy
structure decomposition codes, and for paving the way for more
elaborate, orientation-based dust screening models that may be
implemented into the SAM at a later stage. 

\subsection{IGM absorption models}
\label{sec:igm}

The spectra of galaxies short-ward of 1216~\AA\ in the rest-frame are
primarily affected by photoelectric absorption by the neutral hydrogen
associated with damped \lya\ absorbers (DLAs), Lyman Limit Systems
(LLSs), optically thin systems, and resonance line scattering by the
\lya\ forest along the line of sight\footnote{In our model
  approximations we neglect the much smaller contribution from
  intergalactic metals and He absorption.}. This absorption affects
the magnitudes and colours of galaxies observed in bands corresponding
to these rest wavelengths. The strength and shape of this so-called
`Lyman Break' depend on the redshift of the source, the intrinsic
source spectrum, and the distributions in redshifts and optical depths of the 
absorbing systems. In order to ensure that these effects are properly
accounted for in the MRObs lightcones, at least in a statistical
manner, we have implemented three different models for the IGM
absorption. We include two models based on the recent IGM transmission
calculations by \citet[][`MEIKSIN']{meiksin06} and
\citet[][`INOUE-IWATA']{inoue08} that are conveniently made
available in code form by
\citet[][\texttt{IGMtransmission}]{harrison11}. We also include the
IGM transmission model of \citet[][`MADAU']{madau95} that is still
the most widely used in the literature today even though it has been
shown to significantly over-predict the absorption in the
912--1216\AA\ range compared to the updated models
\citep[e.g.][]{bershady99,meiksin06,inoue08}. 

\begin{figure}
\begin{center}
\includegraphics[width=\columnwidth]{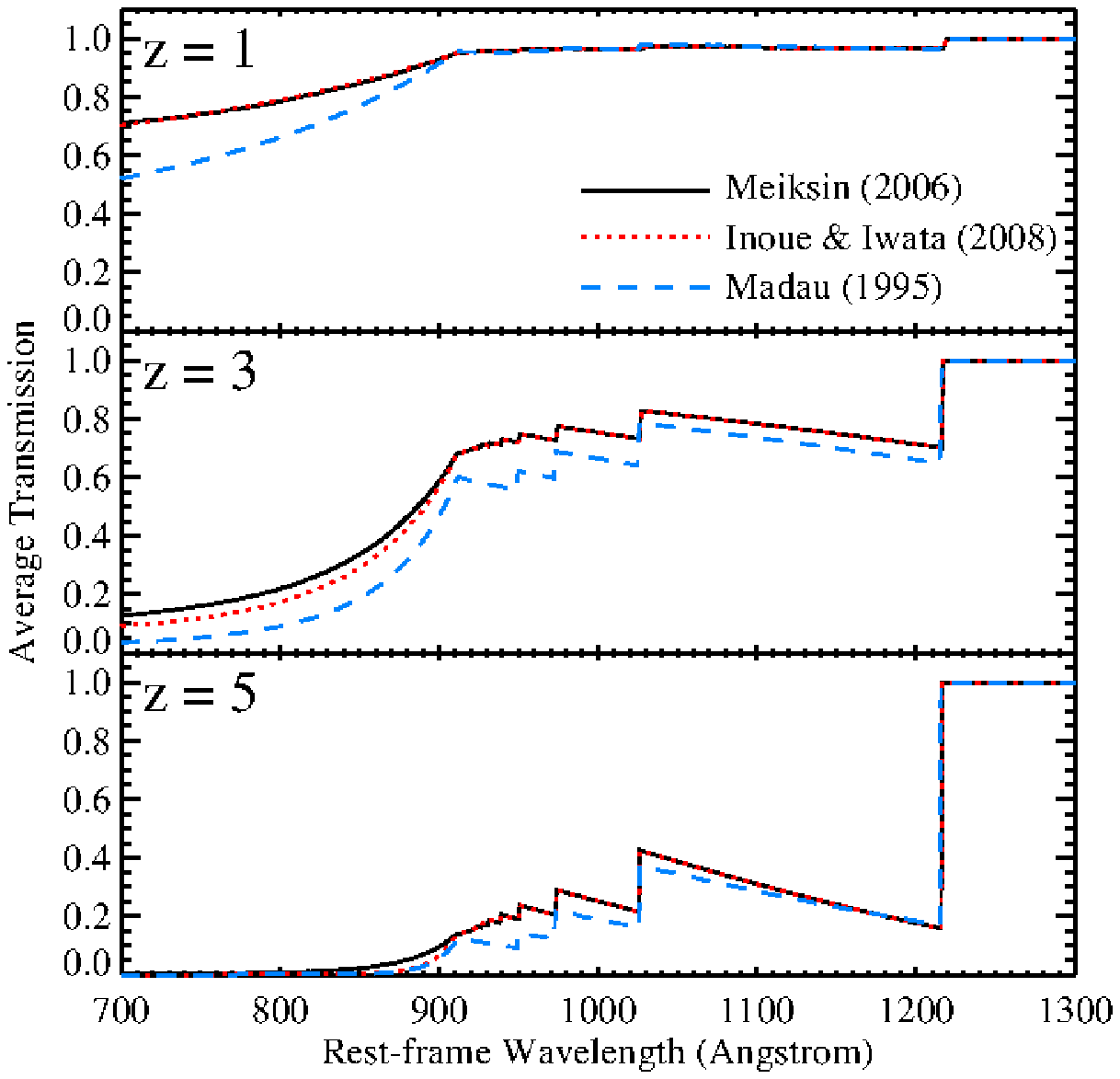}
\end{center}
\caption{\label{fig:IGM_vs_z_models}Attenuation of the UV continuum
  short-ward of \lya\ due to neutral hydrogen along the line of sight,
  which affects the colours of high redshift galaxies. Panels show the
  average transmission of the IGM according to the analytic
  approximation given by \citet[][blue dashed curves]{madau95}, and
  the more recent Monte Carlo modeling techniques of \citet[][black
  solid curves]{meiksin06} and \citet[][red dotted curves]{inoue08}
  for example galaxies at $z=1$ (top panel), $z=3$ (middle panel), and
  $z=5$ (bottom panel). The MRObs offers the choice between the
  different IGM implementations.}
\end{figure}

For an intrinsic galaxy spectrum $f_\lambda$, the attenuated spectrum
observed will be of the form $f_{\lambda,e} = f_\lambda \cdot
e^{-\tau_e(\lambda)}$, where the effective optical depth of the IGM
transmission function can be taken from any of the three IGM
models. In Fig. \ref{fig:IGM_vs_z_models} we show the mean
transmissions for sources at different redshifts. The MADAU model
implies significantly less transmission than the other models. 

In the current release of the MRObs, the IGM absorption
  correction is calculated only after the filter magnitudes are
  computed. We approximate the spectral shape within the filter
  response with that of a 100 Myr old, continuously star-forming,
  solar metallicity stellar population modeled using Starburst99
  \citep{leitherer95}. For most practical purposes (i.e. in the case
  of galaxies with significant far-UV emission from a young stellar
  population), this is a reasonable assumption, and we note that the
  same assumption is typically made by observers when interpreting the
  broad-band colours of high redshift dropout galaxies
  \citep[e.g.][]{bouwens03}. In any case, the corrections do not
  significantly change when assuming a 3 Myr old instantaneous
  low-metallicity starburst model ($<$0.05 mag change). Although
  refinements to this method will become available in the MRObs in the
  future, users can already also apply their own IGM corrections
  directly on the unattenuated lightcone catalogs that we provide and
  then use these as the basis for their own image simulations. Because
  the inclusion of the IGM attenuation is so important for creating
  realistic mock catalogs and images in the MRObs, a brief review of
  the MEIKSIN, INOUE-IWATA. and MADAU modeling recipes is given in
  Appendix A.

\subsection{Construction of the virtual telescope data}
\label{sec:sims}

\subsubsection{Galaxy models}

Now that we have obtained all the necessary information pertaining to
the positions, sizes, viewing angles, bulge-to-disk ratios, and
IGM-corrected magnitudes across different filters, we can populate
simulated images with galaxies. We follow a two-step process. First we
simulate noise-free galaxy profiles projected onto a 2D image plane at
very high pixel resolution using a modified version\footnote{We
  optimized Skymaker for dealing with very large input lists in .csv
  format provided by the MRDB, and for generating extremely
  large images.} of Skymaker \citep{bertin09}. We will refer to the
result of this process as the `perfect' or `pre-observation' image. Once the perfect
image has been made, it is straightforward to apply all the
observational effects such as the PSF, binning, sky background, and
noise for any type of observation. This last step is done using our
own custom code\footnote{Although Skymaker was specifically designed
  to handle point spread function convolution, sky backgrounds, and
  simulating detector noise, for various practical reasons we do not
  currently make use of this functionality but use our own custom IDL and Python codes 
  for these steps of the simulation.}.

In line with the \citet{guo11} semi-analytic predictions, galaxies in the MRObs are composed of an exponential profile for the
disk (D) and a De Vaucouleurs profile (S) for the bulge (if any), each having a 
surface brightness profile $\mu(R)$ in mag arcsec$^{-2}$ given
by:
\begin{eqnarray}
\mu_S(R) &=& m - 2.5\log_{10}(B/T) + 8.3268(R/R_e)^{1/4}\nonumber \\&+& 5\log_{10}(R_e) - 4.9384\\
\mu_D(R) &=& m - 2.5\log_{10}(1-B/T)\nonumber \\&+& 1.0857(R/R_h) + 5\log_{10}(R_h)
+ 1.9955,
\end{eqnarray}
where $m$ is the total magnitude (mag), $B/T$ is the bulge-to-total
ratio, $R_e$ is the bulge half-light radius (arcsec), and $R_h$ the
disk scale height (arcsec). Skymaker builds these profiles as
elliptical shapes at pixel position $x^\prime=x-x_c$, $y^\prime=y-y_c$
projected on the sky with position angle $\theta$ and inclination
$\phi$ according to \citep[see][]{bertin96}:
\begin{equation}
C_{XX}\cdot x^{\prime 2}+C_{YY}\cdot y^{\prime 2}+C_{XY}\cdot x^\prime y^\prime=R^2,\\
\end{equation}
such that the algorithm for calculating the projected light profiles
for disks and bulges becomes:
\begin{eqnarray}
\label{eqn:ids}
I_D[x^\prime,y^\prime] &\propto& e^{-(C_{XX}\cdot x^{\prime 2}+C_{YY}\cdot y^{\prime 2}+C_{XY}\cdot y^\prime x^\prime)^{1/2}}\nonumber\\
I_S[x^\prime,y^\prime] &\propto& e^{-7.6693(C_{XX}\cdot x^{\prime 2}+C_{YY}\cdot y^{\prime 2}+C_{XY}\cdot y^\prime x^\prime)^{1/8}},
\end{eqnarray}
with
\begin{eqnarray}
C_{XX} &=& \frac{\cos^2(\theta)}{A^2} + \frac{\sin^2(\theta)}{B^2}\nonumber\\
C_{YY} &=& \frac{\sin^2(\theta)}{A^2} + \frac{\cos^2(\theta)}{B^2}\nonumber\\
C_{XY} &=& 2\cos(\theta) \sin(\theta)\left(\frac{1}{A^2}-\frac{1}{B^2}\right).\nonumber
\end{eqnarray}
$A$ and $B$ are the projected major and minor axes, with
$A=R_h$ for disks and $A=R_e$ for bulges, and $B=A\cos(\phi)$ with
$\cos(\phi)$ the projected aspect ratio of the system. 

\begin{figure*}
\begin{center}
\includegraphics[width=0.7\textwidth]{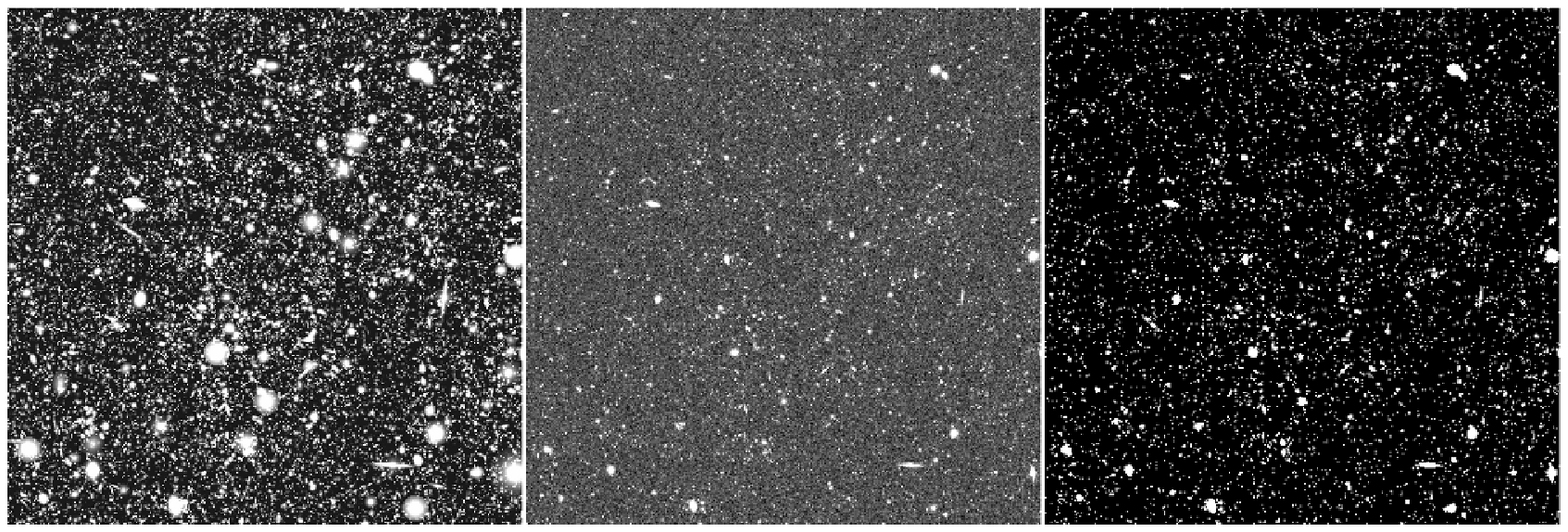}
\includegraphics[width=0.7\textwidth]{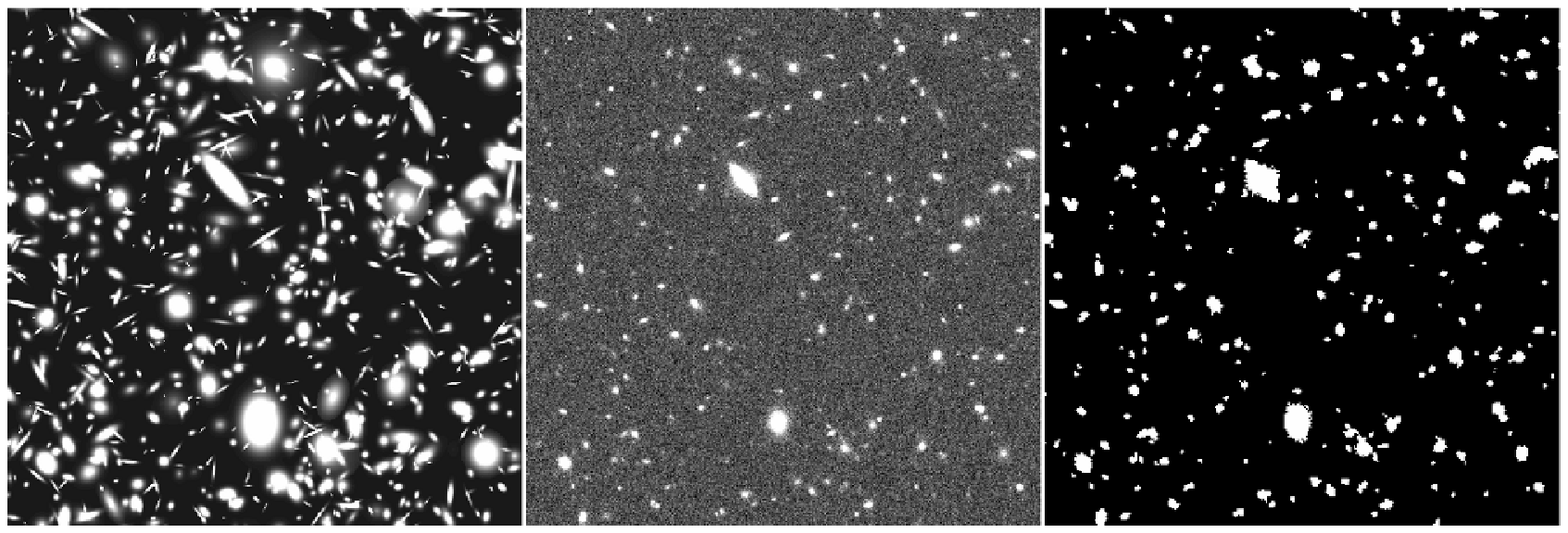}
\end{center}
\caption{\label{fig:skymaker}Single filter simulated image constructed
  from a lightcone. {\it Left panels:} The `perfect' image modeled
  using Skymaker. Galaxies consisting of disks and bulges are placed
  at the proper position, inclination, orientation, brightness and
  apparent size that are all uniquely determined by the
  semi-analytical model and the angles of intersection between the
  lightcone and the simulations volume. The only information that is
  not constrained by the model is the bulge shape, which we set to
  spherical. {\it Middle panels:} The perfect image as seen by our
  telescope simulator. Here we show a mock HST/WFC3 F160W \texttt{.fits} image
  having the same detector properties, point spread function, sky
  background, and signal-to-noise as the ERS observations (a
  $1\farcm5\times1\farcm5$ region is shown at a spatial binning of
  0\farcs09 pixel$^{-1}$). No stars were added to this
  observation. {\it Right panels:} The SExtractor `segmentation'
  image showing the locations and shapes of objects that were detected
  in the simulated image. Panels on the bottom row show a zoom of the
  full images shown in the top row. Although there is a good
  correspondence between objects seen in the simulated telescope image
  and objects detected by SExtractor, the perfect image that was used
  as the input for the image simulation contains many more sources
  that are too faint to be seen in the simulated image.}
\end{figure*}

Our modified version of Skymaker performs this process efficiently for typical MRObs
simulations that are based on lightcones containing several millions
of galaxies per square degree. An example of the `perfect image'
produced is shown in the left panel of Fig. \ref{fig:skymaker}, where
the white shapes indicate the simulated galaxy
images. The
corresponding final (`noisy') telescope image produced following the
process detailed below is shown in the middle panel.
 
\subsubsection{Input parameters: positions, magnitudes, inclinations,
  orientations, sizes, and bulge-to-disk ratios}

The center positions of all objects in the image plane are determined
from the right ascention and declination relative to the lightcone centres and the pixel
scale of the desired instrument. Inclinations and position angles are uniquely determined from
the angular momentum vector of the stellar disks relative to the
orientation direction of the lightcones through the MR volume (Fig. \ref{fig:cones}). Angular sizes are determined from the
physical size and the diameter distance, $D_A$, at the redshift of
each source in the lightcone.

We list the specific parameters required by Skymaker for
simulating each galaxy, and give a brief explanation of how this
parameter follows from our models.

\begin{itemize}

\item $x_c$,$y_c$ : The source position in image pixel coordinates. This
  position is defined by the sky coordinates of a galaxy in the
  lightcones, the desired pixel scale of the image, the field of view,
  and the position of the image center relative to the lightcone
  center.

\item $m$ : The total apparent (AB) magnitude of the source in the
  desired filter. This magnitude includes the attenuation by dust as
  well as the IGM absorption.

\item $B/T$ : The bulge-to-total ratio of the source. This parameter,
  for which we take the ratio of the fluxes predicted for the bulge
  and total in each filter, is needed for assigning magnitudes to the
  bulge ($m_b=m-2.5\log_{10}(B/T)$) and disk
  ($m_d=m-2.5\log_{10}(1-B/T)$) components.

\item $R_{h,disk}$ : The scaleheight of the disk in arcseconds. This
  is defined by \texttt{stellardiskradius}$/3D_A$, with
  \texttt{stellardiskradius} taken from the \texttt{Guo2010a..MR}
  table\footnote{The \texttt{Guo2010a..MR}
  table stores the galaxy catalogue obtained by applying the SAM from
  \citet[]{guo11} to the MR halo merger trees.} in the MRDB and is in units of kpc. 

\item $R_{e,bulge}$ : The equivalent (or half-light) radius of the
  bulge measured in arcseconds. This is calculated as
  \texttt{bulgesize}$/D_A$, where \texttt{bulgesize} is taken from the
  \texttt{Guo2010a..MR} table and is in units of kpc.

\item $\cos(\phi_{disk})$ : The projected aspect ratio of the disk
  that is uniquely determined by the angles of intersection of the
  lightcone with the MR volume and the intrinsic spin axis of the
  galaxy stellar disk.

\item $\theta_{disk}$ : The position angle of the disk, defined by the
  angles of intersection of the lightcone with the MR volume and the
  intrinsic spin axis of the galaxy stellar disk.

\item $\cos(\phi_{bulge})$ : The projected aspect ratio of the
  bulge. Because all bulges in \citet{guo11} are spherical, we set this value
  to 1.0.

\item $\theta_{bulge}$ : The projected aspect ratio of the
  bulge. Because all bulges in \citet{guo11} are spherical, we set this value
  to 0.0.

\end{itemize}

\subsubsection{The virtual telescope model (sky, PSF, noise, and all that)}
\label{sec:od}

The MRObs produces realistic telescope data by applying an
`observation description' (OD) to the perfect image created in the
previous step. The OD consists of a set of instructions that
completely defines a particular observation to be mimicked, e.g., 
telescope, detector, filter, exposure time, number of sub-exposures,
dither strategy, and sky conditions. Although the exact modeling
method may vary depending on the details of a specific instrument or
survey, here we list the basic observational effects typically being
added in sequence:

\noindent
(1) The first step is to scale the perfect image populated by our
bulge+disk surface brightness simulations to their proper fluxes
measured in detector electrons by multiplying the models in
Eq. \ref{eqn:ids} by the factor
\begin{equation}
F_{e-} = 10^{-0.4(m_{AB}+ZP)}   \cdot T_{\textrm{exp}} \cdot G / \sum_{i\in S} \sum_{j\in S}      I[x_i^\prime,y_j^\prime],
\end{equation}
where $m_{AB}$ is the AB magnitude of the disk/bulge, ZP is the
zero-point in AB magnitudes that gives a detector count rate of 1 ADU
$s^{-1}$, $T_{\textrm{exp}}$ the image exposure time in seconds, $G$ is the
detector gain in $e^-$ ADU$^{-1}$, and $x_i^\prime$ and $y_j^\prime$
are the coordinates of pixel $i,j$ belonging to each source. 

\noindent
(2) We add a sky background. The value of the background is usually
kept constant across the field (we use gnomonic projections) based on
the average conditions at a particular site or telescope, or is based
on the sky background level measured in a particular survey that is
being modeled.

\noindent
(3) The image is convolved with a point spread function (PSF). The PSF
can have various origins: it can be taken from a PSF simulator
(e.g. TinyTim in the case of HST), from (a stack of) stars extracted
from a fully reduced observation, or modeled with a simple function
(e.g., a Gaussian).

\noindent
(4) The image is rebinned to the desired pixel scale. If the PSF is
taken from an actual observation and is not available at sub-pixel
resolution, the rebinning step is performed before the PSF convolution
step.

\noindent
(5) Detector dark current is added to the image.

\noindent
(6) Poisson noise is calculated for each pixel value. 

\noindent
(7) Gaussian-distributed readout-noise is added.

\noindent
(8) WCS astrometry is added to the image header based on the pixel
scale and the astrometric system of the lightcone.

\noindent
(9) Scientific images in \texttt{.fits} format are created, optionally
with corresponding background and noise maps. Complex observations
having the proper noise characteristics can be created from co-adds of
multiple exposures made following the same above
procedures.\footnote{Although the original version of Skymaker is
  capable of simulating bleeding, blooming and saturation, we do not
  currently include these effects for practical purposes. However, we
  do release the object input lists and the `perfect' images for all
  our simulations (\S\ref{sec:publicrelease}), such that interested users can perform their own
  modeling of these (or other) effects.}

The middle panels of Fig. \ref{fig:skymaker} show a mock HST/WFC3
$H_{160}$-band image corresponding to the perfect image shown on the
left. The mock HST image was modeled after the $H_{160}$-band
observations of the GOODS ERS survey of \citet{windhorst11}.

\subsubsection{Galactic extinction and stars}

Optionally, we apply Galactic foreground extinction to the input
galaxy models by specifying the amount of reddening in units of
$E(B-V)$ and assuming the \citet{cardelli89} attenuation curve with
$R_V=3.1$. If desired, Galactic stars can be added to the image, either
based on a user-specified input distribution or based on an accurate
Milky Way model \citep[e.g., TRILEGAL;][]{girardi05}.

\subsection{Source extractor}
\label{sec:sextractor}

With the synthetic images produced in the previous section, it is
straightforward to analyze the data analogous to real observations.
Sources in the images are detected by using the Source Extractor
(SExtractor) software \citep{bertin96}, which efficiently decomposes a
pixel image into `objects' detected at some specified threshold of
flux above the image background. Photometry and other basic
measurements are performed on all the detected objects yielding a
source catalog corresponding to the image. The exact way in which
objects are defined and how measurements are performed depend on the
setting of various of the parameters in SExtractor, while the total
number of objects that can be recovered from the image and the errors
on their photometry largely depend on the image quality itself. The
MRObs makes it convenient to test the different detection and
photometry techniques available in the literature, especially because
the properties of the galaxies that were used to create the mock image
are exactly known (as opposed to galaxies in real observations).

\begin{figure}
\includegraphics[width=0.9\columnwidth]{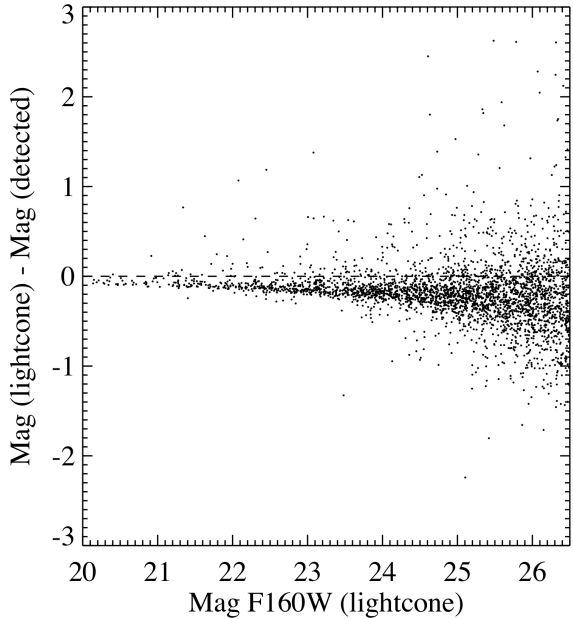}
\caption{\label{fig:candelstrumpet}`Trumpet' diagram showing the
  difference between the input magnitudes (from the MR lightcones) and
  the magnitudes measured by running SExtractor on a simulated
  HST/WFC3 H-band image.}
\end{figure}

We have run SExtractor on the mock HST/WFC3 $H_{160}$-band image shown
in Fig. \ref{fig:skymaker} (middle panels). Panels on the right show
the SExtractor `segmentation image', indicating all the objects that
were detected in the mock image. While there is good correspondence
between the two (nearly all objects seen in the mock image are also
seen in the detection image), the perfect (input) image shown on the
left contains many more galaxies, most of which are too faint to be
detected in the mock observation.  By cross-correlating the positions
of detected objects listed in the SExtractor output catalogs with the
positions of objects in the underlying lightcone (both available in
the MRDB), we can find out which of the semi-analytic galaxies
(identified by their \texttt{GALAXYID}) were detected in the image. This
enables us to perform various diagnostic tests between measurements
extracted from the synthetic observations and the corresponding
intrinsic physical properties from the lightcone. One such test is to
study how well the real magnitudes are recovered from the synthetic
images by SExtractor.  In Fig. \ref{fig:candelstrumpet} we show a
so-called `trumpet' diagram indicating the difference in magnitude
between the `true' input value and the total magnitude given by
SExtractor. The test shows, quantitatively, both how the amount of
flux lost due to missed light, and how the photometric scatter due to
increased noise increases toward fainter magnitudes.

Because the cross-match between the SExtractor catalog and the
lightcone catalog gives us the \texttt{GALAXYID} of each galaxy in the
images, this provides us also with a direct link to all the available
physical quantities in the semi-analytic snapshot catalogs, the dark
matter halo catalogs, and the underlying dark matter density fields,
such that it becomes possible to perform numerous experiments related
to how well we can extract such physical parameters starting from any kind of 
observation that can be modeled using the MRObs.

\section{Example: simulating CANDELS data}
\label{sec:mockcandels}

Large extra-galactic surveys often have complicated tiling patterns,
exposure time variations, and masked regions across their total field
of view that complicate the analysis. It can be convenient to include
these kind of effects into the image simulation. This ensures that the
signal-to-noise properties and the geometriy of the real and mock data
sets are comparable. Here we will illustrate the technique that we use
to accomplish this by performing a mock image simulation of the
ongoing multi-cycle treasury program Cosmic Assembly Near-Infrared
Deep Extragalactic Legacy Survey \citep[CANDELS;][HST Programs
12060--12064, 12440; PI: S. M. Faber]{grogin11,koekemoer11} that
primarily observes with the WFC3 on HST.

\subsection{The CANDELS observations}

\begin{figure}
\begin{center}
\includegraphics[width=\columnwidth]{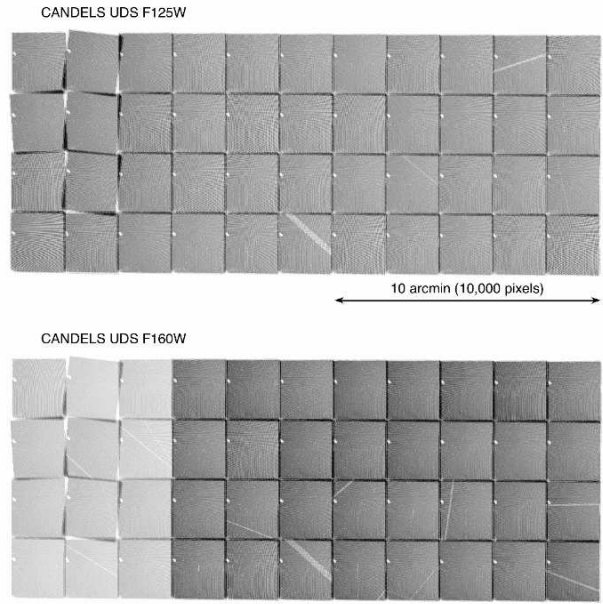}
\end{center}
\caption{\label{fig:candelswht}CANDELS UDS inverse variance weight
  maps. The total field of view measures 23\arcmin$\times$10\arcmin,
  and is constructed from 44 individual tiles observed with HST/WFC3
  in the filters F125W and F160W.}
\end{figure}

\begin{figure}
\begin{center}
\includegraphics[width=0.9\columnwidth]{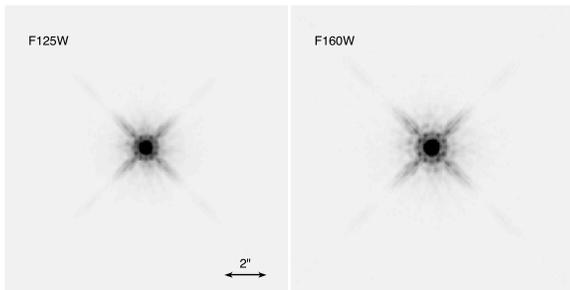}
\end{center}
\caption{\label{fig:candelspsf}Images of the PSF in the filters F125W
  and F160W of the CANDELS UDS field. Our mock `perfect' images are
  convolved with these PSFs.}
\end{figure}

Part of the ongoing HST CANDELS program, the UKIDDS Deep Survey (UDS)
field measures approximately 23\arcmin$\times$10\arcmin\ in the
filters F125W ($J_{125}$) and F160W ($H_{160}$). This field of
view is covered with 44 individual pointings with HST/WFC3 resulting
in the tiling pattern shown in Fig. \ref{fig:candelswht}. For each
tile, four exposures were obtained in both filters, resulting in
average total exposure times across the field of 1900 s in F125W and
3300 s in F160W. The data were combined onto a common output frame
measuring about 22,000$\times$10,000 pixels with a pixel scale of
$0\farcs06$ using the \texttt{MULTIDRIZZLE} software
\citep{koekemoer03,fruchter09}. The resulting PSF in the drizzled data measures
0\farcs12 (F125W) and 0\farcs18 (F160W) in FWHM
(Fig. \ref{fig:candelspsf}). How well can we simulate these kind of
data based on cosmological simulations using the MRObs?

\begin{figure}
\begin{center}
\includegraphics[width=\columnwidth]{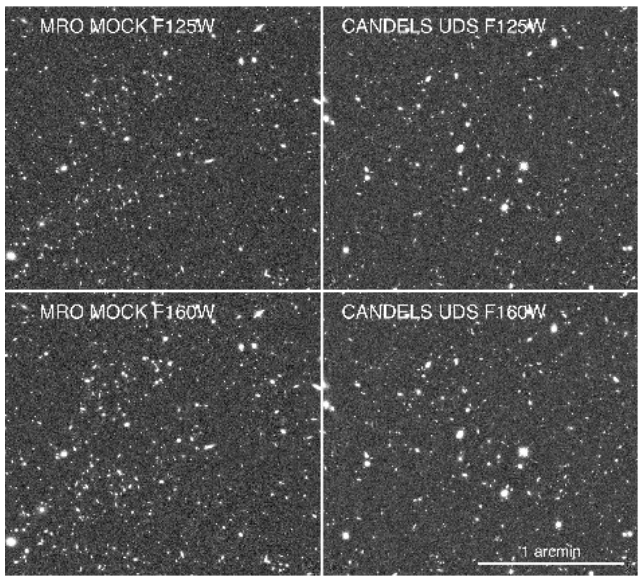}
\end{center}
\caption{\label{fig:candelsimages}Simulated and real CANDELS UDS data
  in the filters F125W (top) and F160W (bottom). At a qualitative
  level the images appear already remarkably similar. Note that this
  is not only the result of our accurate image simulation technique,
  but also because our input galaxy population apparently has a
  striking resemblance compared with the observed one (e.g., in terms
  of number density, clustering, size and shape distributions, and
  brightness). Shown here is a region of about
  $2\arcmin\times2\arcmin$\ extracted from the wider
  $23\arcmin\times10\arcmin$\ UDS field.}
\end{figure}

\subsection{The CANDELS simulation}

Using the procedures outlined in \S\ref{sec:sims} we can produce
highly accurate mock `CANDELS' data in a number of complementary
ways.

(1) The first and most cumbersome method would be to produce each
individual CANDELS tile at the correct telescope position and roll
angle, and then to process the entire data set through
\texttt{MULTIDRIZZLE} analogous to the processing performed on the
real data. While this is certainly possible, for many scientific
applications a good match between the simulated and real data sets can
already be obtained by side-stepping the laborious drizzling process.

(2) The simplest and most straightforward way is to directly generate
mock images the size of the entire UDS field based on our model for
the HST/WFC3 camera, the main UDS survey parameters, and a mock
lightcone as input. This method produces mock UDS images for which the
properties (e.g., noise, resolution) are, on average, very similar to
those of the real survey. This is an extremely fast method for
generating mock data sets that are approximately similar to the
observations that are being modeled. It is also a powerful method to
simulate images for a survey that has not (yet) been performed, or for
simulating a survey at an arbitrary depth or field size.

(3) Our third method, the one that we will use for our demonstration,
is an extremely powerful technique for generating a more precise
simulation in which the pixel-to-pixel noise variations and geometry
of the simulated images can be exactly matched to those of the real
data. For this method we make use of `weight maps' associated with
the science data for many surveys. The CANDELS UDS weight maps (shown
in Fig. \ref{fig:candelswht}) record the inverse variance of each
pixel calculated during the image reduction process
\citep{koekemoer11}. The HST inverse variance images (in units of $(e^- s^{-1})^{-2}$) are usually
calculated as follows
\begin{equation}
\mathrm{Inverse~Variance} \approx \frac{(ft)^2}{(D+fB)+\sigma_{\mathrm{ron}}^2},
\end{equation}
where $f$ is the inverse flat-field, $t$ is the exposure time, $D$ is
the accumulated dark current, $B$ is the accumulated background, and
$\sigma_{\mathrm{ron}}$ is the read-out noise \citep{koekemoer11}. The weight
map includes all sources of instrumental and background noise, but not
that of the science objects themselves to allow proper photometry with
tools like SExtractor. As a first step we therefore produce simulated
images that include the PSF-convolved objects (including the
Poissonian object noise) but not the simulated background and
read-noise we would normally apply. Instead, we add in these sources
of noise by directly taking them from the inverse variance maps.  As a
final step we need to take into account that in the real CANDELS
images the noise is spuriously correlated as a result of the drizzling
process used to combine the many individual exposures. The amount of
noise correlation depends on the multidrizzle parameters, which for
the CANDELS UDS data amounts to a pixel rms noise reduction of a
factor of 2 \citep{casertano00}. We introduce this noise correlation
in our mock images by smoothing the mock images with a small Gaussian
kernel (of about 1.5 pixels FWHM, in this case).
\begin{figure}
\begin{center}
\includegraphics[width=0.85\columnwidth]{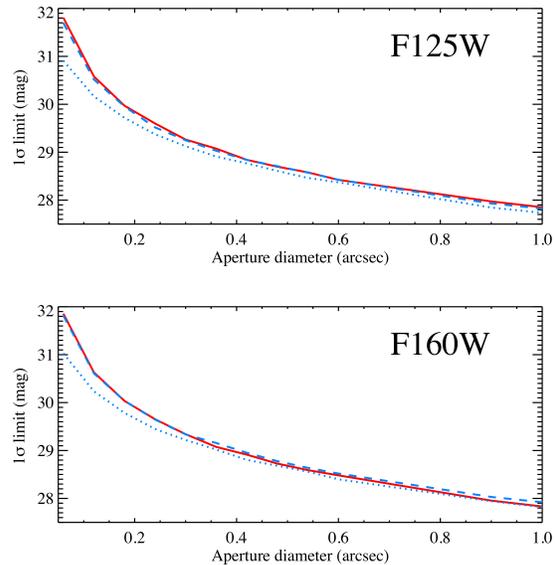}
\end{center}
\caption{\label{fig:candelsdepth}The background noise versus aperture
  size in the real CANDELS images (red solid lines) and our mock
  CANDELS images (blue dashed lines). The noise in the mock and real
  data is nearly identical. The true image noise in the absence of
  correlated noise introduced by the drizzling process is somewhat
  higher (blue dotted lines).}
\end{figure}

\begin{figure}
\begin{center}
\includegraphics[width=0.85\columnwidth]{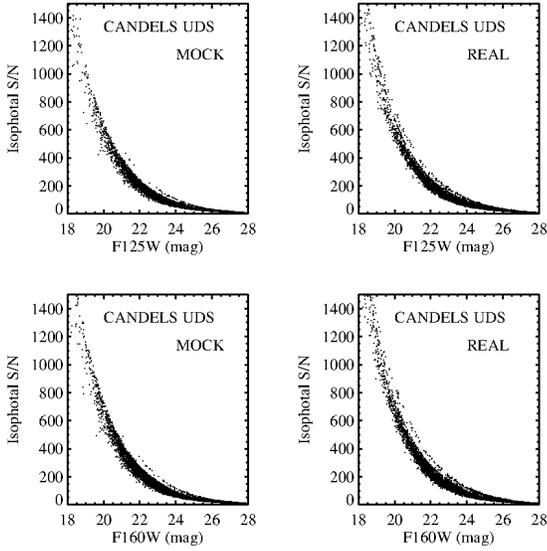}
\end{center}
\caption{\label{fig:candelssn}The isophotal signal-to-noise ratio in $J_{125}$
  and $H_{160}$ versus the total magnitude of detected objects as measured by
  SExtractor. The signal-to-noise ratio distributions in the mock and
  the real CANDELS images are very similar.}
\end{figure}

In the left panels of Fig. \ref{fig:candelsimages} we show a portion of the final
simulated CANDELS images in the filters F125W and F160W. In the
panels on the right, we show a region of the real CANDELS UDS images,
displayed at the same zoom level and at the same colour stretch as the
mock images shown on the left. At a qualitative level the images are
remarkably similar. Note that this is not only the result of our
accurate image simulation technique, but also because our input galaxy
population apparently has a striking resemblance to the observed one
(e.g., in terms of number density, clustering, size and shape
distributions, and brightness). However, before we can compare the
galaxy populations in the simulated and the real data, we need to
ensure that the image properties of our simulated data are indeed
quantitatively similar to the real data. In
Fig. \ref{fig:candelsdepth} we show the measured background noise
fluctuations as a function of aperture diameter as measured in the
real CANDELS images (red solid lines) versus that measured in our
simulated data set. The blue dotted lines indicate the (true) noise
level in the absence of correlated noise. When we introduce the
correlated noise resulting from the drizzling process, we get a near
perfect match between the simulated (blue dashed lines) and real (red
solid lines) CANDELS UDS images. As a second test, we look at the
distribution of signal-to-noise (S/N) for objects detected in the real
and simulated images. We ran SExtractor using identical detection
parameters on the real and simulated images, and plot the isophotal
S/N versus the measured magnitudes. The result is shown in
Fig. \ref{fig:candelssn} for the mock data (left panels) and the real
data (right panels). Again, the S/N distributions are very similar
between the real and simulated data, indicating that our image
simulations are accurate.

In \S\ref{sec:counts} we show an application of these CANDELS
simulations by comparing the galaxy number counts in our semi-analytic
mock lightcones with those extracted from our mock images, and with
those in the real CANDELS images. The simulated CANDELS data produced
here are part of our first scientific data release as announced In
\S\ref{sec:publicrelease}.

\section{Examples of applications}

\subsection{Galaxy number counts in CANDELS}
\label{sec:counts}

\begin{figure*}
\begin{center}
\includegraphics[width=0.9\columnwidth]{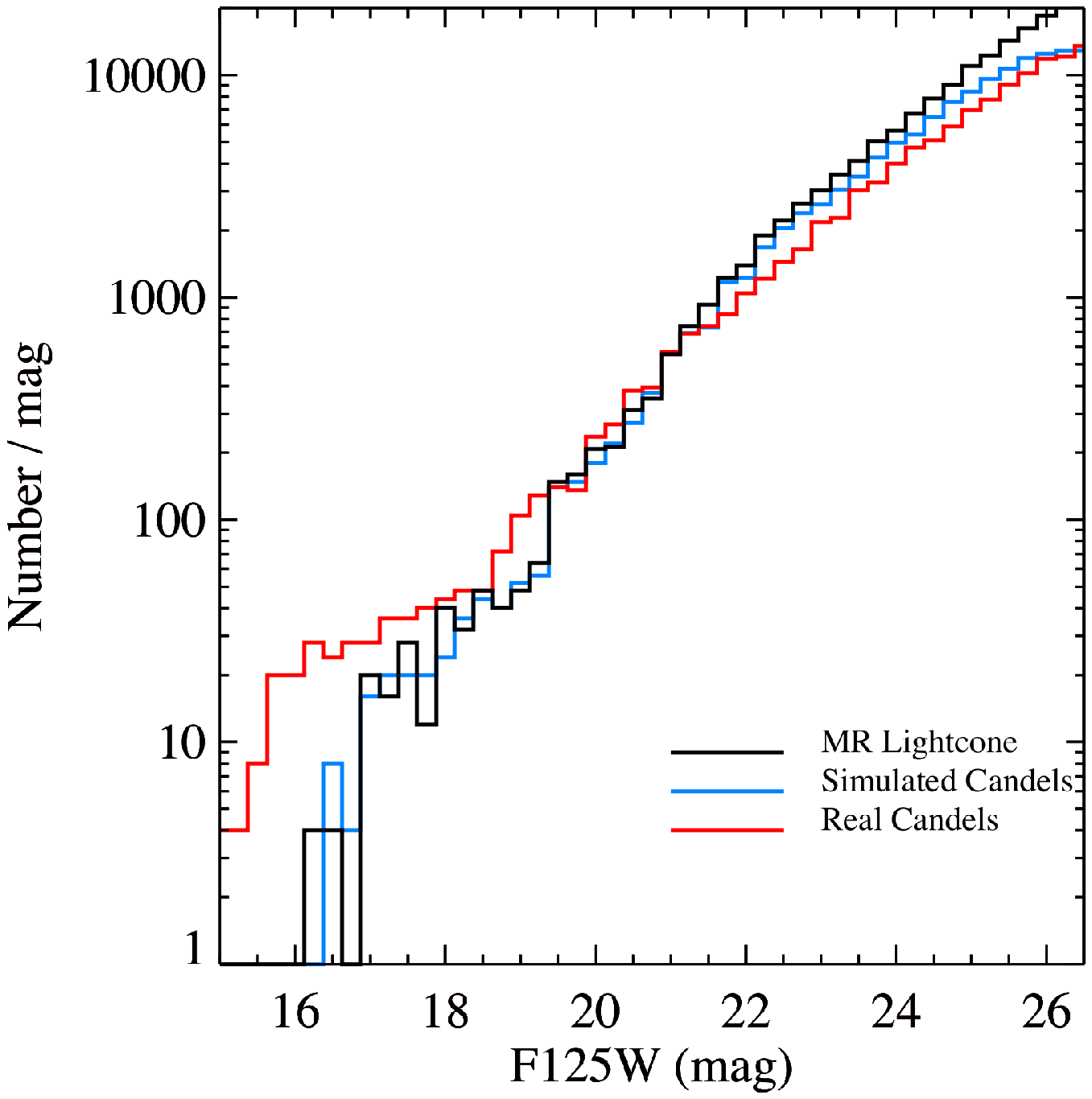}\hspace{1cm}
\includegraphics[width=0.9\columnwidth]{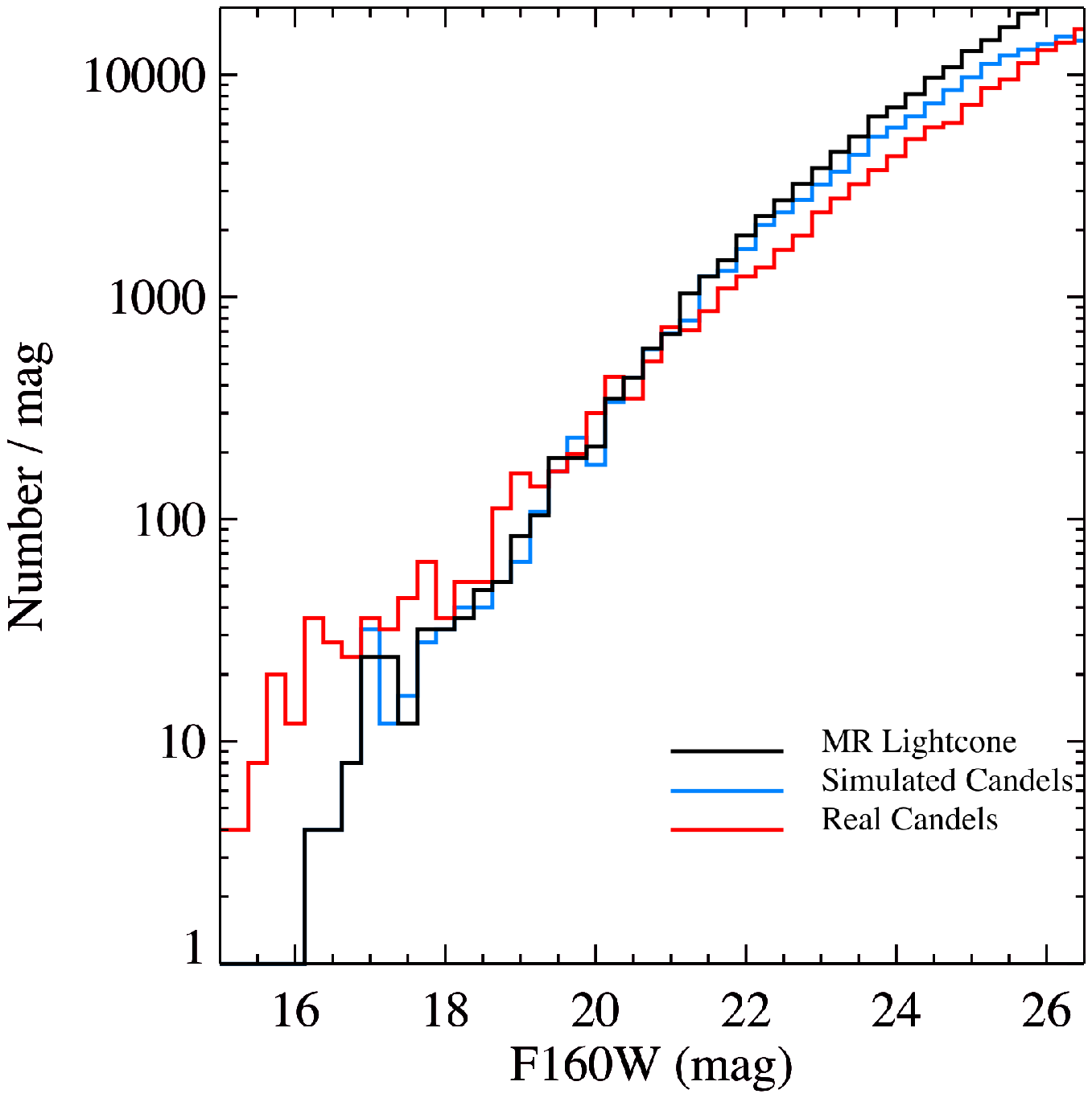}
\end{center}
\caption{\label{fig:candelscounts}Galaxy number counts in $J_{125}$
  and $H_{160}$ as a function
  of magnitude in the lightcone (black solid line), the real CANDELS
  UDS data (red solid line), and those extracted from the simulated
  CANDELS UDS data (blue solid line). While the lightcone data is
  known to over-predict the observed number counts to some extent, the
  discrepancy between the observations and the model predictions is
  significantly reduced after folding the lightcones through the MRObs
  and performing object detection and photometry from the mock images
  as performed on the real images. The large difference between the
  real and simulated data at the bright end is due to Galactic stars
  that are absent in our simulations.}
\end{figure*}

One of the most basic tests that are used to test the accuracy of
semi-analytic model predictions is to compare the number counts of
galaxies observed as a function of apparent magnitude in some band
with those predicted by a mock lightcone observation constructed from
the semi-analytic model as described in \S\ref{sec:lightcones}.
However, as discussed in the introduction, these light-cones do not
suffer from any of the observational effects afflicting real
observations.

The MRObs approach to modeling discussed in sections \ref{sec:igm} to
\ref{sec:sextractor} allows us to make comparisons
between observations and semi-analytic predictions in a highly
realistic way. By simulating
  a mock survey matched to the real observations that one wants to
  compare with, and then running source extraction and photometry
  software on the mock and real images in identical manner, we can in
  principle assess how certain observational biases affect the
  interpretation of the real data, and how this impacts on our
  comparison of those real data with the simulations. Of course, one
  has to be aware that other biases may be introduced simply due to
  the fact that the models are likely no perfect match to the real
  data. 

\begin{figure*}
\begin{center}
\includegraphics[width=0.9\textwidth]{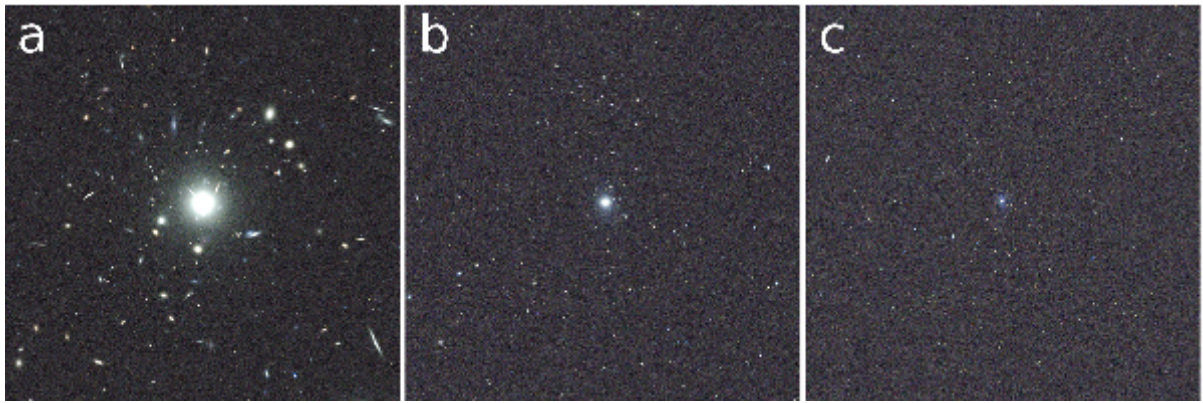}
\end{center}
\caption{\label{fig:sdss}A high-mass galaxy cluster as it would appear
  at redshifts of $z\approx0.02$ (panel a), $z\approx0.09$ (panel b),
  and $z\approx0.21$ (panel c) in an SDSS-type survey. These cluster
  images are based on our lightcone aiming technique described in
  \S\ref{sec:aiming}.}
\end{figure*}

\begin{figure*}
\begin{center}
\includegraphics[height=0.45\textwidth]{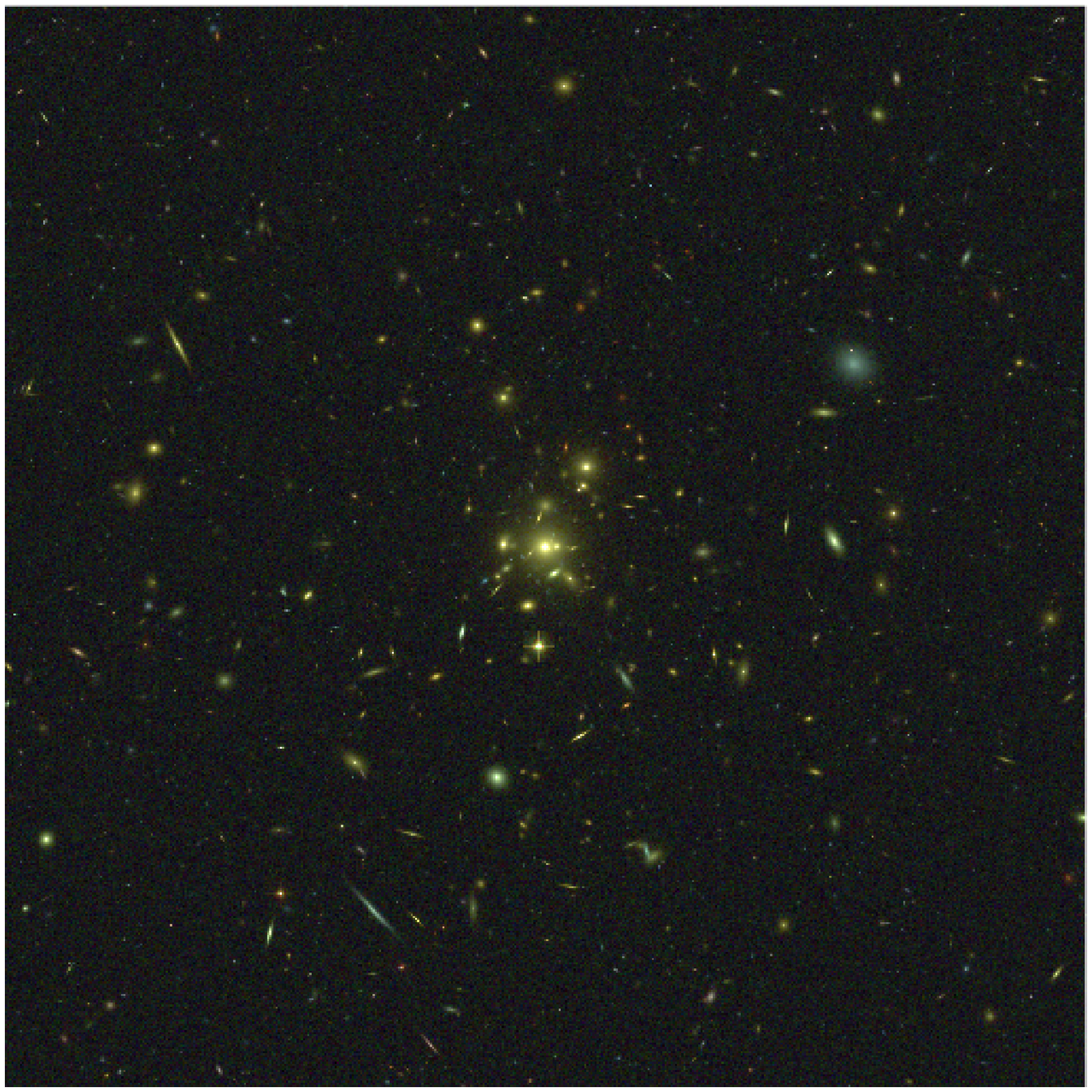}
\includegraphics[height=0.45\textwidth]{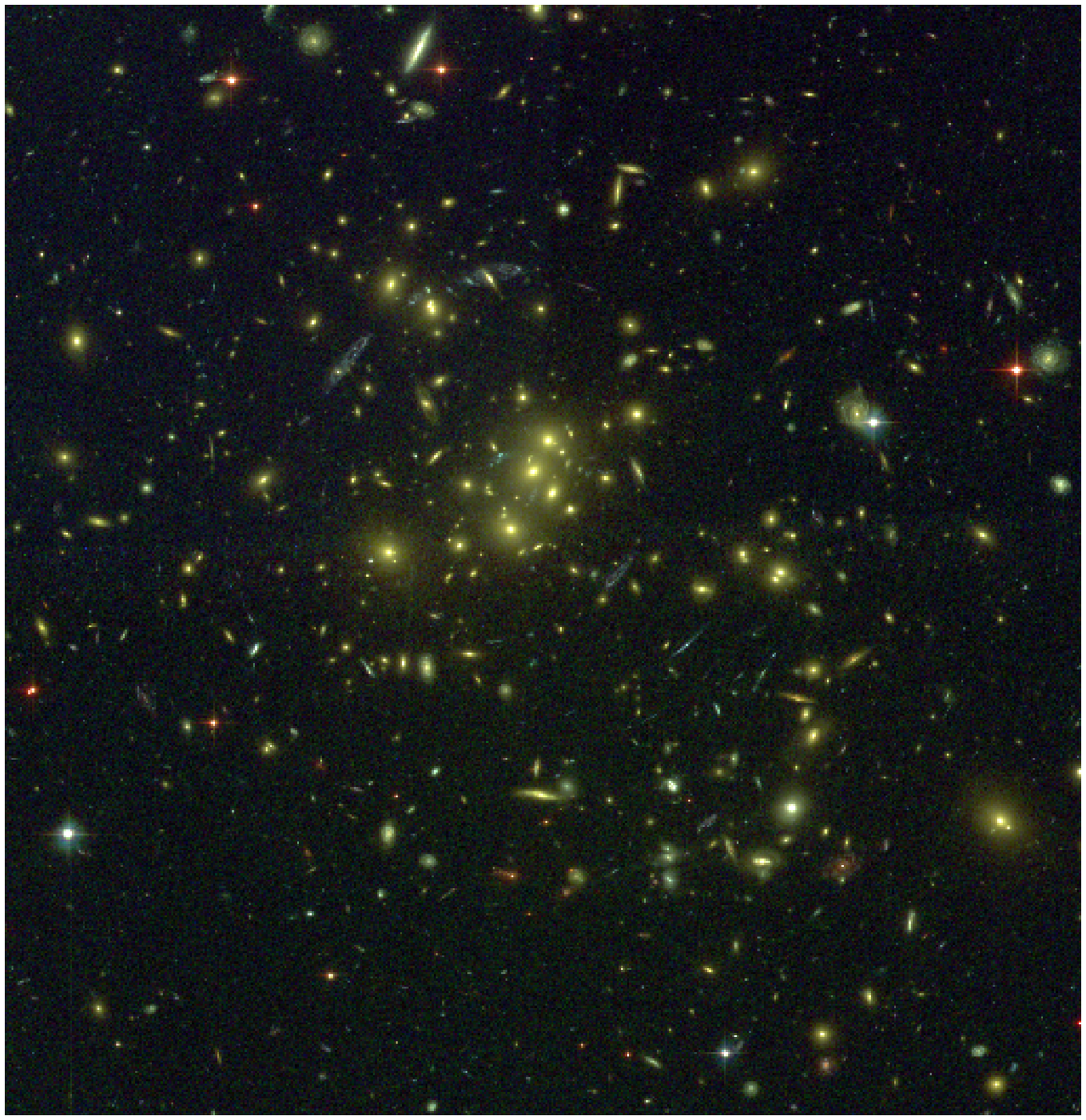}
\end{center}
\caption{\label{fig:cl0024}Mock HST/ACS $g_{475}r_{625}z_{850}$ image of a massive
  galaxy cluster in the MR simulations seen at $z=0.4$ (left) versus a
  real HST/ACS $g_{475}r_{625}z_{850}$ image of the galaxy cluster Cl0024 at $z=0.4$
  \citep{jee07} (right). These cluster images are based on our
  lightcone aiming technique described in \S\ref{sec:aiming}.}
\end{figure*}

\begin{figure*}
\begin{center}
\includegraphics[width=0.9\textwidth]{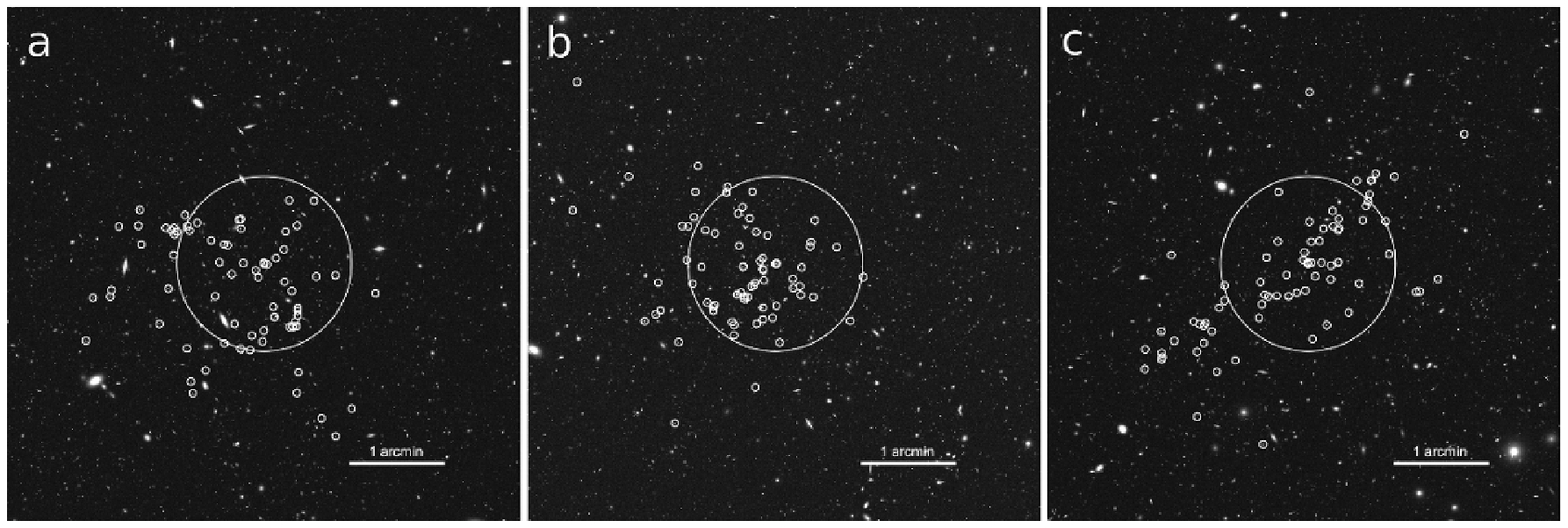}
\end{center}
\caption{\label{fig:hizcluster}Mock HST/ACS $V_{606}i_{775}z_{850}$ colour composite
  image of a massive galaxy cluster at $z\approx1.07$ viewed from
  three different directions. While the projected distribution of
  cluster galaxies appears spherical in the first two orientations
  (left and middle panels), it appears highly filamentary in the third
  orientation (right panel), indicating that projection effects can be
  important. The virial radius of the central halo is marked by a
  yellow circle.}
\end{figure*}

As a first example, we compare the number counts measured from a mock lightcone to those
derived from our mock images based on that
lightcone. Fig. \ref{fig:candelscounts} shows the two types of number counts from our
simulated CANDELS UDS data in blue compared to the plain light cone
data in black. The counts extracted from the simulated image were not
corrected for completeness. At bright magnitudes ($J_{125},H_{160}\lesssim22$ mag)
the counts are in good agreement, but they diverge toward fainter
magnitudes counts detected in the images compared to the lightcone on
which the mock images are based. The extracted counts are about a
factor of 2 lower than the lightcone counts at $J_{125},H_{160}\sim26.5$ mag.

The red lines in the figure show number counts measured in the real
CANDELS UDS data (no completeness corrections applied). At the faint
end, the lightcone substantially over-predicts the observed counts,
similar to discrepancies between semi-analytic predictions and
observations found in earlier studies. However, it is very interesting
to note that the difference between the semi-analytic predictions and
the real number counts becomes smaller when we compare the real data
to our mock data. Simply by `observing' the lightcone we already lose
a significant number of galaxies that would not be detected in a real
observation (if the lightcone was an accurate reflection of reality).

The results presented in Fig. \ref{fig:candelscounts} suggest that it
is important to take observational effects into account when comparing
real data with simulations. These effects need to be quantified before
one can change the parameters in a semi-analytic model to better match
the observations. With the mock data produced by and published through
the MRObs these tests can now be performed easily. A more detailed
analysis of the number counts in synthetic observations as predicted
by the MRObs compared to those predicted by ordinary semi-analytical
models will be presented in a follow-up paper.

\begin{figure*}
\begin{center} 
\includegraphics[width=\textwidth]{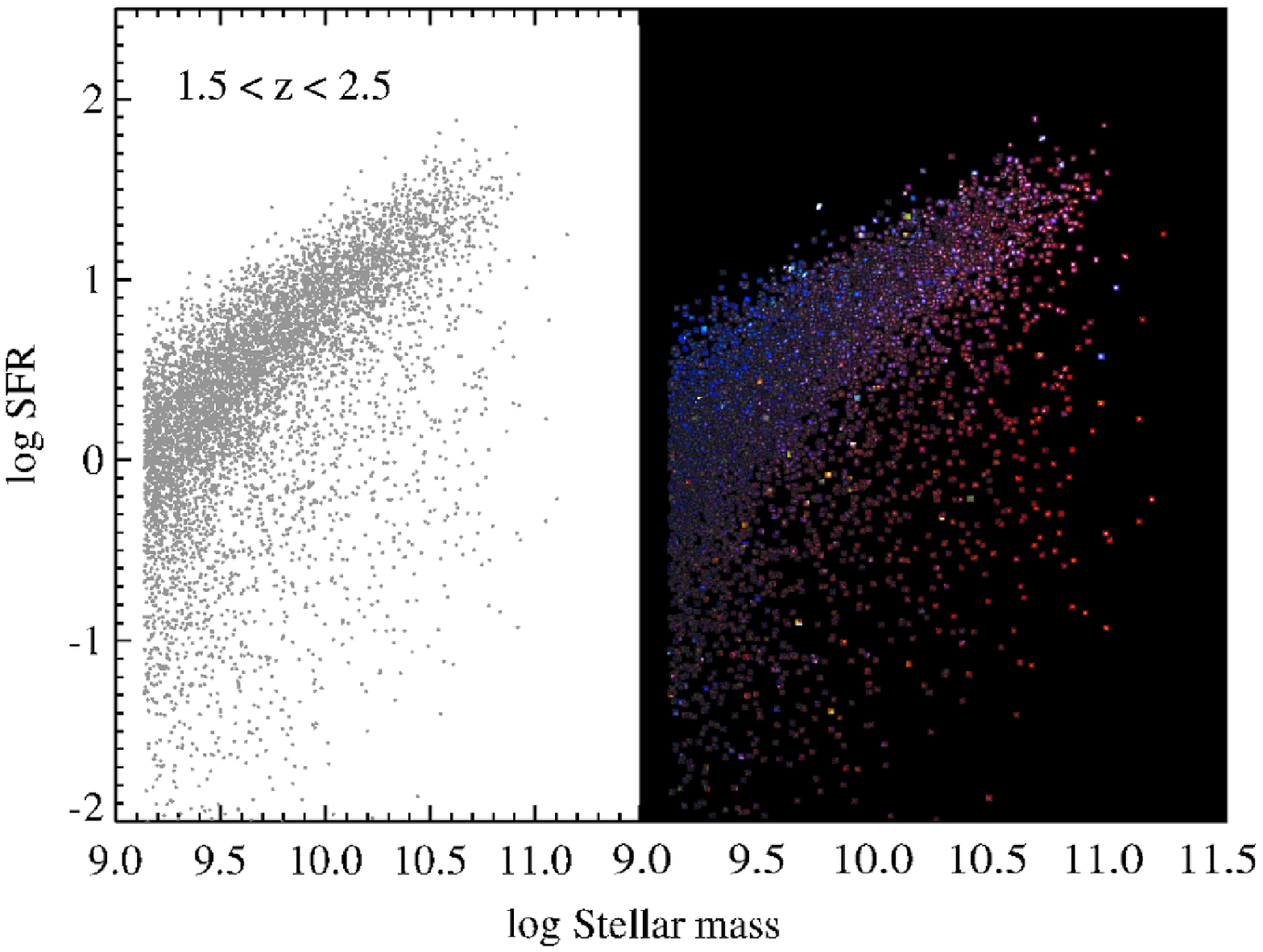} 
\end{center} 
\caption{\label{fig:msfr}MRObs visualisation of the
  structural properties of galaxies as predicted by the semi-analytic model. In
  the left panel we show the stellar mass versus SFR diagram for
  galaxies between $z=1.5$ and $z=2.5$ made directly from the
  lightcone catalog. In the panel on the right, we show the actual
  morphologies of the galaxies as predicted by the SAM by
  showing $1\arcsec\times1\arcsec$ postage stamps extracted from a
  simulated HST image (the image stamps were taken from a mock $B_{435},i_{775},H_{160}$ 
  colour-composite image based on the HST/ERS survey). Quiescent
  objects that lie below the main star-forming sequence appear both
  redder and more compact compared to objects on the star-formation
  sequence. The colours, sizes and structural properties of these type
  of galaxy images can be compared directly to those of galaxies found
  in real data.}
\end{figure*}

\subsection{The properties of galaxy clusters at low and high redshift}
\label{sec:examples_clusters}

Our new lightcone `aiming' technique described in \S\ref{sec:aiming}
offers an efficient way for predicting the detailed observational
properties of, for example, galaxy clusters. Here we present mock SDSS
and HST observations of a massive galaxy cluster at different
redshifts and orientations. The cluster was selected from the roughly
3,000 clusters in the MR, and has a total dark matter mass of
$\sim7\times10^{14}$ $M_\odot$ at $z=0$. The selection was performed
using the table of friends-of-friends groups in the MRDB.  After finding FOF groups in the right mass range, a random
selection was made of a cluster. That cluster was traced backwards in
time using the table with halo merger trees.  At desired redshifts the
position of the cluster's main progenitor was returned.  That
position, together with a direction and using the co-moving distance
corresponding to the redshift, was used to define a light cone that
had the cluster at its center and at exactly the correct redshift. 
This cone was then observed using a few different virtual
telescope configurations.

In the first example, we have produced mock SDSS images in $g'r'i'$
showing what this cluster would look like at redshifts from $z=0.02$
to $z=0.21$ (Fig. \ref{fig:sdss}). These mock data can be compared
directly with real clusters found in the SDSS.  It is clear from
Fig. \ref{fig:sdss} that the study of galaxy clusters in the SDSS
survey becomes challenging already at moderately high redshifts.  As a
second example, we therefore show a mock image of the same cluster,
now seen at $z=0.4$ and observed with HST/ACS in the filters
$g_{475}r_{625}z_{850}$ (Fig. \ref{fig:cl0024}, left panel). In the right panel we show an
actual HST image of the well-studied $z=0.4$ cluster Cl0024 with a
comparable dark matter mass \citep{jee07,harsono09}. The images were
produced using identical parameter settings in the software that
produces the colour images from the \texttt{.fits} files (see
\S\ref{sec:pyramids}). Although both clusters have a dominant
population of red sequence galaxies that appear almost identical in
these HST/ACS colours, the real cluster appears to have a greater
number of cluster galaxies than the mock cluster. The MR contains
thousands of these type of galaxy clusters suitable for
data-mining. Users will be able to use mock observations such as these
to compare the properties of simulated and real clusters in a
quantitative manner. Our third example highlights another unique
feature of our improved lightcone technique, which allows us to
produce observations of structures seen from different
directions. Each light cone is created following the same principle as
above, the only difference being that the cluster is observed from
different directions.  In Fig.~\ref{fig:hizcluster} we show mock HST
images in $V_{606}i_{775}z_{850}$ of the same cluster shown before, but now at
$z\approx1.07$. Panels show the exact same cluster viewed from three
different directions, with (proto-)cluster galaxies having
log$M_*>10M_\odot$ marked with white circles. The large yellow circle
marks the virial radius of the central halo. While the projected
distribution of cluster galaxies appears roughly spherical in the
first two orientations (left and middle panels), it is much more
filamentary in the third orientation (right panel). The line of sight
velocity dispersions in the three cases are 807, 704, and 568 km
s$^{-1}$. This example illustrates that projection effects are
important to take into account when studying the assembly of galaxy
clusters, especially at high redshift where both the samples of
clusters and the number of identified cluster galaxies are relatively
small. The multi-wavelength nature of the MRObs data allows for the
detailed testing, calibrating and tuning cluster detection algorithms
using physically-motivated cluster samples.

\begin{figure*}
\begin{center}
\includegraphics[width=0.25\textwidth]{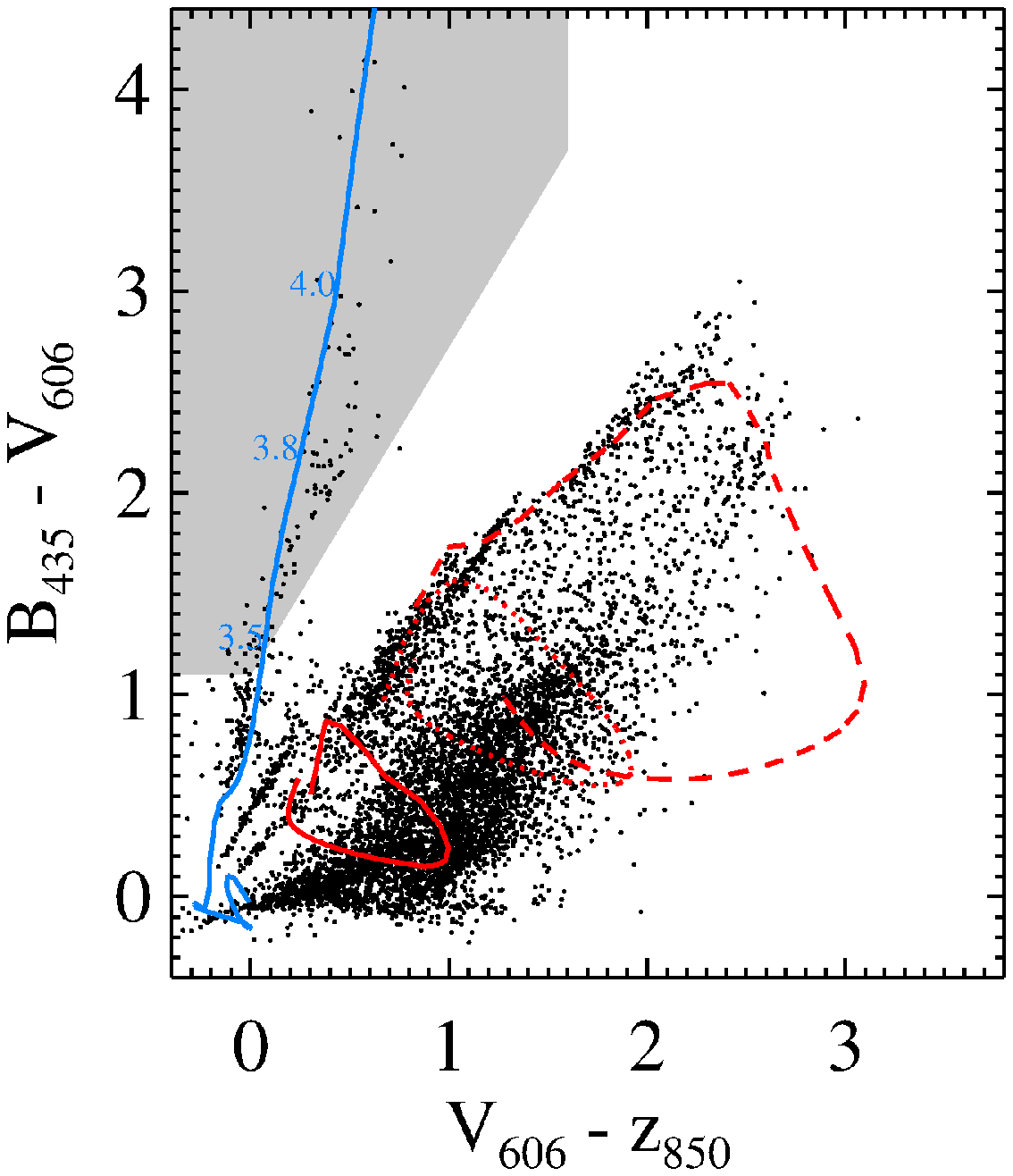}\hspace{0.05\textwidth}
\includegraphics[width=0.25\textwidth]{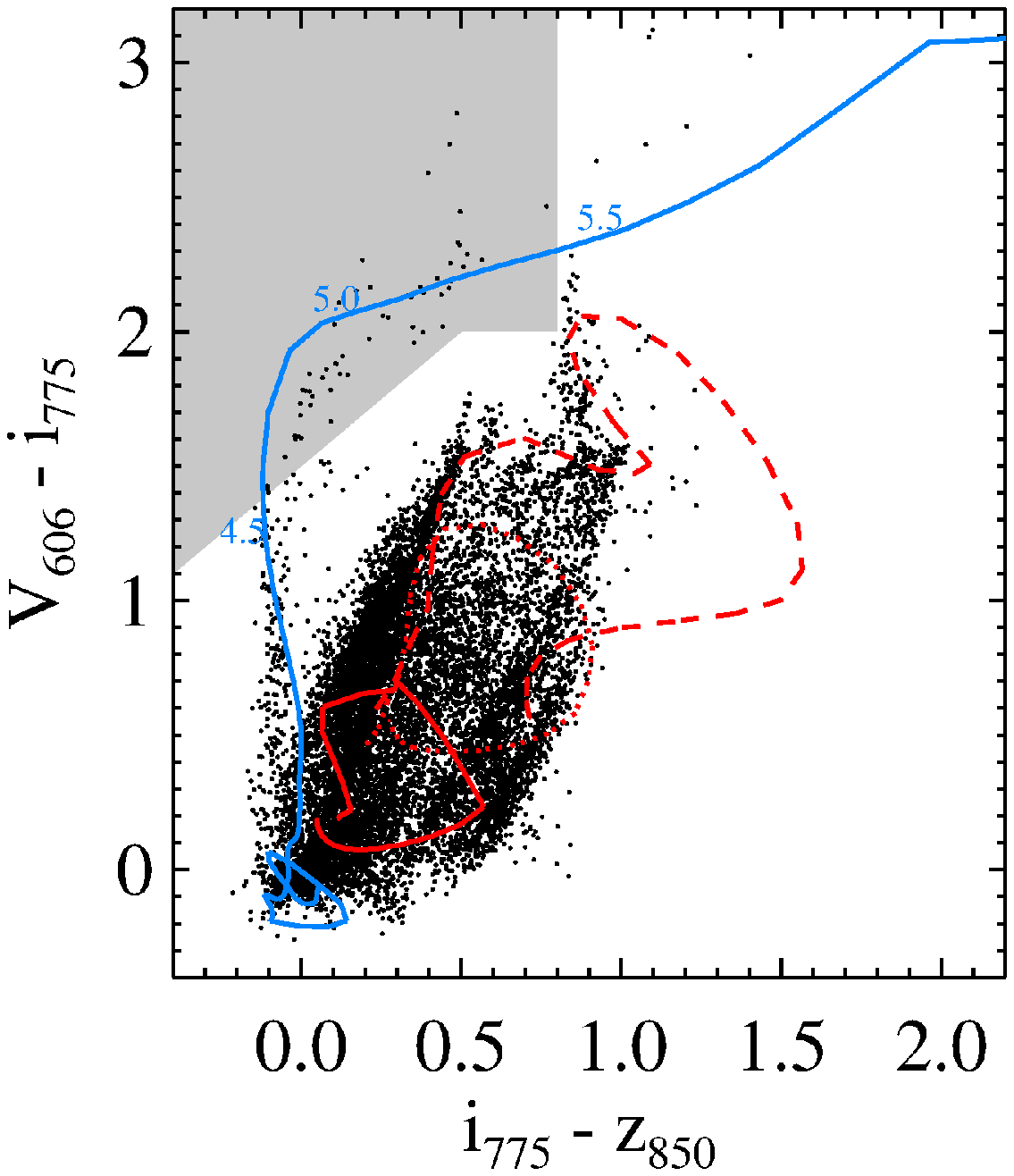}\hspace{0.05\textwidth}
\includegraphics[width=0.25\textwidth]{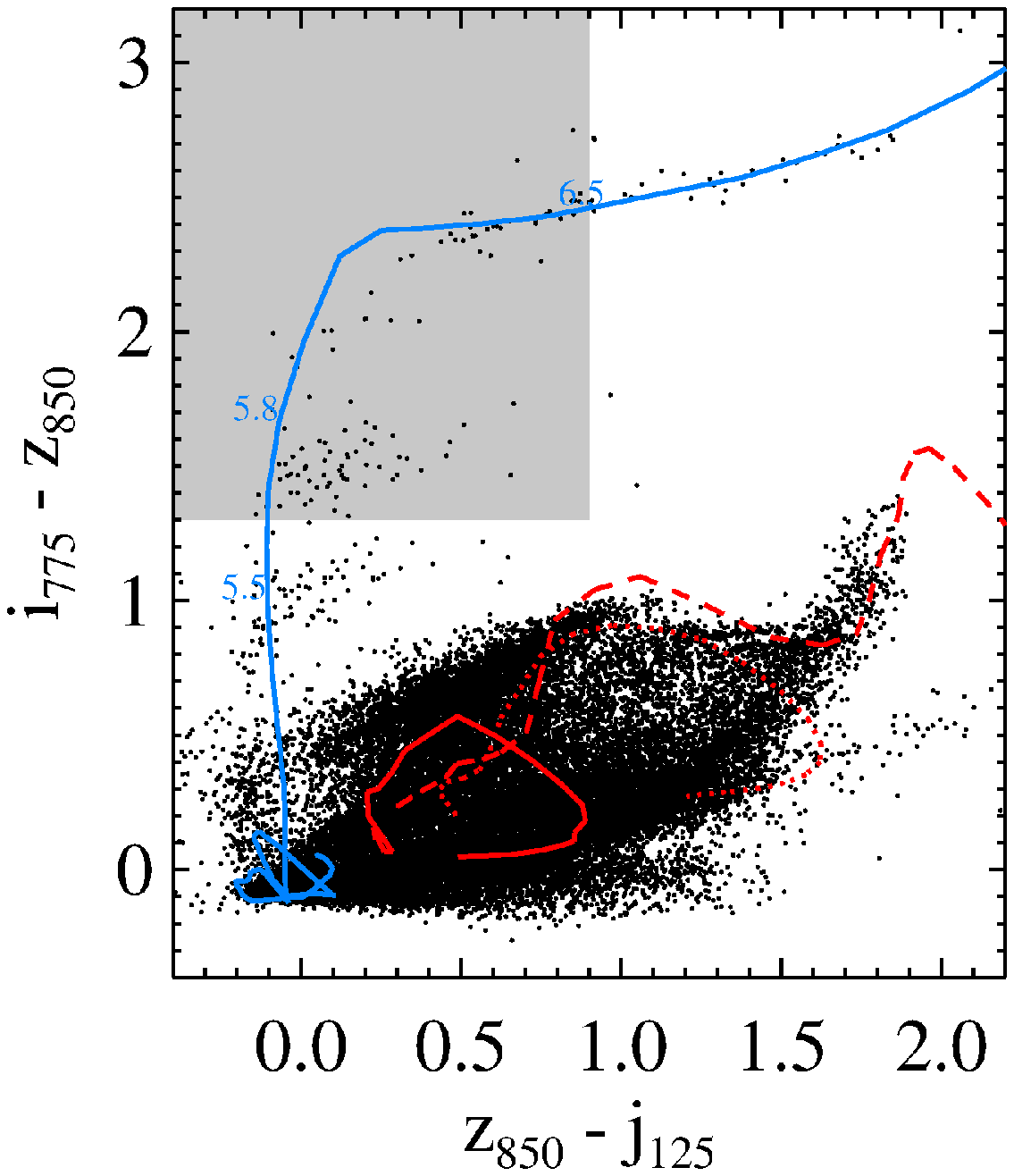}\\
\vspace{5mm}
\includegraphics[width=0.25\textwidth]{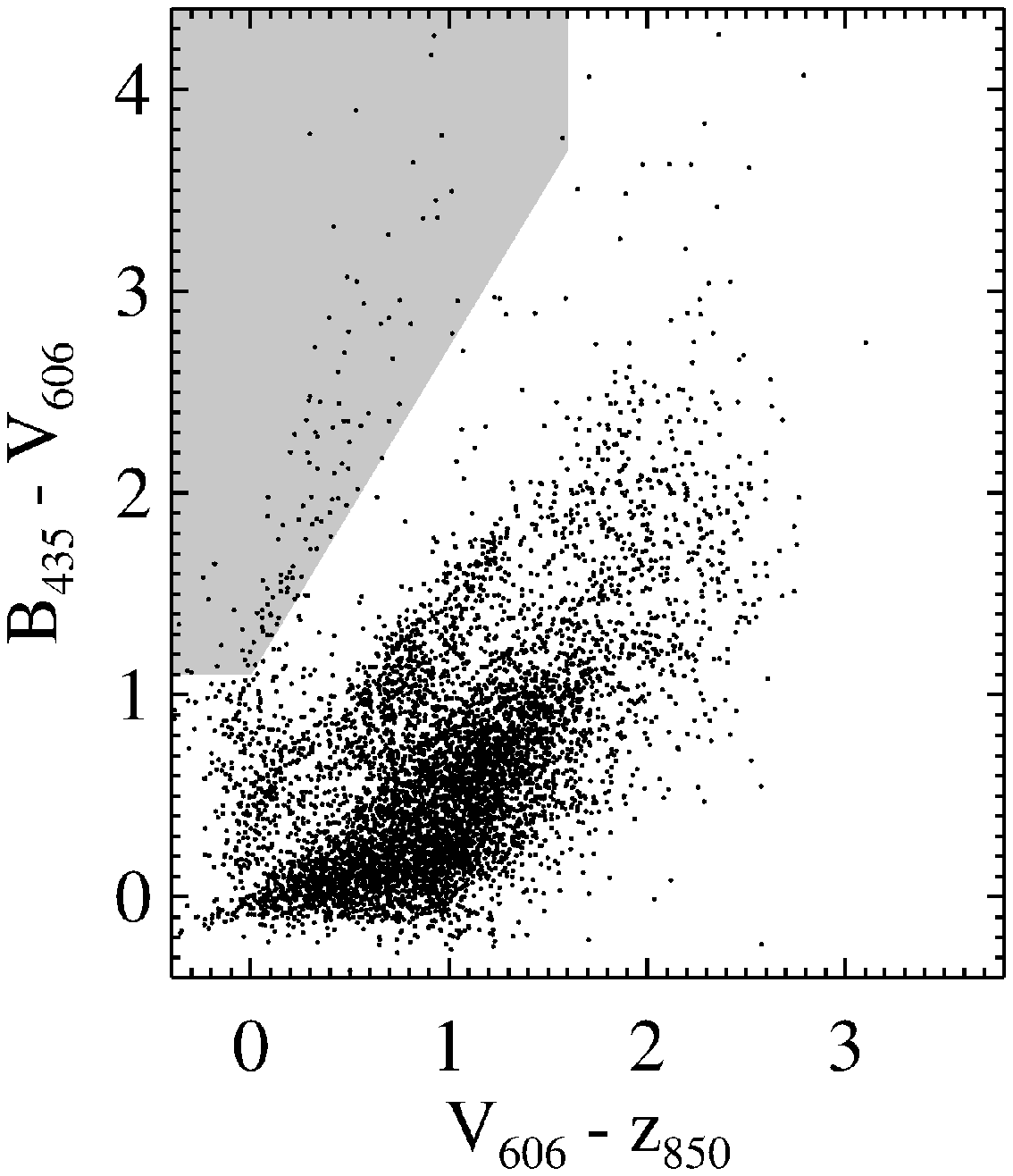}\hspace{0.05\textwidth}
\includegraphics[width=0.25\textwidth]{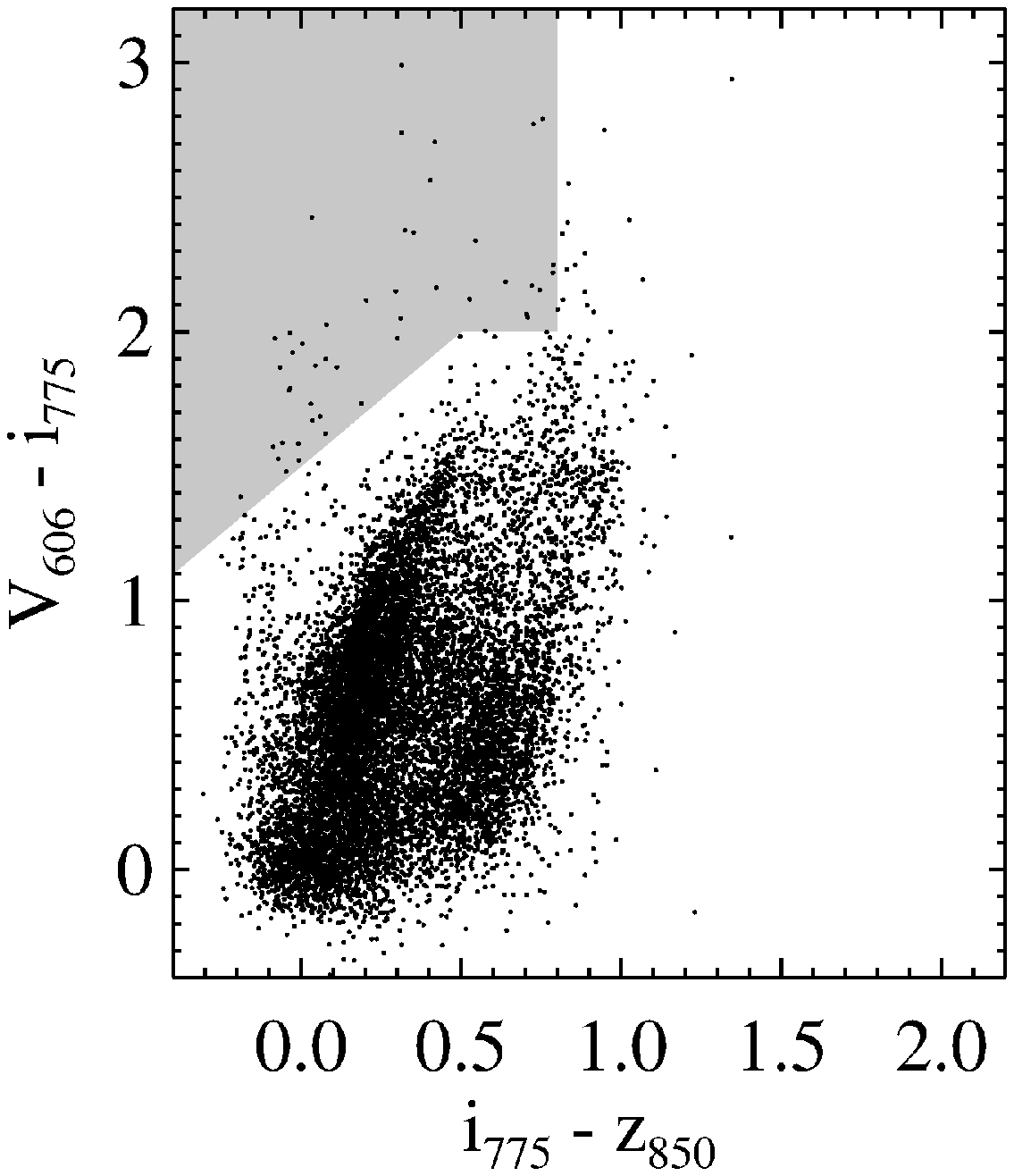}\hspace{0.05\textwidth}
\includegraphics[width=0.25\textwidth]{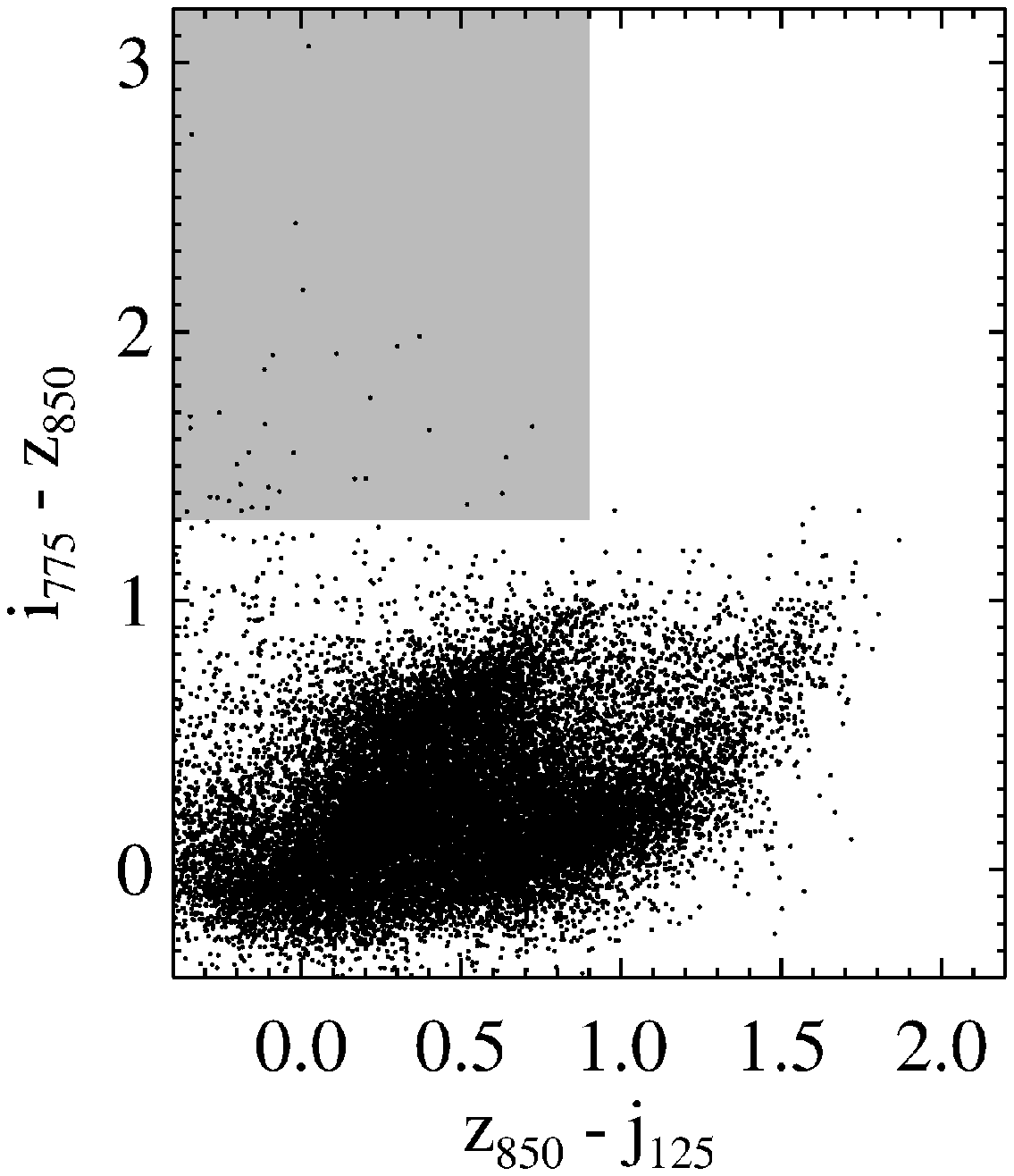}
\end{center}
\caption{\label{fig:dropouts}Colour-colour diagrams commonly used to
  select galaxy samples at $z\sim4$ ($B_{435}$-dropouts, left panels),
  $z\sim5$ ($V_{606}$-dropouts, middle panels)), and $z\sim6$ ($i_{775}$-dropouts,
  right panels). Panels on the top show the colour distributions of all
  objects in the lightcone. Panels on the bottom show the colour
  distribution of the objects detected in mock images based on the
  same lightcone. Tracks indicate the typical colours of simple galaxy
  templates for various low redshift populations (red lines;
  irregulars: solid, Sbc: dotted, Elliptical: dashed) and high
  redshift dropouts (blue, with redshifts marked along the
  tracks). Shaded regions mark the colour-colour selection windows
  commonly used to select high redshift dropout candidates.}
\end{figure*}

\subsection{Colours and structural properties of galaxies}
\label{sec:msfr}

Another new test facilitated by the MRObs is comparison of the
structural properties of galaxies in the semi-analytic model to those
in real observations. In Fig. \ref{fig:msfr} we show a stellar mass
versus SFR diagram for galaxies between $z=1.5$ and $z=2.5$ selected
from one of our mock lightcone catalogs.  In the panel on the right,
we show $1\arcsec\times1\arcsec$ postage stamps indicating the 
appearance of these galaxies in our simulated HST data (the image stamps are drawn from a mock
$B_{435},i_{775},H_{160}$ colour-composite image based on the HST/ERS survey). These
mock data can be used to measure galaxy structural properties (e.g.,
Sersic index, bulge-to-disk ratio, inclination), sizes and colours in
exactly the same way as typically performed on real data only. By
comparing measurements made based on the mock images with the exact
physical quantities given by the semi-analytic model users could test
how well such values can be recovered for a given data set, or for a
given galaxy population. It also allows users to compare
the structural properties of mock and real
galaxies in a relatively unbiased way.

\subsection{Selection of high redshift dropout galaxies}
\label{sec:dropouts}

The last example we show here is the use of the MRObs in the selection
of high redshift dropout galaxies from deep multi-wavelength imaging
surveys and predictions thereof \citep[e.g.][]{baugh98,blaizot04,guo09,overzier09a}. 
In Fig. \ref{fig:dropouts} we show the colour-colour diagrams
typically used to isolate galaxy samples at $z\sim4$ ($B_{435}$-dropouts),
$z\sim5$ ($V_{775}$-dropouts), and $z\sim6$ ($i_{775}$-dropouts). Objects at these
high redshift suffer severe attenuation from the IGM in their spectra
blue-ward of \lya\ (see \S\ref{sec:igm}). Consequently, these
objects can be isolated from lower redshift galaxy population, as
their Lyman break redshifts through a strategically chosen set of
filters. Panels on the top show colour distributions for all objects
found in one of our mock lightcones. The bottom panels show those
objects that were detected in a mock survey based on the same
lightcone.  The limiting magnitudes used for the lightcone and for the
extracted catalog were the same. This figure highlights some of the
main differences between a pure semi-analytic model prediction (top
panels) and what an observer actually sees (bottom panels). The colours
of galaxies extracted from mock images are significantly scattered
compared to their true (input) colours, making it harder to distinguish
between low and high redshifts, or to derive their physical properties
(e.g., redshift, mass, dust, star formation history, SFR) based on
fitting their observed colours to a set of spectral synthesis
models. It is straightforward to study and quantify such effects
through the use of these kind of mock data. 
In the MRDB SQL queries can be performed to cross-match the SExtractor output
catalogs to the lightcone or semi-analytic input catalogs, allowing one
to investigate in detail the offsets between intrinsic and apparent
properties, and to study which galaxies are included and excluded by
certain observational selection criteria (e.g. colour-colour
selections).

\section{Public access to the MRObs data}
\label{sec:publicrelease}

\subsection{MRObs database}

As described above, the MRObs builds upon and extends the popular
MRDB. Apart from the images, all the
datasets produced by the MRObs and described in this paper are stored
in a database that is accessible through the same interface as the
MRDB itself\footnote{See http://gavo.mpa-garching.mpg.de/Millennium
  for a publicly accessible website giving access to the
  milli-Millennium database and information on how to gain access to
  the full database.} and can be directly joined to the existing data
sets.  Here we give a summary description of the database and access
methods, focusing on the new data products and how they are linked to
the existing ones.

The MRDB is a relational database\footnote{This is not the place to
  describe relational databases in detail, there is sufficient
  information available online.}, where data sets are stored in tables
(relations).  A table generally stores objects of a particular type,
with properties of these objects stored in columns.  For example we
have tables storing the the positions and velocities of particles from
an N-body simulation, albeit a small one.  We have tables with FOF
groups and sub-halos as well as galaxies and many more.  The web site
giving access to these tables provides all information about the
structure of the database.

An important feature of relational database design in general, and the
MRDB in particular, is the possibility to manifest
relations or links between objects in different tables.  For example, a
galaxy in the Munich semi-analytical model is always embedded in a
subhalo.  This relation is stored in the tables with galaxies as a
column storing the (unique) identifier of the corresponding halo.  The
MRDB has a particularly rich set of such relations,
especially where it deals with the relations between objects of the
same type at different times \citep{lemson06}.

Recent additions to the database were the results of the latest
version of the Munich SAM from \cite{guo11} and pencil-beam and all
sky light-cones derived from these in \cite{henriques12}.  The images
produced by the MRObs from such light-cones do not lend themselves
easily for storing in a database.  However the SExtractor catalogues
extracted from the images have been stored and we also have tables
storing the different IGM absorption models described in
\S\ref{sec:igm}. More information
and examples on how to apply and cross-correlate the various MRDB and MRObs
data sets are documented at the URL given below. 

\subsection{Data products of the MRObs}

The MRObs delivers a number of entirely new data products to the
community that are useful for independent analysis, or for serving as
the starting point for new simulations. Here we will briefly describe
the different types of new products.

\subsubsection{Multi-wavelength lightcone catalogs with structural properties} 

The random field lightcones released as part of this paper are
identical to the 24 multiwavelength lightcones measuring
$1.4\degr\times1.4\degr$ on the sky from \citet{henriques12}, but with
structural information added. The new structural information (sizes of
the disk and bulge components, inclinations and position angles) is
crucial for building the accurate galaxy models predicted by the MR
simulations. These lightcones can be used, for example, to compare
structural properties measured off the simulated images to the true
input values. They can also be used as the starting point for users
wishing to perform their own image simulations using realistic input
catalogs based on the MR. In addition to the `random' lightcones, we
also release entirely new lightcones that specifically target galaxy
clusters at a range of redshifts (see \S\ref{sec:aiming} and
\S\ref{sec:examples_clusters}). All these lightcone catalogs are made
available through the MRDB.

\subsubsection{IGM tables}

We provide tables that list the mean IGM attenuation as a function of
redshift for a range of models (see \S\ref{sec:appendix}). The IGM tables are applied to the
lightcones to predict accurate colours and magnitudes of galaxies with
redshift.

\subsubsection{Object lists}

Information from the structural light cones, the IGM tables, and a
plate scale are combined to generate the input to the SkyMaker code
that we use to create our synthetic `pre-observation' images. These
object lists may be used by other synthetic image simulators.

\subsubsection{Pre-observation maps (`perfect' model images)}

As described in \S\ref{sec:sims}, for each filter we build a so-called
pre-observation or `perfect' image that is based on the input object list. These
images can be seen as a representation of the sky free of noise, PSF,
or background. As such, they are easily convolved, rebinned, and
scaled to match an arbitrary observation (typically a combination of a
given telescope, camera, and exposure).

\subsubsection{Simulated images}

The `perfect' images are turned into synthetic images that simulate  
real observational data. These images
can be downloaded for further analysis. We also provide the PSF images
that were used to convolve the perfect images to the instrument
resolution, as well as documentation providing full details of how the
images were produced.

\subsubsection{SExtractor products} 

The simulated images are processed using SExtractor to produce the
so-called segmentation maps identifying which image pixels correspond
to which detected object, as well as the standard SExtractor output
photometry catalogs. The SExtractor catalogs are made available
through the MRDB where they can be searched or cross-matched with
other data, such as lightcone catalogs, semi-analytic snapshots, dark
matter halos, or density fields. The segmentation images are available
for download.

\subsection{Simulated surveys currently available in the MRObs}

In its current deployment, the MRObs offers a number of data sets
conveniently matched to some of the most popular extra-galactic
surveys (e.g. the SDSS, CFHT-LS Wide and Deep, GOODS, UDF, GOODS/ERS,
and CANDELS) for use by the community.  Updates and future data
releases will be announced through the MR web portal (URL given
below), and in forthcoming publications.

\subsection{The MRObs image browser}
\label{sec:browser}

A special feature of the MRObs is that many of the data sets can also
be accessed directly by means of our interactive MRObs image
browser. This is an online tool that allows users to scan over and
zoom into the synthetic images. These images are linked to the backend
database (the MRDB) through a simple point-and-click function that
allows retrieval of detailed information about the galaxies that are
displayed. This is useful, for example, for familiarizing oneself with
the relation between physical and observed properties of different
types of galaxies or galaxies at different redshifts, for selecting
interesting objects from the MR simulations for subsequent analysis,
for comparing the quality expected for different types of data sets or
telescopes, and for didactical and outreach purposes. Here we describe
the main features of the MRObs browser in brief.

\begin{figure}
\begin{center}
\includegraphics[width=\columnwidth]{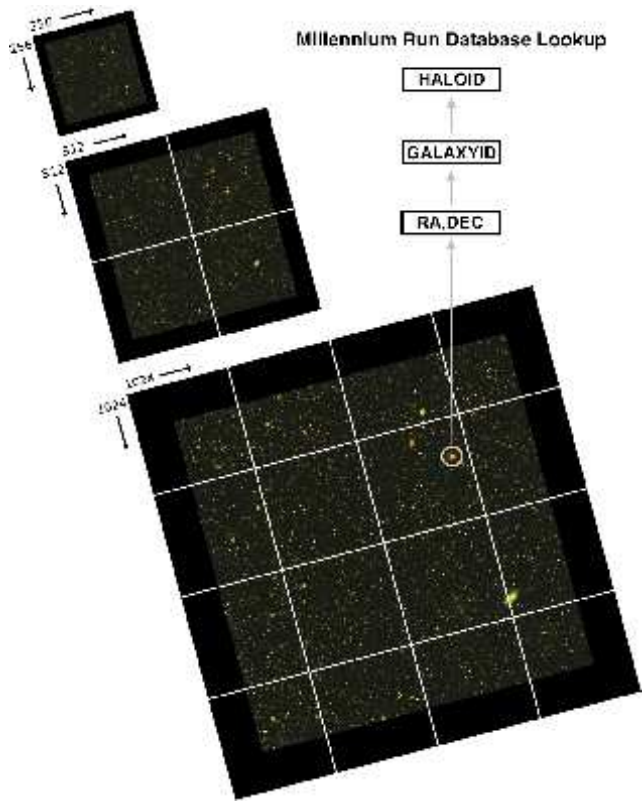}
\end{center}
\caption{\label{fig:pyramid}An image pyramid consisting of three
  levels is shown. At its lowest resolution level, the image consists
  of a single $256\times256$ pixels tile having 1/4th of the true
  image resolution. At the base of the pyramid, the image is divided
  into $4\times4$ full-resolution tiles each measuring $256\times256$
  pixels. The MRObs deep zoom image browser makes heavy use of these
  kinds of tilings for an efficient viewing of the data. The browser
  also makes a translation between pixel coordinates and WCS
  coordinates within each viewport. This conveniently enables the user
  to retrieve high-level properties of any object found within the
  image by matching its sky coordinates to the underlying lightcone
  catalog, and by querying the galaxy or halo catalogs stored on the
  MRDB server based on the \texttt{GALAXYID} or
  \texttt{HALOID} of any matches found.}
\end{figure}

\subsubsection{Deep zoom RGB image pyramids}
\label{sec:pyramids}

The images produced by the MRObs are typically very large. For
example, a simulated HST survey covering an area of
$30\arcmin\times30\arcmin$ at a (drizzled) pixel scale of $0\farcs09$
already measures 20,000$\times$20,000 pixels (400 Megapixels), and in
principle the MRObs could create much larger fields at much higher
resolution than this. These images therefore do not fit on a standard
computer screen.  Using a technology similar to, e.g., Google Maps, the
MRObs browser allows users to efficiently pan around and zoom in such
large, high resolution images. We here describe in some detail how we
have implemented this truly virtual telescope.

First the simulated, multi-wavelength filter images are combined into
false-colour RGB composites. We use the publicly available code
\texttt{STIFF}\footnote{http://astromatic.net/software/stiff} that
handles the conversion from arbitrarily large scientific FITS input
images to standard TIFF format output images
\citep{bertin12}. \texttt{STIFF} automatically (or manually) applies
contrast and brightness adjustments, colour balance and saturation,
and gamma corrections producing colour images that are highly
informative of the level of detail present in the mono-chromatic input
fits images.  When we have multiple bands available for any of the
three RGB channels (for example when making colour composites of data
sets based on more than three filters), we reduce the number of input
images to three by creating variance-weighted averages and use those
as the input for each channel.

From this high-resolution image we then create a so-called `image
pyramid' consisting of representations of this high resolution image
at ever decreasing resolution.  The method is illustrated in
Fig. \ref{fig:pyramid}. The top of the pyramid (level 0) consists of a
single $s\times s$ pixels low resolution image that is a heavily
rebinned version of the original or full-resolution $N\times N$ pixels
image. The next level contains $p^1\times p^1$ image tiles each of $p$
times higher resolution compared to the previous level. At the $n$th
level (corresponding to the base of the pyramid), there will be
$p^n\times p^n$ tiles each representing only a small portion of the
original image but now at its highest resolution.

The browser software\footnote{We use the Deep Zoom technology
  developed by Seadragon/Microsoft embedded in custom written java
  script libraries.} uses this data format to download only those
tiles that at the current zoom level are required to show the image.
This significantly reduces the download time and creates a smooth
transition between the different levels or different regions of the
image when viewed in a web browser.  For example, if we adopt a factor
of $p=2$ scalings between levels, tiles of $s=256$ pixels, and an
original image of $N=32,768$ pixels, the last level (level 7) will
consist of $128\times128$ tiles of $256\times256$ pixels.  This means
that only about 0.1\% of data needs to be downloaded at any time to
display a particular region at its fullest resolution on a
$1024\times1280$ resolution display.

\begin{figure*}
\includegraphics[width=0.7\textwidth]{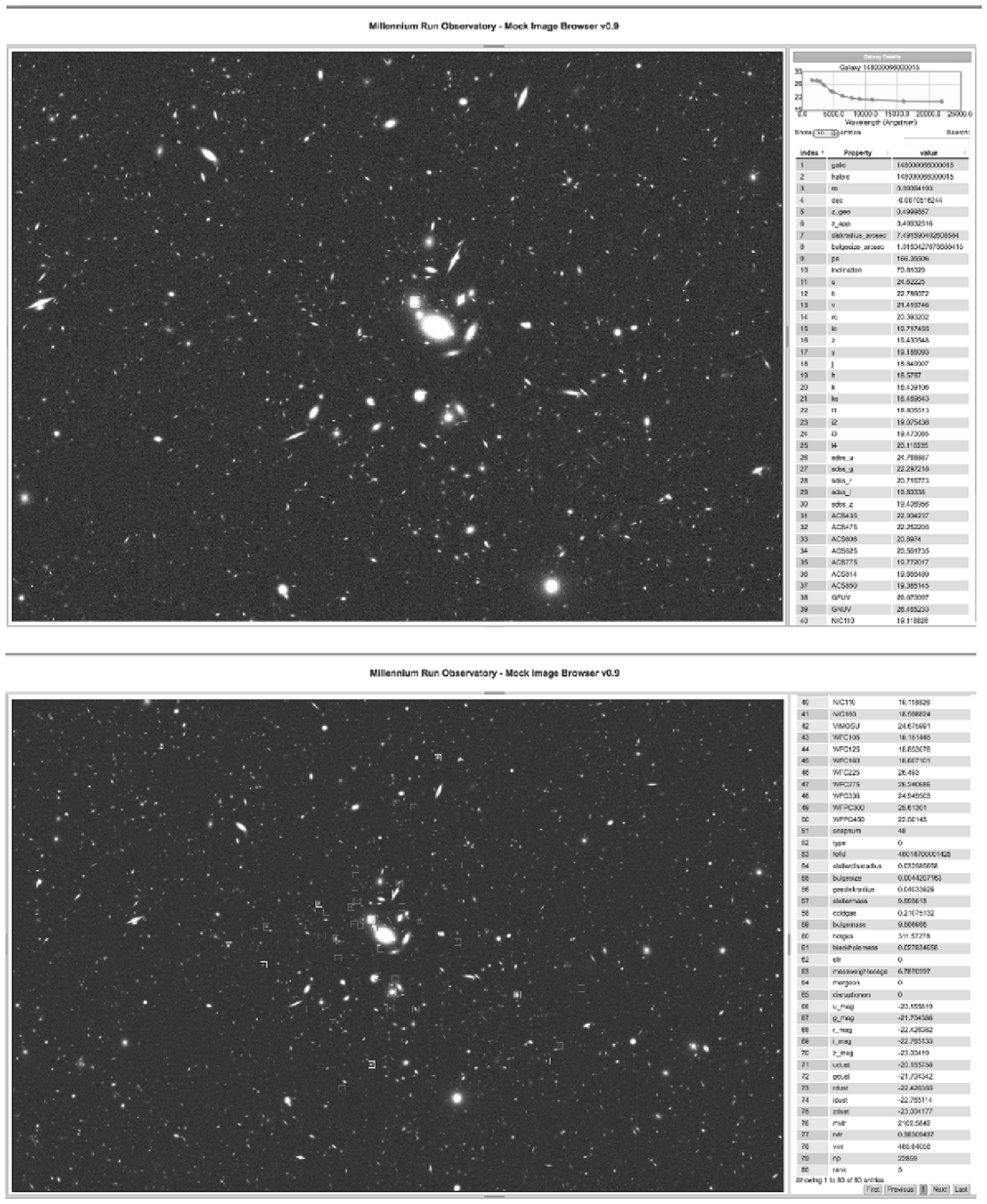}
\caption{\label{fig:mrobs}Screenshots of the MRObs v0.9 image browser
  available online. Top panel: basic view of the browser showing a
  small region of our synthetic HST GOODS observation (the particular
  colour image pyramid shown here is composed of the filters
  $V_{606},i_{775},z_{850}$). Users can pan around and zoom the synthetic observation,
  and directly query the MRDB by clicking on a galaxy. Information
  about the selected object (marked by a white square) is retrieved
  from the MRDB, and displayed in the information panel on the
  right-hand side of the screen. The MRObs shows a broad-band spectrum
  of the object, as well as about one hundred attributes of this
  object retrieved from the MRDB (e.g., size, SFR, stellar mass, age,
  redshift, magnitudes, black hole mass and dark matter halo virial
  mass and radius). Bottom panel: One can highlight all galaxies
  belonging to the same FOF group as the selected galaxy. In this
  case, the selected galaxy is the central galaxy of a galaxy group at
  $z\approx0.5$ (red squares: galaxies that are orphan (type 2)
  galaxies of the central halo; yellow squares: galaxies that are
  satellite (type 1) galaxies of the central halo; white square: the
  central (type 0) galaxy).}
\end{figure*}

\subsubsection{User interface of the MRObs browser}

The MRObs Browser offers the user the choice of a large number of
image pyramids, based on sets of different mock images, for a variety
of virtual telescopes and with different wavelength bands.  Each mock
image is derived from a light cone stored in the MRDB, and the
MRObs Browser allows interactive querying of these cones. 

Screen-shots of an HST simulation viewed through he MRObs Browser are
shown in Fig.~\ref{fig:mrobs}.  Clicking the image leads to an SQL
query being submitted to the database that searches for the nearest
galaxy to the selected (virtual) sky position, up to a maximum radius
(currently 1\arcsec).  If a galaxy is found a large amount of
information is retrieved and displayed in a table on the screen next
to the image, as shown on the right hand side of the screenshots in
Fig.~\ref{fig:mrobs}. The selected galaxy is indicated on the image
with a little white square (top panel). The table includes information
of the galaxy on the light cone, such as redshift, apparent sizes and
luminosities in up to 40 bands.  The observed-frame SED is shown in
graphical form above the table.  It also includes, through the linking
of the light-cone galaxy to the underlying semi-analytical galaxy
catalogues, information about physical parameters such as stellar
mass, gas mass, metallicities and rest frame magnitudes in the SDSS
bands. The information also includes details about the original
dark-matter subhalo and friends-of-friends group the galaxy belongs
to.  The latter information can in its turn be used to search for all
other galaxies in the image that belong to the same FOF group as the
selected galaxy. In the bottom panel, the positions of all galaxies
that were retrieved are indicated on the screen. The structure turns
out to be a galaxy group at $z\approx0.5$. 

The query capabilities of the MRObs Browser will be extended over time
and will be tied to the plain SQL query capabilities of the MRDB.

\section{Future developments}
\label{sec:future}

Future refinements of the modeling recipes in the MRObs currently in
preparation include the calculation of extinctions based on the
physically motivated inclinations of galaxies
(\S\ref{sec:angles}), as well as the inclusion of strong
gravitational lensing based on the distribution of dark matter halos
along each lightcone. Although the current simulations span the
observed far-UV to the observed near-IR, improvements in the
dust-modeling should eventually allow us to construct images out to
far-infrared and sub-mm wavelengths. Future simulated surveys will be
made available through the MRObs web portal. Ultimately, fully
interactive, online versions of the SAM, the lightcone generator, and
the telescope simulator will allow users of the MRObs even greater flexibility in
performing the simulations that best suit their needs.

\section{Summary}

In order to make predictions in the observational plane and to allow
unbiased comparisons between semi-analytic models and real data, we
have developed the Millennium Run Observatory (MRObs), a new virtual
telescope facility that can be used to simulate realistic
observational data based on the semi-analytic model galaxy catalogues
associated with the dark matter MR simulations. The MRObs
allows one to produce {\it scientific} image data sets in
\texttt{.fits} format. These artificial data can be analyzed using the standard tools
routinely used for analyzing real observations, allowing a relatively
unbiased comparison between SAMs and real data. This contrasts with
previous studies that compare highly idealized SAM predictions to
observational data. The new technique will help -- but is by no means
limited -- to:

\begin{itemize}

\item{Extend the MR simulations project approach by producing data
    products directly corresponding to observations, namely
    synthetic images and extracted source catalogs}

\item{Aid theorists in testing analytical models against observations}

\item{Aid observers in making detailed predictions for observations
    and better analyses of observational data}

\item{Allow the community to subject the models to new kinds of tests}

\item{Allow detailed comparisons with synthetic observations produced
    by other groups performing cosmological simulations}

\item{Allow calibration of observational analysis methods  by
    making available synthetic data for which the entire underlying
    physical `reality' is known}.

\item{Extend the realism with which semi-analytic
    models can address questions such as what is the probability that a $z\sim10$ galaxy
    will be detected within a particular observational data set?}

\item{Allow one to explore the uniqueness of certain physical models that are
  based on the analysis of astronomical images, as different models or different parameter sets may produce
  indistinguishable synthetic observations.}

\item{Provide a framework for future virtual theoretical
    observatories}

\end{itemize}

One of the great advantages provided by our extended modeling approach
is that for the synthetic observations produced by the MRObs, the
physical properties (e.g., dark matter halo mass, SFR, stellar mass,
size, redshift) and photometric properties (e.g., magnitudes and
colours) of every galaxy are precisely known, in contrast to real
observations where one does not know the exact or
`true' answer. This makes the MRObs an ideal facility for
calibrating many of the measurement and analysis techniques that are
applied to real observations. The MRObs will allow observers and
theorists to approach a problem from different directions with freedom
in deciding where to meet.

We have introduced a modified lightcone technique that allows us to
create lightcones aimed at selected objects or regions placed at any
desired position or orientation. The new technique is useful for
extending the range of questions that can be asked of the MRObs, such
as what would be the appearance of a particular galaxy cluster at
$z\sim1$? What does this same cluster look like at $z\sim6$ or at
$z=0$? How is the interpretation of observations of such structures
affected by viewing angle or chance superpositions? Special cones
centred on clusters at a range of redshifts have been added to the
MRDB for studies of cluster evolution.

Attenuation by the IGM is applied to the lightcones statistically
using the baseline model from \citet{madau95}, as well as two more
recent implementations based on Monte Carlo modeling of the IGM by
\citet{harrison11}. Our IGM attenuation tables have also been added to
the MRDB such that they can be used to apply `on the fly' IGM
absorption corrections to lightcones (see the MRObs URL for a tutorial on
how to apply the IGM absorption corrections to the lightcones from
\citet{henriques12} also available in the MRDB). This is essential for
making comparisons with high redshift observations.

In order to allow the community to use our predictions as the basis
for other mock observation experiments, we provide not only our final
image products, but also the intermediate steps such as the input
object lists and the pre-observation model images.

In order to introduce the communities of theoretically and
observationally inclined researchers to the `added value' of the
MRObs modeling approach, we have provided the following four example
user cases:\\

(1) We compared the galaxy number counts in the CANDELS/UDS survey
with the predicted counts taken directly from the semi-analytic
lightcone and with the counts extracted from synthetic CANDELS images
(Fig. \ref{fig:candelscounts} and \S\ref{sec:counts}). Interestingly,
the counts recovered from the synthetic images are lower than those
predicted by the lightcone that was used to construct the synthetic
observation, suggesting that the discrepancy between semi-analytic
model predictions and observations may be smaller than previously
claimed. The implications of this will be detailed in a followup paper
(Overzier et al., in prep.).
\\

(2) We simulated images of galaxy clusters seen with SDSS and HST at a
range of redshifts (Figs. \ref{fig:sdss}--\ref{fig:hizcluster} and
\S\ref{sec:examples_clusters}). We also showed synthetic images of the
same galaxy cluster at $z=1.1$ from three different directions,
illustrating that orientation effects can be important when
interpreting the visual appearance of large-scale structure at high
redshift.
\\

(3) We showed how the MRObs allows one to study the detailed
structural properties of semi-analytic galaxies in synthetic images
(Fig. \ref{fig:msfr} and \S\ref{sec:msfr}). In these synthetic images
one can measure colours, sizes, bulge-to-disk ratios and profile shapes
using standard observational techniques. The outcome of these
measurements can then be compared to the intrinsic values provided by
the MRObs, or to measurements performed on real galaxies.
\\

(4) We showed how the MRObs images can be used to search for high
redshift dropout galaxies in a manner that is directly analogous to
that used for real high redshift doprout searches
(Fig. \ref{fig:dropouts} and \S\ref{sec:dropouts}). This enables a
much more realistic comparison with the data, and allows us to assess
how well we are able to derive the intrinsic physical properties from
the observations.
\\

Extending the successful open-access approach of the MR project, we
make available new data products for use by the community. As part of
our first data release, we have produced simulated data that emulates
a number of key surveys, including SDSS, CFHT-LS (Wide and Deep),
GOODS, HUDF, GOODS/ERS, and CANDELS (UDS, COSMOS and GOODS-S). The
data sets are modeled using two different cosmologies (WMAP1 and
WMAP7), two spectral synthesis models (BC03 and M05), and three IGM
absorption models (MADAU, MEIKSIN, and INOUE-IWATA). In specific
cases, we provide synthetic images that have the exact same geometric
and identical noise properties as the reference survey. The MRObs data
can furthermore be explored using an online image browser that allows
users to interactively explore the available mock observations. The
browser graphically links objects (galaxies) in the synthetic images
to various types of information available in catalogs in the MRDB. For
each synthetic galaxy, this information includes the physical
properties of their dark matter halos, the intrinsic properties of the
galaxy itself, the absolute and apparent photometric properties, and
the observed properties recovered from the synthetic images using
SExtractor.

The public data and the MRObs browser can be accessed at the following URL:
\\

\centerline{\url{http://galformod.mpa-garching.mpg.de/mrobs/}}

\medskip
In conclusion, the MRObs allows us to study our  simulated
universes through the eyes of our telescopes. We
hope that the methods and data presented in this paper will encourage
others to take advantage of the new opportunities offered by this
approach.

\section*{Acknowledgments}

We thank Rychard Bouwens, Yi-Kuan Chiang, Marijn Franx, Guinevere
Kauffmann, Jaewon Kim, Ben Metcalfe, and Masami Ouchi for helpful
discussions.  We thank Chris Harrison for making his code
IGMTransmission publicly available and for his clarifications. We
thank Russell Ryan, Seth Cohen, Rogier Windhorst and other members of
the GOODS ERS team for discussions and supplying us with their
empirical PSFs. We thank Daniel Harsono for providing the HST/ACS
images of Cl0024 used in Fig. \ref{fig:cl0024}. We thank Matthias
Egger for his excellent technical support. The Millennium Simulation
databases used in this paper and the web application providing online
access to them were constructed as part of the activities of the
German Astrophysical Virtual Observatory. This work was supported by
Advanced Grant 246797 `GALFORMOD' from the European Research
Council. The authors acknowledge the Texas Advanced Computing Center
(TACC) at The University of Texas at Austin for providing the 307
Mpixel tiled-display system (Stallion) and the 12.4 Mpixel multi-touch
display (Lasso) visualization resources that have contributed to
developing the methods and results reported within this paper. We
thank Ben Urick and Brandt Westing of TACC for their time and expert
support.

\begin{table}
\begin{scriptsize}
\begin{center}
\caption{\label{tab:filters}List of filters currently available in the
  MRObs and the \texttt{FILTERID} under which they are known in the MR
  database.}
\begin{tabular}{ll|ll}
\hline
\hline
\texttt{FILTERID} & Description  & \texttt{FILTERID} & Description \\
  & & & \\
\hline
U 	  & Johnson $U$& ACS435 &HST/ACS-WFC F435W\\
B         &Johnson $B$& ACS475 &HST/ACS-WFC F475W\\	
V 	  &Johnson $V$& ACS606 &HST/ACS-WFC F606W\\
Rc 	  &Cousins $R$& ACS625 &HST/ACS-WFC F625W\\
Ic 	  &Cousins $I$& ACS775 &HST/ACS-WFC F775W\\
Z 	  &UKIDDS $Z$& ACS814 &HST/ACS-WFC F814W\\
Y 	  &UKIDDS $Y$& ACS850 &HST/ACS-WFC F850LP\\
J 	  &UKIDDS $J$&GFUV   &GALEX FUV\\
H 	  &UKIDDS $H$& GNUV   &GALEX NUV\\
K 	  &Johnson $K$& NIC110&HST/NICMOS F110W\\
Ks 	  &UKIDDS $K_s$& NIC160 &HST/NICMOS F160W\\
i1 	  &Spitzer/IRAC channel 1& VIMOSU &VLT/VIMOS $U$\\
i2 	  &Spitzer/IRAC channel 2& WFC105 &HST/WFC3-IR F105W\\
i3 	  &Spitzer/IRAC channel 3& WFC125 &HST/WFC3-IR F125W\\
i4 	  &Spitzer/IRAC channel 4& WFC160 &HST/WFC3-IR F160W\\
SDSS\_u &SDSS $u^\prime$ &WFC225  &HST/WFC3-UVIS F225W\\ 
SDSS\_g &SDSS $g^\prime$ &WFC275  &HST/WFC3-UVIS F275W\\
SDSS\_r &SDSS $r^\prime$ &WFC336  &HST/WFC3-UVIS F336W\\
SDSS\_i &SDSS $i^\prime$ &WFPC300 &HST/WFPC2 F300W\\
SDSS\_z&SDSS $z^\prime$  &WFPC450 &HST/WFPC2 F450W\\
\hline
\hline
\end{tabular}
\end{center}
\end{scriptsize}
\end{table}

\appendix

\section{Intergalactic medium corrections}
\label{sec:appendix}

The MEIKSIN and INOUE-IWATA models \citep{meiksin06,inoue08} discussed
in \S\ref{sec:igm} are both based on a Monte Carlo approach that
distributes LLSs chosen from a redshift distribution $dN/dz$ and an
optical depth distribution $dN/d\tau$ (both constrained by
observations), and averages over the IGM transmission measured along a
large number of random lines of sight. The IGM effective optical depth
$\tau_e$ at observed wavelength $\lambda$ is taken to be the sum of
the Lyman continuum (LC) optical depth due to LLSs, the optically thin IGM, and the \lya\
forest as follows:

\begin{equation}
  \tau_{e}(\lambda) = \tau_{\mathrm{LC}}^{\mathrm{LSS}}(\lambda) + \tau_{\mathrm{LC}}^{\mathrm{IGM}}(\lambda) + \sum_i \bar\tau_i(\lambda).
\end{equation}

The optical depth due to photoelectric absorption by LLSs along the
line of sight to a source at redshift $z$ is given by
\begin{footnotesize}
\begin{eqnarray}
\lefteqn{\tau_{\mathrm{LC}}^{\mathrm{LSS}}(\lambda)=}\nonumber\\
&&\int_{z_L}^z dz^\prime \int_1^{\infty} d\tau_L \frac{\partial^2N}{\partial\tau_L\partial z^\prime}\left( 1-\exp\left[-\tau_L{\left( \frac{1+z_L}{1+z^\prime} \right)}^3 \right] \right),\nonumber\\
\end{eqnarray}
\end{footnotesize}

\noindent
with $z_L=\lambda/912~\AA\ -1$, and
%$\frac{\partial^2N}{\partial\tau_L\partial z^\prime}$
$\partial^2N/\partial\tau_L\partial z^\prime$ the number of
absorbers per unit redshift and optical depth. In the MEIKSIN model
the LLSs are randomly drawn from the distributions
\begin{eqnarray}
\frac{dN}{dz}&=&0.25(1+z)^{1.5}\\
\frac{dN}{d\tau_L}&\propto&\tau_L^{-1.5}, 
\end{eqnarray}

\noindent
while INOUE-IWATA assumes
\begin{eqnarray}
\frac{dN}{dz}=\frac{A}{688.4} \left\{
\begin{array}{lll}
 \left(\frac{1+z}{1+z_1}\right)^{\gamma_1}&&   (0<z\le z_1)\\
 \left(\frac{1+z}{1+z_1}\right)^{\gamma_2} &&  (z_1<z\le z_2)\\
 \left(\frac{1+z_2}{1+z_1}\right)^{\gamma_2}\left(\frac{1+z}{1+z_2}\right)^{\gamma_3}&&   (z>z_2)\\
\end{array}
\right.
\end{eqnarray}
\begin{equation}
\frac{dN}{d\tau_L}\propto\tau_L^{-1.3},
\end{equation}
with $A=400$, $\gamma_1=0.2$, $\gamma_2=2.5$, $\gamma_3=4$, $z_1=1.2$, and $z_2=4$. The mean IGM transmission due to the LLSs typically stabilizes after averaging over $\sim$10,000 random lines of sight. Following \citet{meiksin06} and \citet{harrison11}, both models include a static contribution from the diffuse or optically thin IGM and the \lya\ forest:
\begin{footnotesize}
\begin{eqnarray}
\tau_{\mathrm{LC}}^{\mathrm{IGM}}(\lambda)& =& C(1+z_L)^{4.4}\left[\frac{1}{(1+z_L)^{3/2}}-\frac{1}{(1+z)^{3/2}}\right],\\
\bar\tau_n(\lambda)&\equiv& -\ln\langle \exp(-\tau_n(\lambda))\rangle,
\end{eqnarray}
\end{footnotesize}
with $C=0.07553$ and the Lyman transitions $n\rightarrow1$ up to $n=31$ are included.  

For completeness, we also give the MADAU modeling approximation \citep{madau95}:
\begin{footnotesize}
\begin{eqnarray}
\tau_{\mathrm{e}}(\lambda) &=& \tau_{\mathrm{LC}}(\lambda) + \sum_{j=1,i} A_j\left(\frac{\lambda}{\lambda_j}\right)^{3.46}\\
\tau_{\mathrm{LC}}(\lambda)&\simeq&0.25x_\mathrm{c}^3(x_{\mathrm{em}}^{0.46}-x_\mathrm{c}^{0.46})+9.4x_\mathrm{c}^{1.5}(x_{\mathrm{em}}^{0.18}-x_\mathrm{c}^{0.18})\nonumber\\
&-&0.7x_\mathrm{c}^3(x_{\mathrm{c}}^{-1.32}-x_{\mathrm{em}}^{-1.32})-0.023(x_{\mathrm{em}}^{1.68}-x_\mathrm{c}^{1.68}),
\nonumber\\
\label{eq:madau}
\end{eqnarray}
\end{footnotesize}

\noindent 
with $A_j=(0.0036,0.0017,0.0012,0.00093)$ for
$\lambda_j=(1216,1016,973,950$)~\AA, $x_\mathrm{c}\equiv 1+z_\mathrm{c}$, $x_{\mathrm{em}}\equiv
1+z_{\mathrm{em}}$, $z_\mathrm{c}=\lambda/\lambda_L-1$, and $z_{\mathrm{em}}$ is the redshift of
the source. Eq. \ref{eq:madau} is the approximation given for Eq. 16
in \citet[][see footnote 3 in that paper]{madau95}.

\bsp

\label{lastpage}


\begin{thebibliography}{}
\bibitem[Angulo \& White(2010)]{angulo10} Angulo, R.~E., \& White, S.~D.~M.\ 2010, \mnras, 405, 143 
\bibitem[Angulo et al.(2012)]{angulo12} Angulo, R.~E., Springel, 
V., White, S.~D.~M., et al.\ 2012, arXiv:1203.3216 
\bibitem[Baugh et al.(1998)]{baugh98} Baugh, C.~M., Cole, S., Frenk, C.~S., \& Lacey, C.~G.\ 1998, \apj, 498, 504 
\bibitem[Baugh et al.(2005)]{baugh05} Baugh, C.~M., Lacey, 
C.~G., Frenk, C.~S., et al.\ 2005, \mnras, 356, 1191 
\bibitem[Benson et al.(2003)]{benson03} Benson, A.~J., Bower, 
R.~G., Frenk, C.~S., et al.\ 2003, \apj, 599, 38 
\bibitem[Bertin \& Arnouts(1996)]{bertin96} Bertin, E., \& Arnouts, S.\ 1996, \aaps, 117, 393 
\bibitem[Bertin(2009)]{bertin09} Bertin, E.\ 2009, \memsai, 80, 422 
\bibitem[Bertin(2012)]{bertin12} Bertin, E.\ 2012, in Proceedings of ADASS XXI P.Ballester and D.Egret editors, in press 
\bibitem[Bertone et al.(2007)]{bertone07}Bertone, S., DeLucia, G., \& Thomas, P.~A.,\ 2007, \mnras, 379, 1143
\bibitem[Bershady et al.(1999)]{bershady99} Bershady, M.~A., Charlton,  J.~C., \& Geoffroy, J.~M.\ 1999, \apj, 518, 103 
\bibitem[Blaizot et al.(2004)]{blaizot04} Blaizot, J., Guiderdoni, B., Devriendt, J.~E.~G., et al.\ 2004, \mnras, 352, 571 
\bibitem[Blaizot et al.(2005)]{blaizot05} Blaizot, J., Wadadekar, Y., Guiderdoni, B., et al.\ 2005, \mnras, 360, 159 
\bibitem[Bouch{\'e} et al.(2010)]{bouche10} Bouch{\'e}, N., Dekel, A., Genzel, R., et al.\ 2010, \apj, 718, 1001 
\bibitem[Bouwens et al.(1999)]{bouwens99} Bouwens, R., Cay{\'o}n, L., \& Silk, J.\ 1999, \apj, 516, 77 
\bibitem[Bouwens et al.(2003)]{bouwens03} Bouwens, R.~J., Illingworth, G.~D., Rosati, P., et al.\ 2003, \apj, 595, 589 
\bibitem[Bouwens et al.(2006)]{bouwens06} Bouwens, R.~J., Illingworth, G.~D., \& Magee, D.~K.\ 2006, Astronomical Data Analysis Software and Systems XV, 351, 145 
\bibitem[Bower et al.(2006)]{bower06} Bower, R.~G., Benson, 
A.~J., Malbon, R., et al.\ 2006, \mnras, 370, 645 
\bibitem[Boylan-Kolchin et al.(2009)]{boylan09} Boylan-Kolchin, M., Springel, V., White, S.~D.~M., Jenkins, A., 
\& Lemson, G.\ 2009, \mnras, 398, 1150 
\bibitem[Bruzual \& Charlot(2003)]{bc03} Bruzual, G., \& Charlot, S.\ 2003, \mnras, 344, 1000
\bibitem[Bruzual(2007)]{bruzual07} Bruzual, G.\ 2007, From Stars to Galaxies: Building the Pieces to Build Up the Universe, 374, 303 
\bibitem[Cardelli et al.(1989)]{cardelli89} Cardelli, J.~A., Clayton, G.~C., \& Mathis, J.~S.\ 1989, \apj, 345, 245 
\bibitem[Carlson \& White(2010)]{carlson10} Carlson, J., \& White, M.\ 2010, \apjs, 190, 311 
\bibitem[Casertano et al.(2000)]{casertano00} Casertano, S., de 
Mello, D., Dickinson, M., et al.\ 2000, \aj, 120, 2747 
\bibitem[Cattaneo et al.(2005)]{cattaneo05} Cattaneo, A., Blaizot, J., Devriendt, J., \& Guiderdoni, B.\ 2005, \mnras, 364, 407 
\bibitem[Cattaneo et al.(2006)]{cattaneo06} Cattaneo, A., Dekel, A., Devriendt, J., Guiderdoni, B., \& Blaizot, J.\ 2006, \mnras, 370, 1651 
\bibitem[Ceverino et al.(2010)]{ceverino10} Ceverino, D., Dekel, A., \& Bournaud, F.\ 2010, \mnras, 404, 2151 
\bibitem[Chilingarian et al.(2010)]{chilingarian10} Chilingarian, I.~V., Di Matteo, P., Combes, F., Melchior, A.-L., \& Semelin, B.\ 2010, \aap, 518, A61 
\bibitem[Cirasuolo et al.(2010)]{cirasuolo10} Cirasuolo, M., McLure, R.~J., Dunlop, J.~S., et al.\ 2010, \mnras, 401, 1166 
\bibitem[Cole et al.(1994)]{cole94} Cole, S., Aragon-Salamanca, A., Frenk, C.~S., Navarro, J.~F., \& Zepf, S.~E.\ 1994, \mnras, 271, 781 
\bibitem[Cole et al.(1998)]{cole98} Cole, S., Hatton, S., Weinberg, D.~H., \& Frenk, C.~S.\ 1998, \mnras, 300, 945 
\bibitem[Cole et al.(2000)]{cole00}Cole, S., Lacey, C.~G., Baugh, C.~M., \& Frenk, C.~S.,\ 2000, \mnras, 319, 168
\bibitem[Connolly et al.(2010)]{connolly10} Connolly, A.~J., Peterson, J., Jernigan, J.~G., et al.\ 2010, \procspie, 7738,  
\bibitem[Conroy et al.(2009)]{conroy09} Conroy, C., Gunn, J.~E., \& White, M.\ 2009, \apj, 699, 486 
\bibitem[Croton et al.(2006)]{croton06} Croton, D.~J., Springel, V., White, S.~D.~M., et al.\ 2006, \mnras, 365, 11 
\bibitem[Cucciati et al.(2012)]{cucciati12} Cucciati, O., De Lucia, G., Zucca, E., et al.\ 2012, arXiv:1205.1517 
\bibitem[De Lucia \& Blaizot(2007)]{delucia07} De Lucia, G., \& Blaizot, J.\ 2007, \mnras, 375, 2 
\bibitem[Davis et al.(1982)]{davis82} Davis, M., Huchra, J., Latham, D.~W., \& Tonry, J.\ 1982, \apj, 253, 423 
\bibitem[Davis et al.(1985)]{davis85} Davis, M., Efstathiou, G., Frenk, C.~S., \& White, S.~D.~M.\ 1985, \apj, 292, 371 
\bibitem[Dekel et al.(2009)]{dekel09} Dekel, A., Sari, R., \& Ceverino, D.\ 2009, \apj, 703, 785 
\bibitem[Diaferio et al.(1999)]{diaferio99} Diaferio, A., Kauffmann, G., Colberg, J.~M., \& White, S.~D.~M.\ 1999, \mnras, 307, 537 
\bibitem[Dobke et al.(2010)]{dobke10} Dobke, B.~M., Johnston, 
D.~E., Massey, R., et al.\ 2010, \pasp, 122, 947 
\bibitem[Dutton et al.(2010)]{dutton10} Dutton, A.~A., van den Bosch, F.~C., \& Dekel, A.\ 2010, \mnras, 405, 1690 
\bibitem[Eminian et al.(2008)]{eminian08} Eminian, C., Kauffmann, G., Charlot, S., et al.\ 2008, \mnras, 384, 930 
\bibitem[Erben et al.(2001)]{erben01} Erben, T., Van Waerbeke, L., Bertin, E., Mellier, Y., \& Schneider, P.\ 2001, \aap, 366, 717 
\bibitem[Finkelstein et al.(2010)]{finkelstein10} Finkelstein, S.~L., Papovich, C., Giavalisco, M., et al.\ 2010, \apj, 719, 1250 
\bibitem[Finlator et al.(2011)]{finlator11} Finlator, K., Oppenheimer, B.~D., \& Dav{\'e}, R.\ 2011, \mnras, 410, 1703 
\bibitem[Fioc \& Rocca-Volmerange(1997)]{fioc97} Fioc, M., \& Rocca-Volmerange, B.\ 1997, \aap, 326, 950 
\bibitem[Forero-Romero et al.(2007)]{forero07} Forero-Romero, J.~E., Blaizot, J., Devriendt, J., van Waerbeke, L., \& Guiderdoni, B.\ 2007, \mnras, 379, 1507 
\bibitem[Fruchter et al.(2009)]{fruchter09} Fruchter, A., Sosey, M., et al.\ 2009, `The MultiDrizzle Handbook', version 3.0, (Baltimore, STScI)
\bibitem[Gibson et al.(2011)]{gibson11} Gibson, R.~R., Ahmad, Z., Bankert, J., et al.\ 2011, Astronomical Data Analysis Software and Systems XX, 442, 329 
\bibitem[Girardi et al.(2005)]{girardi05} Girardi, L., Groenewegen, M.~A.~T., Hatziminaoglou, E., \& da Costa, L.\ 2005, \aap, 436, 895 
\bibitem[Gonz{\'a}lez et al.(2011)]{gonzalez11} Gonz{\'a}lez, V., Labb{\'e}, I., Bouwens, R.~J., et al.\ 2011, \apjl, 735, L34 
\bibitem[Grogin et al.(2011)]{grogin11} Grogin, N.~A., Kocevski, 
D.~D., Faber, S.~M., et al.\ 2011, \apjs, 197, 35 
\bibitem[Guo \& White(2009)]{guo09} Guo, Q., \& White, S.~D.~M.\ 2009, \mnras, 396, 39 
\bibitem[Guo et al.(2011)]{guo11} Guo, Q., White, S., Boylan-Kolchin, M., et al.\ 2011, \mnras, 413, 101 
\bibitem[Guo et al.(2012)]{guo12} Guo, Q., White, S., Angulo, R.~E., et al.\ 2012, \mnras, submitted (arXiv:1206.0052) 
\bibitem[Hamana et al.(2001)]{hamana01} Hamana, T., Colombi, S., \& Suto, Y.\ 2001, \aap, 367, 18 
\bibitem[Harrison et al.(2011)]{harrison11} Harrison, C.~M., Meiksin,
  A., \& Stock, D.\ 2011, arXiv:1105.6208 
\bibitem[Harsono \& De Propris(2009)]{harsono09} Harsono, D., \& De Propris, R.\ 2009, \aj, 137, 3091 
\bibitem[Hatton et al.(2003)]{hatton03} Hatton, S., Devriendt, J.~E.~G., Ninin, S., et al.\ 2003, \mnras, 343, 75 
\bibitem[Henriques et al.(2011)]{henriques11} Henriques, B., Maraston, C., Monaco, P., et al.\ 2011, \mnras, 415, 3571 
\bibitem[Henriques et al.(2012)]{henriques12} Henriques, B.~M.~B., White, S.~D.~M., Lemson, G., et al.\ 2012, \mnras, 421, 2904 
\bibitem[Heymans et al.(2006)]{heymans06} Heymans, C., Van Waerbeke, L., Bacon, D., et al.\ 2006, \mnras, 368, 1323 
\bibitem[Inoue \& Iwata(2008)]{inoue08} Inoue, A.~K., \& Iwata, I.\ 2008, \mnras, 387, 1681 
\bibitem[Jee et al.(2007)]{jee07} Jee, M.~J., Ford, H.~C., 
Illingworth, G.~D., et al.\ 2007, \apj, 661, 728 
\bibitem[Jenkins et al.(1998)]{jenkins98} Jenkins, A., Frenk, C.~S., Pearce, F.~R., et al.\ 1998, \apj, 499, 20 
\bibitem[Jonsson et al.(2010)]{jonsson10} Jonsson, P., Groves, B.~A., \& Cox, T.~J.\ 2010, \mnras, 403, 17 
\bibitem[Jonsson et al.(2006)]{jonsson06} Jonsson, P., Cox, T.~J., Primack, J.~R., \& Somerville, R.~S.\ 2006, \apj, 637, 255 
\bibitem[Kang et al.(2005)]{kang05}Kang, X., Jing, Y.~P., Mo, H.~J., \& B{\"o}rner, G.,\ 2005, \apj, 631, 21
\bibitem[Kauffmann \& White(1993)]{kauffmannwhite93} Kauffmann, G., \& White, S.~D.~M.\ 1993, \mnras, 261, 921 
\bibitem[Kauffmann et al.(1993)]{kauffmann93} Kauffmann, G., White, S.~D.~M., \& Guiderdoni, B.\ 1993, \mnras, 264, 201 
\bibitem[Kauffmann et al.(1999)]{kauffmann99}Kauffmann, G., Colberg, J.~M., Diaferio, A., \& Wite, S.~D.~M.\ 1999, \mnras, 303, 188
\bibitem[Kauffmann \& Haehnelt(2000)]{kauffmann00} Kauffmann, G., \& Haehnelt, M.\ 2000, \mnras, 311, 576 
\bibitem[Kere{\v s} et al.(2009)]{keres09} Kere{\v s}, D., Katz, N., Fardal, M., Dav{\'e}, R., \& Weinberg, D.~H.\ 2009, \mnras, 395, 160 
\bibitem[Kitzbichler \& White(2006)]{kitzbichler06} Kitzbichler, M.~G., \& White, S.~D.~M.\ 2006, \mnras, 366, 858 
\bibitem[Kitzbichler \& White(2007)]{kitzbichler07} Kitzbichler, M.~G., \& White, S.~D.~M.\ 2007, \mnras, 376, 2 
\bibitem[Klypin et al.(2011)]{klypin11} Klypin, A.~A., Trujillo-Gomez, S., \& Primack, J.\ 2011, \apj, 740, 102 
\bibitem[Koekemoer et al.(2003)]{koekemoer03} Koekemoer, A.~M., Fruchter, A.~S., Hook, R.~N., 
\& Hack, W.\ 2003, The 2002 HST Calibration Workshop : Hubble after the Installation of the ACS and the NICMOS Cooling System, Proceedings of a Workshop held at the Space Telescope Science Institute, Baltimore, Maryland, October 17 and 18, 2002.~ Edited by Santiago Arribas, Anton Koekemoer, and Brad Whitmore.~Baltimore, MD: Space Telescope Science Institute, 2003., p.337, 337 
\bibitem[Koekemoer et al.(2011)]{koekemoer11} Koekemoer, A.~M., 
Faber, S.~M., Ferguson, H.~C., et al.\ 2011, \apjs, 197, 36 
\bibitem[Kriek et al.(2006)]{kriek06} Kriek, M., van Dokkum, P.~G., Franx, M., et al.\ 2006, \apjl, 649, L71 
\bibitem[Kriek et al.(2010)]{kriek10} Kriek, M., Labb{\'e}, I., Conroy, C., et al.\ 2010, \apjl, 722, L64 
\bibitem[Lacey \& Cole(1994)]{laceycole1994}Lacey, C. \& Cole, S.\ 1994, \mnras, 271, 676
\bibitem[Lee et al.(2010)]{lee10} Lee, S.-K., Ferguson, H.~C., Somerville, R.~S., Wiklind, T., \& Giavalisco, M.\ 2010, \apj, 725, 1644 
\bibitem[Leitherer \& Heckman(1995)]{leitherer95} Leitherer, C., \& Heckman, T.~M.\ 1995, \apjs, 96, 9 
\bibitem[Lemson \& Virgo Consortium(2006)]{lemson06} Lemson, G., \& Virgo Consortium, \ 2006, arXiv:astro-ph/0608019 
\bibitem[Lotz et al.(2008)]{lotz08} Lotz, J.~M., Jonsson, P., Cox, T.~J., \& Primack, J.~R.\ 2008, \mnras, 391, 1137 
\bibitem[Lotz et al.(2010)]{lotz10} Lotz, J.~M., Jonsson, P., Cox, T.~J., \& Primack, J.~R.\ 2010, \mnras, 404, 575 
\bibitem[Madau(1995)]{madau95} Madau, P.\ 1995, \apj, 441, 18 
\bibitem[Maraston(1998)]{maraston98} Maraston, C.\ 1998, \mnras, 300, 872 
\bibitem[Maraston(2005)]{maraston05} Maraston, C.\ 2005, \mnras, 362, 799 
\bibitem[Maraston et al.(2006)]{maraston06} Maraston, C., Daddi, E., Renzini, A., et al.\ 2006, \apj, 652, 85 
\bibitem[Maraston et al.(2010)]{maraston10} Maraston, C., Pforr, J., Renzini, A., et al.\ 2010, \mnras, 407, 830 
\bibitem[Marchesini et al.(2010)]{marchesini10} Marchesini, D., Whitaker, K.~E., Brammer, G., et al.\ 2010, \apj, 725, 1277 
\bibitem[Marinoni et al.(2005)]{marinoni05} Marinoni, C., Le F{\`e}vre, O., Meneux, B., et al.\ 2005, \aap, 442, 801 
\bibitem[Meiksin(2006)]{meiksin06} Meiksin, A.\ 2006, \mnras, 365, 807 
\bibitem[Mo \& White(1996)]{mo96} Mo, H.~J., \& White, S.~D.~M.\ 1996, \mnras, 282, 347 
\bibitem[Monaco et al.(2007)]{monaco07} Monaco, P., Fontanot, F., \& Taffoni, G.\ 2007, \mnras, 375, 1189 
\bibitem[Neistein \& Dekel(2008)]{neistein08} Neistein, E., \& Dekel, A.\ 2008, \mnras, 383, 615 
\bibitem[Neistein \& Weinmann(2010)]{neistein10} Neistein, E., \& Weinmann, S.~M.\ 2010, \mnras, 405, 2717 
\bibitem[Ouchi et al.(2005)]{ouchi05} Ouchi, M., Hamana, T., Shimasaku, K., et al.\ 2005, \apjl, 635, L117 
\bibitem[Overzier et al.(2006)]{overzier06} Overzier, R.~A., Bouwens, R.~J., Illingworth, G.~D., \& Franx, M.\ 2006, \apjl, 648, L5 
\bibitem[Overzier et al.(2009a)]{overzier09a} Overzier, R.~A., Guo, Q., Kauffmann, G., et al.\ 2009a, \mnras, 394, 577 
\bibitem[Overzier et al.(2009b)]{overzier09b} Overzier, R.~A., Shu, X., Zheng, W., et al.\ 2009b, \apj, 704, 548 
\bibitem[Papovich et al.(2011)]{papovich11} Papovich, C., Finkelstein, S.~L., Ferguson, H.~C., Lotz, J.~M., \& Giavalisco, M.\ 2011, \mnras, 412, 1123 
\bibitem[Pforr et al.(2012)]{pforr12} Pforr, J., Maraston, C., \& Tonini, C.\ 2012, \mnras, 422, 3285 
\bibitem[Pollo et al.(2006)]{pollo06} Pollo, A., Guzzo, L., Le F{\`e}vre, O., et al.\ 2006, \aap, 451, 409 
\bibitem[Prada et al.(2012)]{prada11} Prada, F., Klypin, A.~A., Cuesta, A.~J., Betancort-Rijo, J.~E., \& Primack, J.\ 2012, \mnras, 423, 3018 
\bibitem[Press \& Schechter(1974)]{press74} Press, W.~H., \& Schechter, P.\ 1974, \apj, 187, 425 
\bibitem[Robertson \& Bullock(2008)]{robertson08} Robertson, B.~E., \& Bullock, J.~S.\ 2008, \apjl, 685, L27 
\bibitem[Ruiz et al.(2011)]{ruiz11} Ruiz, A.~N., Padilla, N.~D., Dom{\'{\i}}nguez, M.~J., \& Cora, S.~A.\ 2011, \mnras, 1600 
\bibitem[Sheth et al.(2001)]{sheth01} Sheth, R.~K., Mo, H.~J., \& Tormen, G.\ 2001, \mnras, 323, 1 
\bibitem[Somerville \& Primack(1999)]{somerville99} Somerville, R.~S., \& Primack, J.~R.\ 1999, \mnras, 310, 1087 
\bibitem[Somerville \& Kolatt(1999)]{somervillekolatt99} Somerville, R.~S., \& Kolatt, T.~S.\ 1999, \mnras, 305, 1 
\bibitem[Somerville et al.(2001)]{somerville01} Somerville, R.~S., Primack, J.~R., \& Faber, S.~M.\ 2001, \mnras, 320, 504 
\bibitem[Somerville et al.(2008)]{somerville08} Somerville, R.~S., Hopkins, P.~F., Cox, T.~J., Robertson, B.~E., \& Hernquist, L.\ 2008, \mnras, 391, 481 
\bibitem[Somerville et al.(2012)]{somerville11} Somerville, R.~S., Gilmore, R.~C., Primack, J.~R., \& Dom{\'{\i}}nguez, A.\ 2012, \mnras, 423, 1992 
\bibitem[Sousbie et al.(2008)]{sousbie08} Sousbie, T., Courtois, H., Bryan, G., \& Devriendt, J.\ 2008, \apj, 678, 569 
\bibitem[Spergel et al.(2003)]{spergel03} Spergel, D.~N., Verde, L., Peiris, H.~V., et al.\ 2003, \apjs, 148, 175 
\bibitem[Springel et al.(2001)]{springel01} Springel, V., White, S.~D.~M., Tormen, G., \& Kauffmann, G.\ 2001, \mnras, 328, 726 
\bibitem[Springel et al.(2005)]{springel05} Springel, V., White, S.~D.~M., Jenkins, A., et al.\ 2005, \nat, 435, 629 
\bibitem[Teyssier et al.(2009)]{teyssier09} Teyssier, R., Pires, S., Prunet, S., et al.\ 2009, \aap, 497, 335 
\bibitem[Tinsley(1973)]{tinsley73} Tinsley, B.~M.\ 1973, \apj, 186, 35 
\bibitem[Tinsley(1980)]{tinsley80} Tinsley, B.~M.\ 1980, \fcp, 5, 287 
\bibitem[Tonini et al.(2009)]{tonini09} Tonini, C., Maraston, C., Devriendt, J., Thomas, D., \& Silk, J.\ 2009, \mnras, 396, L36 
\bibitem[Tonini et al.(2010)]{tonini10} Tonini, C., Maraston, C., Thomas, D., Devriendt, J., \& Silk, J.\ 2010, \mnras, 403, 1749 
\bibitem[de la Torre et al.(2011)]{delatorre11} de la Torre, S., Meneux, B., De Lucia, G., et al.\ 2011, \aap, 525, A125 
\bibitem[V{\'a}zquez \& Leitherer(2005)]{vazquez05} V{\'a}zquez, G.~A., \& Leitherer, C.\ 2005, \apj, 621, 695 
\bibitem[White \& Frenk(1991)]{white91} White, S.~D.~M., \& Frenk, C.~S.\ 1991, \apj, 379, 52 
\bibitem[Windhorst et al.(2011)]{windhorst11} Windhorst, R.~A., Cohen, S.~H., Hathi, N.~P., et al.\ 2011, \apjs, 193, 27 
\bibitem[Wuyts et al.(2009)]{wuyts09} Wuyts, S., Franx, M., Cox, T.~J., et al.\ 2009, \apj, 700, 799 
\end{thebibliography}
\end{document}